\documentclass[11pt]{article}
\usepackage{rotating} 
\usepackage{adjustbox,lipsum}
\usepackage{float}
\usepackage{float}
\usepackage{amssymb}
\usepackage{amsthm}
\usepackage{amsmath}
\usepackage{latexsym}
\usepackage{algorithm}
\usepackage[noend]{algpseudocode}
\usepackage{epsfig}
\usepackage{graphicx}
\usepackage{wasysym}
\usepackage{threeparttable}
\usepackage{natbib}
\usepackage{color}
\usepackage{epstopdf}
\usepackage{caption}
\usepackage{subcaption}
\usepackage{booktabs}

\newcommand{\ind}{\stackrel{\mathrm{ind}}{\sim}}

\usepackage{multirow}
\usepackage[breaklinks]{hyperref}

\usepackage{enumitem}
\usepackage{caption}
\usepackage{algorithm}

\def\TTiny{\fontsize{8pt}{8pt}\selectfont}
\def\boxit#1{\vbox{\hrule\hbox{\vrule\kern6pt
          \vbox{\kern6pt#1\kern6pt}\kern6pt\vrule}\hrule}}

\def\bse{\begin{eqnarray*}}
\def\ese{\end{eqnarray*}}
\def\be{\begin{eqnarray}}
\def\ee{\end{eqnarray}}
\def\bq{\begin{equation}}
\def\eq{\end{equation}}
\def\bse{\begin{eqnarray*}}
\def\ese{\end{eqnarray*}}

\usepackage{mathptmx}
\usepackage{bm}

 {\begin{list}{}%
         {\setlength{\leftmargin}{#1}}%
         \item[]%
 }
 {\end{list}}
\usepackage{setspace}

\def\bu{\textbf{u}}

\def\by{\textbf{y}}

\begin{document}
	
	\chapter{}
	\makeatletter
	\setlength{\@fptop}{0pt}
	\makeatother
	
\thispagestyle{empty} \baselineskip=28pt

\begin{center}
{\LARGE{\bf Spatio-Temporal Models for Big Multinomial Data using the Conditional Multivariate Logit-Beta Distribution}}
\end{center}

\baselineskip=12pt

\vskip 2mm
\begin{center}
Jonathan R. Bradley\footnote{(\baselineskip=10pt to whom correspondence should be addressed) Department of Statistics, Florida State University, 117 N. Woodward Ave., Tallahassee, FL 32306-4330, jrbradley@fsu.edu},
Christopher K. Wikle\footnote{\baselineskip=10pt  Department of Statistics, University of Missouri, 146 Middlebush Hall, Columbia, MO 65211-6100}, and
Scott H. Holan$^{2,}$ \footnote{\baselineskip=10pt U.S. Census Bureau, 4600 Solver Hill Road, Washington D. C. 20233-9100}\\
\end{center}
%
%
%
%
\vskip 4mm

\begin{center}
\large{{\bf Abstract}}
\end{center}
We introduce a Bayesian approach for analyzing high-dimensional multinomial data that are referenced over space and time. In particular, the proportions associated with multinomial data are assumed to have a logit link to a latent spatio-temporal mixed effects model. This strategy allows for covariances that are nonstationarity in both space and time, asymmetric, and parsimonious. We also introduce the use of the conditional multivariate logit-beta distribution into the dependent multinomial data setting, which leads to conjugate full-conditional distributions for use in a collapsed Gibbs sampler. We refer to this model as the multinomial spatio-temporal mixed effects model (MN-STM). Additionally, we provide methodological developments including: the derivation of the associated full-conditional distributions, a relationship with a latent Gaussian process model, and the stability of the non-stationary vector autoregressive model. We illustrate the MN-STM through simulations and through a demonstration with public-use Quarterly Workforce Indicators (QWI) data from the Longitudinal Employer Household Dynamics (LEHD) program of the U.S. Census Bureau.
\baselineskip=12pt

%
%
%

\baselineskip=12pt
\par\vfill\noindent
{\bf Keywords:} Bayesian hierarchical model; Big data; P\'olya-Gamma; Markov chain Monte Carlo; Generalized Linear Mixed Model; Gibbs sampler.
\par\medskip\noindent
\clearpage\pagebreak\newpage \pagenumbering{arabic}
\baselineskip=24pt
\section{Introduction} 
High-dimensional multinomial data referenced over space and time are ubiquitous among several disciplines. For example,  machine learners are often interested in analyzing text corpora \citep[or large sets of texts;][]{blei2003}, demographers are interested in assessing populations among several categories across the U.S. \citep{carl}, ecologists often track the population of tree species over space and time \citep[e.g., see][Chp. 3]{Hooten}, and epidemiologists monitor cancer rates by race over the U.S. \citep{cancertime}. To better estimate proportions associated with these categories, one can leverage spatio-temporal dependence \citep[e.g. see][among others]{cressie-wikle-book,nipps}; however, spatio-temporal dependence can lead to difficult methodological and computational challenges. \textit{Thus, the primary goal of this article is to develop a computationally feasible dynamic multinomial spatio-temporal model for high-dimensional datasets, with the purpose of estimating proportions. }

We are partially motivated by data from the Longitudinal Employer Household Dynamics (LEHD) program's public-use Quarterly Workforce Indicators (QWIs). LEHD has become a leading authority on U.S. economics data (e.g., see \citet{abowd}, \citet{abowdlmm}, and the references therein). We are motivated by the public-use QWI dataset (\href{https://lehd.census.gov/data}{https://lehd.census.gov/data}) partially because it is high-dimensional, has complex dependencies, and is an important tool for economists. Henceforth, we will use the phrase ``complex dependencies'' to indicate nonstationarity in both space and time, and asymmetry, which are well-known properties in the spatio-temporal literature (see, \citet{cressie-wikle-book} and Appendix A of the Supplementary Material of this article for a discussion of these terms). Additionally, many of the QWIs are not made available because some states do not sign the required Memorandum of Understanding (MOU) every year \citep[][Sections 5.5.1 and 5.6]{abowd}. This suggests a need for small area estimation. Also, public-use LEHD currently does not release associated margins of error, and hence, there is an opportunity for to use a statistical model for uncertainty quantification. 

LEHD provides data over all North American Industry Classification System (NAICS) sectors, all 3,145 U.S. counties, and over every yearly quarter. The public-use QWI dataset is very similar to high-dimensional panel data, or sometimes called longitudinal time series data \citep[e.g., for recent references see][and the references therein]{belloni2016inference,kohn2}. However, a key difference is that models for panel data do not explicitly model spatial correlations, which are present in the LEHD dataset. We are particularly interested in estimating the proportion of individuals employed at the beginning of a quarter over each of the 20 NAICS sectors at each U.S. county. The size of this correlated dataset is quite large, at 2, 247, 586 observations.

It is important to be precise by what we mean by ``big data.'' For example, we are \textit{not} considering the ``$p$ greater than $n$'' problem, which is an important type of ``big data'' problem \citep{htf}. Instead, we are interested in the methodological difficulties involved with defining covariances when the sample size is large (e.g., millions of observations), which has become a common problem in the ``big spatial data'' literature \citep[e.g., see][for reviews]{reviewmethods,bradley2014_comp,heaton2017methods}.

There are several methods that one might adapt to analyze the high-dimensional LEHD dataset. In terms of dependent (spatial, spatio-temporal, or multivariate) multinomial data, there are many tools currently available. In particular, latent Dirichlet allocation \citep[LDA;][]{blei2003} is a well-known model often used to solve the text corpora problem. LDA has been developed in the context of dependent (i.e., possibly in space and time) data, and is sometimes referred to as the correlated topic model \citep[CTM;][]{blei2006}, which models dependencies by assuming that the logit of the proportions follow a multivariate normal distribution. The first implementation of the CTM \citep{blei2006} used non-conjugate optimizations \citep[e.g., see][]{Andrieu,Murray}. As a result, several recent papers have capitalized on the fact that the binomial distribution with a logit-link can be augmented using a P\'{o}lya-Gamma representation. This leads to easy to sample from conjugate full-conditional distributions \citep[e.g., see][]{Zhou,ChenCTM,polson2,polson,nipps,glynn} for use in a Gibbs sampler.  Additionally, this P\'{o}lya-gamma augmentation approach has been used to model dynamics \citep{blei2006_2}, through the incorporation of a vector autoregressive model \citep{blei2006_2,nipps}.

The current state-of-the-art in the CTM literature requires one to augment the multinomial random vector with P\'{o}ly-gamma random variables. These P\'{o}ly-gamma random random variables are simulated using an algorithm with iterative calculations and an accept/reject step \citep{polson2}, which is not computationally advantageous when doing repeated simulations within a Gibbs sampler. \textit{Thus, the first contribution of this article is to propose the use of the conditional multivariate logit-beta distribution (MLB) from \citet{BradleyLCM} to model dependent multinomial data.} The MLB distribution gives conjugate full-conditional distributions that are straightforward to simulate from, and allows one to {avoid} computationally less efficient data augmentation. We provide technical a result that develops the relationship between the MLB distribution and other multivariate distributions. Most notably, one can view our hierarchical model as a special type of latent Gaussian process model. This MLB distribution has been recently introduced by \citet{BradleyLCM} and has been used to model Bernoulli, binomial, and negative binomial data in the multivariate/spatial settings. However, the MLB distribution has not yet been used in the multinomial spatio-temporal setting.

The computational advantages of the MLB distribution for modeling multinomial data are especially powerful when considering the current state of the literature for non-Gaussian dependent data \citep[e.g., see][for standard early references.]{diggle}. There are many Bayesian methods available to analyze non-Gaussian data. Quite often these algorithms require difficult to tune Metropolis-Hasting steps nested within a Gibbs sampler  \citep{Shaby,Tweedie,RM}. There are methods that aid in tuning a Markov chain Monte Carlo (MCMC) algorithm, such as pseudo-marginal MCMC \citep{andrieu2009pseudo}; however, these approaches have not been extended to the multinomial spatio-temporal setting. Tuning steps are completely avoided by our proposed approach, and by the method in \citet{nipps}. However, the method in \citet{nipps} has not been developed for high-dimensional settings (e.g., it does not incorporate dimension reduction). Other algorithms such as Hamiltonian Markov chain Monte Carlo \citep[e.g., see][]{Neal,dang2017hamiltonian}, particle MCMC \citep{gunawan2017efficient}, and splice sampling \citep{Murray} are also known to be difficult to implement when using high-dimensional datasets \citep[e.g., see discussions in][]{rue,bradleyPMSTM}.

There are, of course, other perspectives for non-Gaussian dependent data that exist outside the realm of a fully Bayesian analysis, which are not considered in this manuscript. For example, in the spatial/non-Gaussian setting, empirical Bayesian methods are available. \citet{sengupta}, \citet{sengupta2013empirical}, and \citet{sengupta2016predictive} employ an empirical Bayesian algorithm that involves a Newton-Raphson algorithm, nested within an expectation maximization algorithm, which is then nested within a Markov chain Monte Carlo algorithm. \citet{sengupta} provide empirical results that demonstrates that their empirical Bayesian analysis can lead to improvements in prediction over a specific fully Bayesian model and is computationally advantageous. However, empirical Bayesian approaches do not explicitly model the variability introduced by estimating process model parameters. Approximate Bayesian methods are also available for spatial and spatio-temporal settings including integrated nested Laplace approximations (INLA) \citep[e.g., see][and the references therein]{rue,bakka2018spatial} and variational Bayes methods \citep{ren2011variational,nathoo2014variational}. However, these models have not been extended to the high-dimensional multinomial spatio-temporal setting that we are considering.

In addition to our new use of the MLB distribution, we also provide an efficient parameterization of the spatio-temporal covariance. In particular, we use two commonly used techniques to parameterize spatial/spatio-temporal covariances. The first technique, is to allow different random variables to share the same
 random effect. These techniques have been used in the spatial \citep[e.g., see][among others]{banerjee,johan,lindgren-2011,hughes,nychkaLK,sengupta,sengupta2013empirical,sengupta2016predictive,BradleyLCM}, multivariate-spatial
 \citep{royle1999,finley,finley2}, and spatio-temporal \citep{stcar,wikle2001,bradleySTCOS} settings. The second commonly used technique is to partition the joint likelihood into a product
 of more manageable conditional likelihoods. For example, see the early papers on conditional autoregressive (CAR) models by \citet{besag-74}, \citet{besag-86}, and \citet{besag-91} in the spatial setting;  \citet{berlinerroyle}, \citet{mardia}, and \citet{billheimer} for the multivariate
 spatial setting; and \citet{wiklecress_spt}, \citet{cressie-shi-kang-2010}, \citet{katzfuss_1}, and \citet{katzfuss2012} in the spatio-temporal setting.
 
 Both techniques are used to define the multivariate spatio-temporal mixed effects (MSTM) introduced in \citet{bradleyMSTM}. In particular, to incorporate multivariate-spatial
 dependencies the MSTM lets the different categories and spatial regions share the same random effects at
 time $t$ (i.e., the first technique). Then, to incorporate temporal dependence the MSTM uses an order one vector autoregressive model VAR(1) (i.e., the second technique). The MSTM was first introduced for Gaussian data in \citet{bradleyMSTM}, and later adapted to Poisson data \citep[][]{bradleyPMSTM}. \textit{Our second contribution is to adapt the MSTM to the multinomial data setting.} The main methodological development needed here is to extend prior distributions of precision parameters (i.e., inverse of the covariance matrix) when using random effects distributed according to the MLB distribution. Previous MSTMs specify the model based precision parameter of the random effect to be ``close'' (in terms of the Frobenious norm) to the precision parameter implied by a CAR model. We use the same strategy in this article, and call our proposed model the \textit{multinomial spatio-temporal mixed effects model} (MN-STM).

 
An important motivator of the MSTM is that it allows for dimension reduction, which aids in analyzing high-dimensional data (such as the LEHD dataset). \textit{Thus, our third contribution is to use a reduced rank model for multinomial spatio-temporal data, which has not yet been proposed in the CTM literature.} Dimension reduction methods are often used when there is a computational bottleneck when computing a high-dimensional likelihood. To address this issue, latent high-dimensional random vectors are replaced by low-dimensional random vectors; in multivariate analysis, this is similar to principal component analysis \citep[e.g., see][among others]{Jolliffe,Cox}. Dimension reduction in the spatial/spatio-temporal setting has a mature literature associated with it. For example, see  \citet{johan-2006}, \citet{johan}, \citet{cressie-tao}, \citet{banerjee}, \citet{kang-cressie-2011}, \citet{lindgren-2011}, \citet{nychkaLK} for the univariate-spatial setting; and \citet{wiklecress_spt}, \citet{cressie-shi-kang-2010}, \citet{katzfuss_1}, and \citet{katzfuss2012} in the spatio-temporal setting. Also see, \citet{wikleHandbook} for a discussion on spatial and spatio-temporal dimension reduction modeling. 
 
 In addition to dimension reduction, the MSTM also adapts aspects of the model suggested by \citet{hughes} from the spatial only setting to the multivariate-spatio-temporal setting. Specifically, the propagator matrix associated with a VAR(1) model is specified so that random effects are not confounded \citep{bradleyMSTM,bradleyPMSTM}. This propagator matrix is referred to as the Moran's I (MI) propagator matrix because of a connection to the MI statistic from \citet{MoranI}. This motivation is similar to the specification of the MI basis functions used in \citet{griffith2000}, \citet{griffith2002}, \citet{griffith2004}, \citet{griffith2007}, \citet{hughes}, \citet{aaronp}, and \citet{burden2015sar}, which we review in this manuscript. The dynamic properties associated with the MI propagator matrix still need development. In particular, when introducing a VAR(1) model for forecasting, it is especially important to verify that the Wold representation of the VAR(1) model exists \citep[e.g., see][for a standard reference]{andersonts}. However, \citet{bradleyMSTM} and \citet{bradleyPMSTM} did not investigate this. \textit{Thus, our fourth methodological contribution is to provide the conditions in which the Wold representation of the VAR(1) model (with MI propagator matrices) exists.}

The remainder of this article is organized as follows. In Section~2, we introduce the use of the MLB distribution to model multinomial data with fixed and random effects. We end Section 2 with an illustrative example of logistic regression with MLB random effects, which is used to motivate the MN-STM to model data with more complex features (e.g., LEHD). In Section~3, we define the MN-STM and present the necessary technical development. We also provide a small technical result showing a type of equivalence between the use of the MLB distribution and the Gaussian distribution. This is done in an effort to ``de-mystify'' our use of new distribution theory. Empirical results are discussed in Section 4. This includes a simulation study that illustrates the computational performance of the MN-STM relative to a latent Gaussian process model and the the Poisson multivariate spatio-temporal mixed effects model (PMSTM) from \citet{bradleyPMSTM}. Additionally, an example analysis using a dataset obtained from LEHD is offered. Finally, Section~5 contains a discussion. For convenience of exposition, some statements and proofs of technical results are given in the appendices.

\section{Logistic Regression with Latent Multivariate Logit-Beta Random Effects} Consider multinomial data that are recorded at $N$ small areas $\{A_{i}: i = 1,\ldots,N\}$ and at $T$ equally spaced discrete time-points, where we assume that $A_{i} \cap A_{j}$ is empty for $i \ne j$. Let $\textbf{y}_{it} = \left(Y_{1it},\ldots, Y_{Kit}\right)^{\prime}$ be the multinomial data vector with $K$ categories, observed at time $t$ and areal unit $A_{i}$, where each $Y_{kit}$ is integer-valued from zero to $m_{it} = \sum_{k = 1}^{K}Y_{kit}$. For example, $\textbf{y}_{it}$ might consist of counts of the number of people employed in $K$ different NAICS sectors at county $A_{i}$ and yearly-quarter $t$. The goal of our analysis is to estimate the probability of each category for every $A_{i}$ and $t$. Let $D_{t}\subset \{A_{i}: i = 1,\ldots,N_{t}\}$ represent the counties at time $t$ where a multinomial vector $\textbf{y}_{it}$ is observed, and suppose there are $N_{t}\le N$ areal units in $D_{t}$ and let $n = (K-1)\sum_{t = 1}^{T}N_{t}$.

\subsection{Multinomial Data} We assume that $\textbf{y}_{it}$ follows a multinomial distribution. That is,
\begin{equation*}
f(\textbf{y}_{it}\vert \left\lbrace\pi_{kit}\right\rbrace, \left\lbrace m_{it}\right\rbrace) = \frac{m_{it}!}{Y_{1it}!\ldots  Y_{K,it}!}\pi_{1it}^{Y_{1it}}\ldots \pi_{Kit}^{Y_{Kit}};\hspace{5pt} t = 1,\ldots, T,\hspace{5pt} A_{i}\in D_{t},
\end{equation*}
\noindent
where it is assumed that $\pi_{kit} \in (0,1)$ and $\sum_{k = 1}^{K}\pi_{kit} = 1$ for every $i$ and $t$. Now, write the pmf of the multinomial distribution as the product of $K-1$ different binomial probability mass functions (pmfs) as follows \citep[e.g., see][among others]{nipps}:
\begin{equation}\label{binomialversion}
f(\textbf{y}_{it}\vert \{p_{kit}\}, \left\lbrace m_{it}\right\rbrace) = \prod_{k = 1}^{K-1} {n_{kit} \choose y_{kit}} p_{kit}^{Y_{kit}}(1-p_{kit})^{n_{kit} - Y_{kit}};\hspace{5pt} t = 1,\ldots, T,\hspace{5pt} A_{i}\in D_{t},
\end{equation}
\noindent
where $n_{kit} = m_{it} - \underset{j < k}{\sum}Y_{jit}$, and $p_{kit} = \frac{\pi_{kit}}{1 - \underset{j < k}{\sum}\pi_{jit}}$. Let $\mathrm{logit}(p_{kit}) \equiv \mathrm{log}\left\lbrace p_{kit}/(1-p_{kit})\right\rbrace \equiv v_{kit}$, so that Equation (\ref{binomialversion}) can be re-expressed as
\begin{align}\label{binomialLogitversion}
\nonumber
f(\textbf{y}_{it}\vert p_{it}, \left\lbrace m_{it}\right\rbrace) &= \prod_{k = 1}^{K-1} {n_{kit} \choose y_{kit}} \frac{\left\lbrace\mathrm{exp}(\nu_{kit})\right\rbrace^{Y_{kit}}}{\left\lbrace 1 + \mathrm{exp}(\nu_{kit})\right\rbrace^{n_{kit}}} \\
& = \prod_{k = 1}^{K-1} {n_{kit} \choose y_{kit}} \mathrm{exp}\left[Y_{kit}\nu_{kit} - n_{kit}\mathrm{log}\left\lbrace 1 + \mathrm{exp}(\nu_{kit})\right\rbrace \right];\hspace{5pt} t = 1,\ldots, T,\hspace{5pt} A_{i}\in D_{t}.
\end{align}
Hence, the pmf for the $n$-dimensional vector $\textbf{y}_{t} = \left(\textbf{y}_{it}^{\prime}: i = 1,\ldots, N_{t}, t = 1,\ldots, T\right)^{\prime}$ is given by,
\begin{align}
\nonumber
&f_{MN}(\textbf{y}\vert \bm{\nu}, \left\lbrace m_{it}\right\rbrace) = \prod_{t = 1}^{T}\prod_{i = 1}^{N_{t}}f(\textbf{y}_{it}\vert p_{it}) = \prod_{t = 1}^{T}\prod_{i = 1}^{N_{t}}\prod_{k = 1}^{K-1} {n_{kit} \choose y_{kit}} \mathrm{exp}\left[Y_{kit}\nu_{kit} - n_{kit}\mathrm{log}\left\lbrace 1 + \mathrm{exp}(\nu_{kit})\right\rbrace \right] \\
\label{dm}
&\underset{\nu}{\propto} \mathrm{exp}\left[\textbf{y}^{\prime}\bm{\nu} - \textbf{n}^{\prime}\mathrm{log}\left\lbrace\bm{1} + \mathrm{exp}\left(\bm{\nu}\right)\right\rbrace\right],
\end{align}
where the $n$-dimensional vector $\bm{\nu} = \left(\nu_{kit}: k = 1,\ldots, K-1, A_{i} \in D_{t}, t = 1,\ldots, T\right)^{\prime}$, the $n$-dimensional vector $\textbf{n} = \left(n_{kit}: k = 1,\ldots, K-1, A_{i} \in D_{t}, t = 1,\ldots, T \right)^{\prime}$, and the symbol ``$\underset{\nu}{\propto}$'' can be read as ``proportional to as a function of $\bm{\nu}$.''

\subsection{The Multivariate Logit-Beta Distribution} The derivation of the MLB distribution starts with the univariate logit-beta random variable, which is defined as
\begin{equation}\label{simp_lin}
	q\equiv \mathrm{logit}(\gamma)\equiv \mathrm{log}\left(\frac{\gamma}{1-\gamma}\right),
\end{equation}
\noindent
where $\gamma$ is a beta random variable with first shape parameter $\alpha>0$ and second shape parameter $\kappa-\alpha>0$. The shape parameter $\kappa$ is parameterized so that it is strictly larger than $\alpha$. This choice of parameterization will simplify subsequent expressions. The transformation in (\ref{simp_lin}) leads to the following probability density function (pdf) for $q$,
\begin{equation}
	\label{univ_LG}
	f(q\vert \alpha,\kappa) = \left\lbrace\frac{\Gamma(\kappa)}{\Gamma(\alpha)\Gamma(\kappa-\alpha)}\right\rbrace\mathrm{exp}\left[\alpha q - \kappa\mathrm{log}\left\lbrace 1 + \mathrm{exp}(q)\right\rbrace\right]; \hspace{5pt} q \in \mathbb{R}, \kappa>\alpha, \alpha>0,
\end{equation}
\noindent
where recall that $f$ is used to denote a generic pdf. Then, the MLB distribution is found through the following transformation:
\begin{equation}\label{chov}
	\textbf{q} = \bm{\mu} + \textbf{V}^{-1}\textbf{w},
\end{equation}
\noindent
where $\bm{\mu}$ is an $M$-dimensional real-valued vector, $\textbf{V}$ is an $M\times M$ real-valued invertible matrix, $\textbf{w} = \left(w_{1},\ldots, w_{M}\right)^{\prime}$, and the $w_{i}$'s are independent logit-beta random variables with shapes $\alpha_{i}>0$ and $\kappa_{i}>\alpha_{i}$, respectively. Here, $\bm{\mu}$ is a vector-valued location parameter and $\textbf{V}$ is a matrix-valued precision parameter. Again, straightforward change-of-variables of the transformation in (\ref{chov}) yields the following expression for the pdf of $\textbf{q}$ \citep{BradleyLCM}:
\begin{align}\label{mlb}
	&f(\textbf{q}\vert \bm{\mu}, \textbf{V}, \bm{\alpha}, \bm{\kappa})= \mathrm{det}(\textbf{V})\left\lbrace \prod_{i = 1}^{M}\frac{\Gamma(\kappa_{i})}{\Gamma(\alpha_{i})\Gamma(\kappa_{i}-\alpha_{i})}\right\rbrace \mathrm{exp}\left[\bm{\alpha}^{\prime} \textbf{V}(\textbf{q} - \bm{\mu}) - \bm{\kappa}^{\prime}\mathrm{log}\left\lbrace \bm{1}_{M} + \mathrm{exp}(\textbf{V}(\textbf{q} - \bm{\mu}))\right\rbrace\right],
\end{align}
\noindent
where $\bm{1}_{M}$ is a $M$-dimensional vector of ones, $\bm{\alpha} = (\alpha_{1},\ldots, \alpha_{M})^{\prime}$, and $\bm{\kappa} = \left(\kappa_{1},\ldots, \kappa_{M}\right)^{\prime}$. The pdf in (\ref{mlb}) has a similar functional form as the likelihood associated with a multinomial probability mass function with a logit-link in (\ref{dm}), and is the aforementioned MLB distribution. This relationship leads to conjugacy between the MLB distribution and the multinomial distribution.

The conditional MLB distribution also has a similar form as (\ref{mlb}). Denote the partioning of the vectors $\textbf{q}= \left(\textbf{q}_{1}^{\prime},\textbf{q}_{2}^{\prime}\right)^{\prime}$ and $\bm{\mu}= \left(\bm{\mu}_{1}^{\prime},\bm{\mu}_{2}^{\prime}\right)^{\prime}$, such that $\textbf{q}_{1}^{\prime}$ and $\bm{\mu}_{1}^{\prime}$ are $r$-dimensional, and $\textbf{q}_{2}^{\prime}$ and $\bm{\mu}_{2}^{\prime}$ are $(M-r)$-dimensional. Also, partition $\textbf{V} = \left[\textbf{H}, \textbf{B}\right]$ into the $M\times r$ matrix $\textbf{H}$ and the $M\times (M-r)$ matrix $\textbf{B}$. Then the conditional MLB distribution is given by
\begin{align}\label{cmlb}
	\nonumber
	&f(\textbf{q}_{1}\vert \textbf{q}_{2} = \textbf{d},\bm{\mu}, \textbf{V}, \bm{\alpha}, \bm{\kappa}) \propto f(\textbf{q}_{1},\textbf{q}_{2}= \textbf{d}\vert \bm{\mu}, \textbf{V}, \bm{\alpha}, \bm{\kappa})\\
	 &\propto \mathrm{exp}\left[\bm{\alpha}^{\prime} \textbf{H}(\textbf{q}_{1} - \bm{\mu}_{1}) 
- \bm{\kappa}^{\prime}\mathrm{log}\left\lbrace \bm{1}_{M} + \mathrm{exp}(\textbf{H}(\textbf{q}_{1} - \bm{\mu}_{1}) + \textbf{B}(\textbf{d} - \bm{\mu}_{2}))\right\rbrace\right],
\end{align}
\noindent
where $\textbf{d}$ is a fixed real-valued $(M-r)$-dimensional vector. Define the $M$-dimensional vector $\textbf{c} = \textbf{H}\bm{\mu}_{1} +\textbf{B}\bm{\mu}_{2} - \textbf{B}\textbf{d}$. For ease of notation, we will let
\begin{align}\label{cmlbhelp}
	&f_{MLB}(\textbf{q}_{1}\vert \textbf{c}, \textbf{V}, \bm{\alpha}, \bm{\kappa}) = f(\textbf{q}_{1}\vert \textbf{q}_{2} = \textbf{d},\bm{\mu}, \textbf{V}, \bm{\alpha}, \bm{\kappa}) \propto  \mathrm{exp}\left[\bm{\alpha}^{\prime}( \textbf{H}\textbf{q}_{1} - \textbf{c}) - \bm{\kappa}^{\prime}\mathrm{log}\left\lbrace \bm{1}_{M} + \mathrm{exp}(\textbf{H}\textbf{q}_{1} - \textbf{c})\right\rbrace\right].
\end{align}
Here, $\textbf{c}$ can be equal to any $M$-dimensional real-valued vector $\textbf{d}_{c}\in \mathbb{R}^{M}$. Specifically, re-parameterize $\bm{\mu}$ so that $\bm{\mu} = \textbf{V}^{-1}(\textbf{d}_{c} + \textbf{B}\textbf{d})$ and $\textbf{c} = \textbf{H}\bm{\mu}_{1} + \textbf{B}\bm{\mu}_{2} - \textbf{B}\textbf{d} = \textbf{d}_{c}$. This will be useful later on, as the full-conditional distributions associated with our model will have this re-parameterized form. Furthermore, we drop $\textbf{q}_{2} = \textbf{d}$ and $\bm{\mu}$ and include $\textbf{c}$ in our notation, and signify this notational convenience with the subscript ``MLB'' in $f_{MLB}$, and refer to (\ref{cmlbhelp}) as the conditional MLB distribution. We refer to $\textbf{c}$ as the location parameter.

It is difficult to directly simulate from a conditional MLB. However, in a Bayesian context one can augment the likelihood in a manner so that updates within a Gibbs sampler can be computed using marginal distributions from a MLB. In particular, we will use the following result.\\

\noindent
\textit{Proposition 1: Let $\textbf{q}_{1}^{*}\vert \textbf{c}^{*}, \textbf{V}^{*} = (\textbf{H}^{*},\textbf{B}^{*}), \bm{\alpha}^{*},\bm{\kappa}^{*} \sim f_{MLB}(\textbf{q}_{1}^{*}\vert \textbf{c}^{*}, \textbf{V}^{*} = (\textbf{H}^{*},\textbf{B}^{*}), \bm{\alpha}^{*},\bm{\kappa}^{*})$, where $\textbf{H}^{*} \in \mathbb{R}^{M}\times \mathbb{R}^{r}$ is full column rank, $\bm{\alpha}^{*} = (\alpha_{1}^{*},\ldots, \alpha_{M}^{*})^{\prime}$, $\bm{\kappa}^{*} = (\kappa_{1}^{*},\ldots, \kappa_{M}^{*})^{\prime}$, $\alpha_{i}^{*}>0$, and $\kappa_{i}^{*}>\alpha_{i}^{*}$ for $i = 1,\ldots, M$. Assume a re-parameterized value (see discussion below (\ref{cmlbhelp})) of $\textbf{c}^{*} = -\textbf{B}^{*}\textbf{q}_{2}^{*}+\bm{\mu}^{*}$, and the improper prior $g(\textbf{q}_{2}^{*}\vert \textbf{c}^{*}, \textbf{V}^{*} = (\textbf{H}^{*},\textbf{B}^{*}), \bm{\alpha}^{*},\bm{\kappa}^{*})\propto 1$, where $\textbf{q}_{2}$ is $(M-r)$-dimensional. Also let $\textbf{B} \in \mathbb{R}^{M}\times \mathbb{R}^{M-r}$ be the orthonormal basis for the null space of $\textbf{H}^{*}$, $\textbf{q}^{*} = (\textbf{q}_{1}^{*\prime},\textbf{q}_{2}^{*\prime})^{\prime}$, $\bm{\mu}^{*}\in \mathbb{R}^{M}$, $\textbf{I}_{n}$ be an $n\times n$ identity matrix, and let $\textbf{w}^{*}\sim f_{MLB}(\textbf{q}_{1}^{*}\vert \bm{\mu}^{*}, \textbf{I}_{n}, \bm{\alpha}^{*},\bm{\kappa}^{*})$. Then,}
\begin{align}
\nonumber
	\int f_{MLB}(\textbf{q}_{1}^{*}\vert \textbf{c}^{*} = -\textbf{B}^{*}\textbf{q}_{2}^{*}&+\bm{\mu}^{*}, \textbf{V}^{*} = (\textbf{H}^{*},\textbf{B}^{*}), \bm{\alpha}^{*},\bm{\kappa}^{*})d\textbf{q}_{2}^{*}\\
	\label{cond_pdf4}
	&\propto \int \mathrm{exp}\left[\bm{\alpha}^{*\prime}\textbf{V}^{*}\textbf{q}^{*}-\bm{\kappa}^{*\prime}\mathrm{log}\left\lbrace\bm{1}_{M}+\mathrm{exp}(\textbf{V}^{*}\textbf{q}^{*}-\bm{\mu}^{*})\right\rbrace\right]d\textbf{q}_{2}^{*},
\end{align}
\textit{where the integrand in (\ref{cond_pdf4}) is proportional to $f(\textbf{q}^{*}\vert \textbf{V}^{*-1}\bm{\mu}^{*}, \textbf{V}^{*} = (\textbf{H}^{*},\textbf{B}^{*}), \bm{\alpha}^{*}, \bm{\kappa}^{*})$ for $f$ defined in (\ref{mlb}). Furthermore, the affine transformation, }
\begin{equation}
\label{trans}
(\textbf{H}^{*\prime}\textbf{H}^{*})^{-1}\textbf{H}^{*\prime}\textbf{w}^{*},
\end{equation}\
\textit{is a draw from the density in (\ref{cond_pdf4}).}\\

\noindent
\textit{Proof:} See Appendix A.\\		

\noindent
Proposition 1 will allow us to augment the likelihood and update multinomial parameters using a collapsed Gibbs sampler \citep{liu1994collapsed}. That is, when a full-conditional distribution is proportional to a $f_{MLB}$ we will collapse across the mean, which will be assumed to have an improper prior. We provide an example of this in a simplified setting in Section 2.3.

\subsection{Example: Logistic Regression with Latent Multivariate Logit-Beta Random Effects} We now provide an illustration of Proposition 1 for implementing a logistic regression model with latent multivariate logit-beta random effects. Consider $T = 1$ and the following hierarchical model:
\begin{gather}
	\nonumber
		\textbf{y}\vert \bm{\beta}, \bm{\eta},\{m_{i1}\} \sim f_{MN}(\by\vert \bm{\nu}, \{m_{i1}\}), \\
	\nonumber
	\bm{\nu} = \textbf{X}\bm{\beta} + \bm{\Phi}\bm{\eta} + \bm{\xi}\\
	\nonumber
	\bm{\beta} \sim f_{MLB}\left\lbrace \bm{\beta}\hspace{5pt}\Bigg\vert \hspace{5pt}
	 \bm{0}_{n+p},\left(\begin{array}{c}\sigma\textbf{X}\\
		\textbf{I}_{p}\end{array}\right),\left(\begin{array}{c}\bm{\epsilon}/\sigma\\
		\alpha\bm{1}_{p}\end{array}\right),\left(\begin{array}{c}\bm{\delta}\\
		\kappa\bm{1}_{p}\end{array}\right)\right\rbrace,\\
\nonumber
	\bm{\eta} \sim f_{MLB}\left\lbrace \bm{\eta}\hspace{5pt}\Bigg\vert \hspace{5pt}
	\bm{0}_{n+r},\left(\begin{array}{c}\sigma\bm{\Phi}\\
	\textbf{I}_{r}\end{array}\right),\left(\begin{array}{c}\bm{\epsilon}/\sigma\\
	\alpha_{\eta}\bm{1}_{r}\end{array}\right),\left(\begin{array}{c}\bm{\delta}\\
	\kappa_{\eta}\bm{1}_{r}\end{array}\right)\right\rbrace\\
		\label{noaugmentedexample}
			\bm{\xi} \sim f_{MLB}\left\lbrace \bm{\xi}\hspace{5pt}\Bigg\vert \hspace{5pt}
			\bm{0}_{2n},\left(\begin{array}{c}\sigma\textbf{I}_{n}\\
			\textbf{I}_{n}\end{array}\right),\left(\begin{array}{c}\bm{\epsilon}/\sigma\\
			\alpha_{\xi}\bm{1}_{n}\end{array}\right),\left(\begin{array}{c}\bm{\delta}\\
			\kappa_{\xi}\bm{1}_{n}\end{array}\right)\right\rbrace,
\end{gather}

\noindent
where $\sigma>0$, the $\bm{\beta}$ is a $p$-dimensional vector of covariate effects, $\bm{\eta}$ is a $r$-dimensional vector of random effects, $\bm{\xi}$ is a $n$-dimensional vector of random effects, $\kappa>\alpha>0$, $\kappa_{\eta}>\alpha_{\eta}>0$, and $\kappa_{\xi}>\alpha_{\xi}>0$. Let $\textbf{I}_{n}$ be an $n\times n$ identity matrix, $\bm{0}_{n}$ be a $n$-dimensional vector of zeros, let the $n$-dimensional vector $\bm{\epsilon}$ consist of strictly positive elements, and the $n$-dimensional vector $\bm{\delta}$ have elements strictly larger than the corresponding elements of $\bm{\epsilon}$. For illustration, let $\textbf{X}$ be an $n\times p$ matrix of \textit{orthonormal} covariates and let $\bm{\Phi}$ be an $n\times r$ orthonormal ``design matrix'' \citep{RPLM}. The prior distributions in (\ref{noaugmentedexample}) depends on the sample size (e.g., the precision parameter for the prior of $\bm{\beta}$ is $(n+p)$-dimensional). Thus, we introduce $\sigma$ and choose it to be small to mitigate changes in the prior as more samples are collected. The need for both $\bm{\epsilon}$ and $\bm{\delta}$ is discussed at the end of Section 2.3.

 This hierarchical model might be implemented with the Gibbs sampler, which is outlined in Pseudo-Code (\ref{euclid}). To implement Pseudo-Code (\ref{euclid}) we first find the full conditional distribution for $\bm{\beta}$ as follows,
 \begin{algorithm}[H]\caption{Pseudo-Code 1: Gibbs sampler for the model in (\ref{noaugmentedexample})}\label{euclid}
 	\begin{algorithmic}[1]
 		\item Set $b = 1$ and initialize $\bm{\beta}^{[0]}$, $\bm{\eta}^{[0]}$, and $\bm{\xi}^{[0]}$.
 		\item Sample $\bm{\beta}^{[b]}$ from $f(\bm{\beta}\vert \by, \bm{\eta}^{[b-1]},\bm{\xi}^{[b-1]})$
 		\item Sample $\bm{\eta}^{[b]}$ from $f(\bm{\eta}\vert \by, \bm{\beta}^{[b]},\bm{\xi}^{[b-1]})$.
 		\item Sample $\bm{\xi}^{[b]}$ from $f(\bm{\eta}\vert \by, \bm{\beta}^{[b]},\bm{\eta}^{[b]})$.
 		\item Repeat Steps 2, 3, and 4 until $b = B$ for a prespecified value of $B$.
 	\end{algorithmic}
 \end{algorithm}
\begin{align*}
	\nonumber
	f(\bm{\beta}\vert \by, \bm{\eta},\bm{\xi}) &\propto \mathrm{exp}\left[\bm{\epsilon}^{\prime}\textbf{X}\bm{\beta} + \alpha \bm{1}_{p}^{\prime}\bm{\beta}- \bm{\delta}^{\prime}\mathrm{log}\left\lbrace \bm{1}_{n} + \mathrm{exp}(\sigma\textbf{X}\bm{\beta})\right\rbrace - \kappa\bm{1}_{p}^{\prime}\mathrm{log}\left\lbrace \bm{1}_{p} + \mathrm{exp}(\bm{\beta})\right\rbrace\right]\\
	&\times  \mathrm{exp}\left[\textbf{y}^{\prime}\textbf{X}\bm{\beta} - \textbf{n}^{\prime}\mathrm{log}\left\lbrace\bm{1} + \mathrm{exp}\left(\textbf{X}\bm{\beta} + \bm{\Phi}\bm{\eta} + \bm{\xi}\right)\right\rbrace\right]\\
	&\propto {\TTiny f_{MLB}\left\lbrace \bm{\beta}\hspace{5pt}\Bigg\vert \hspace{5pt}
\textbf{c}_{\beta} = \left(\begin{array}{c}-\bm{\Phi}\bm{\eta}-\bm{\xi}\\
\bm{0}_{n}\\\bm{0}_{p}\end{array}\right),\textbf{H}_{\beta}^{*} = \left(\begin{array}{c}\textbf{X}\\
\sigma\textbf{X}\\
	\textbf{I}_{p}\end{array}\right), \bm{\alpha}_{\beta}^{*} = \left(\begin{array}{c}\rho\by + \bm{\epsilon}_{1}\\
	(1-\rho)\by/\sigma + \bm{\epsilon}_{2}/\sigma\\\alpha\bm{1}_{p}\end{array}\right) ,\bm{\kappa}_{\beta}^{*} =\left(\begin{array}{c}\textbf{n}\\
\bm{\delta}\\\kappa\bm{1}_{p}\end{array}\right)\right\rbrace},
\end{align*}

\noindent
where $\rho \in (0,1)$ and $\bm{\epsilon} = \bm{\epsilon}_{1} + \bm{\epsilon}_{2}$. Similarly, the full-conditional distributions for $\bm{\eta}$ and $\bm{\xi}$ are given by,
\begin{align}
\nonumber
{\TTiny f(\bm{\eta}\vert \by, \bm{\beta},\bm{\xi})}
&{\TTiny\propto f_{MLB}\left\lbrace \bm{\eta}\hspace{5pt}\Bigg\vert \hspace{5pt}
\textbf{c}_{\eta} = \left(\begin{array}{c}-\textbf{X}\bm{\beta}-\bm{\xi}\\
\bm{0}_{n}\\\bm{0}_{r}\end{array}\right),\textbf{H}_{\eta}^{*} =\left(\begin{array}{c}\bm{\Phi}\\
\sigma\bm{\Phi}\\
\textbf{I}_{r}\end{array}\right),\bm{\alpha}_{\eta}^{*} =\left(\begin{array}{c}\rho\by + \bm{\epsilon}_{1}\\
(1-\rho)\by/\sigma + \bm{\epsilon}_{2}/\sigma\\\alpha_{\eta}\bm{1}_{r}\end{array}\right) ,\bm{\kappa}_{\eta}^{*} =\left(\begin{array}{c}\textbf{n}\\
\bm{\delta}\\\kappa_{\eta}\bm{1}_{r}\end{array}\right)\right\rbrace}\\
\nonumber
{\TTiny f(\bm{\xi}\vert \by, \bm{\beta}, \bm{\eta})}
&{\TTiny\propto f_{MLB}\left\lbrace \bm{\xi}\hspace{5pt}\Bigg\vert \hspace{5pt}
	\textbf{c}_{\xi} = \left(\begin{array}{c}-\textbf{X}\bm{\beta}-\bm{\Phi}\bm{\eta}\\
	\bm{0}_{n}\\\bm{0}_{n}\end{array}\right),\textbf{H}_{\xi}^{*} =\left(\begin{array}{c}\textbf{I}_{n}\\
\sigma\textbf{I}_{n}\\
	\textbf{I}_{n}\end{array}\right),\bm{\alpha}_{\xi}^{*} =\left(\begin{array}{c}\rho\by + \bm{\epsilon}_{1}\\
	(1-\rho)\by/\sigma + \bm{\epsilon}_{2}/\sigma\\\alpha_{\eta}\bm{1}_{r}\end{array}\right) ,\bm{\kappa}_{\xi}^{*} =\left(\begin{array}{c}\textbf{n}\\
	\bm{\delta}\\\kappa_{\xi}\bm{1}_{n}\end{array}\right)\right\rbrace}.
\end{align}
As discussed in Section 2.2, it is difficult to directly sample from these full-conditional distributions; hence we offer a data augmentation scheme that leads to full-conditional distributions that are easy to sample from. To do this, we augment the likelihood in (\ref{noaugmentedexample}) with the $2n$-dimensional random vector $\textbf{q}_{\beta}$, the $2n$-dimensional random vector $\textbf{q}_{\eta}$, and the $2n$-dimensional random vector $\textbf{q}_{\xi}$.
Now, consider the following hierarchical model:
\begin{gather}
\nonumber
\textbf{y} \sim f_{MN}(\by\vert \bm{\nu}, \{m_{i1}\},\textbf{q}_{\beta},\textbf{q}_{\eta},\textbf{q}_{\xi})\mathrm{exp}\left\lbrace\by^{\prime}\textbf{B}_{2,\beta}\textbf{q}_{\beta}+\by^{\prime}\textbf{B}_{2,\eta}\textbf{q}_{\eta}+\by^{\prime}\textbf{B}_{2,\xi}\textbf{q}_{\xi}
+\bm{\epsilon}^{\prime}\textbf{B}_{1,\beta}\textbf{q}_{\beta}+ \bm{\epsilon}^{\prime}\textbf{B}_{1,\eta}\textbf{q}_{\eta}+ \bm{\epsilon}^{\prime}\textbf{B}_{1,\xi}\textbf{q}_{\xi}
\right\rbrace,\\
\nonumber
\bm{\nu} = \textbf{X}\bm{\beta} + \bm{\Phi}\bm{\eta}+\bm{\xi}+ \textbf{B}_{1,\beta}\textbf{q}_{\beta} +\textbf{B}_{2,\eta}\textbf{q}_{\eta}+\textbf{B}_{3,\xi}\textbf{q}_{\xi}\\
\nonumber
\bm{\beta} \sim f_{MLB}\left\lbrace \bm{\beta}\hspace{5pt}\Bigg\vert \hspace{5pt}
\left(\begin{array}{c}-\textbf{B}_{2,\beta}\textbf{q}_{\beta}\\
-\textbf{B}_{3,\beta}\textbf{q}_{\beta}\end{array}\right),\left(\begin{array}{c}\sigma\textbf{X}\\
\textbf{I}_{p}\end{array}\right)\left(\begin{array}{c}\bm{\epsilon}/\sigma\\
\alpha\bm{1}_{p}\end{array}\right),\left(\begin{array}{c}\bm{\delta}\\
\kappa\bm{1}_{p}\end{array}\right)\right\rbrace,\\
\nonumber
\bm{\eta} \sim f_{MLB}\left\lbrace \bm{\eta}\hspace{5pt}\Bigg\vert \hspace{5pt}
\left(\begin{array}{c}-\textbf{B}_{2,\eta}\textbf{q}_{\eta}\\
-\textbf{B}_{3,\eta}\textbf{q}_{\eta}\end{array}\right),\left(\begin{array}{c}\sigma\bm{\Phi}\\
\textbf{I}_{r}\end{array}\right)\left(\begin{array}{c}\bm{\epsilon}/\sigma\\
\alpha_{\eta}\bm{1}_{r}\end{array}\right),\left(\begin{array}{c}\bm{\delta}\\
\kappa_{\eta}\bm{1}_{r}\end{array}\right)\right\rbrace\\
\nonumber
\bm{\xi} \sim f_{MLB}\left\lbrace \bm{\xi}\hspace{5pt}\Bigg\vert \hspace{5pt}
\left(\begin{array}{c}-\textbf{B}_{2,\xi}\textbf{q}_{\xi}\\
-\textbf{B}_{3,\xi}\textbf{q}_{\xi}\end{array}\right),\left(\begin{array}{c}\sigma\textbf{I}_{n}\\
\textbf{I}_{n}\end{array}\right)\left(\begin{array}{c}\bm{\epsilon}/\sigma\\
\alpha_{\xi}\bm{1}_{r}\end{array}\right),\left(\begin{array}{c}\bm{\delta}\\
\kappa_{\xi}\bm{1}_{r}\end{array}\right)\right\rbrace\\
\nonumber
\textbf{q}_{\beta} \sim 1\\
\nonumber
\textbf{q}_{\eta}\sim 1\\
\label{augmentedexample}
\textbf{q}_{\xi} \sim 1,
\end{gather}
where $\textbf{B}_{\beta} = (\textbf{B}_{1,\beta}^{\prime},\textbf{B}_{2,\beta}^{\prime},\textbf{B}_{3,\beta}^{\prime})^{\prime}$ is the $(2n+p)\times 2n$ orthonormal basis of the $(2n+p)\times p$ orthogonal matrix $\textbf{H}_{\beta}^{*}$, where $\textbf{B}_{1,\beta}^{\prime}$ is $n\times (2n)$, $\textbf{B}_{2,\beta}^{\prime}$ is $n\times (2n)$, and $\textbf{B}_{3,\beta}^{\prime}$ is $p\times (2n)$. Similarly, $\textbf{B}_{\eta} = (\textbf{B}_{1,\eta}^{\prime},\textbf{B}_{2,\eta}^{\prime},\textbf{B}_{3,\eta}^{\prime})^{\prime}$ is the $(2n+r)\times 2n$ orthonormal basis of the $(2n+r)\times r$ orthogonal matrix $\textbf{H}_{\eta}^{*}$, where $\textbf{B}_{1,\eta}^{\prime}$ is $n\times (2n)$, $\textbf{B}_{2,\eta}^{\prime}$ is $n\times (2n)$, and $\textbf{B}_{3,\eta}^{\prime}$ is $r\times (2n)$. Likewise, $\textbf{B}_{\xi} = (\textbf{B}_{1,\xi}^{\prime},\textbf{B}_{2,\xi}^{\prime},\textbf{B}_{3,\xi}^{\prime})^{\prime}$ is the $(3n)\times 2n$ orthonormal basis of the $(3n)\times n$ orthogonal matrix $\textbf{H}_{\xi}^{*}$, where $\textbf{B}_{1,\xi}^{\prime}$ is $n\times (2n)$, $\textbf{B}_{2,\xi}^{\prime}$ is $n\times (2n)$, and $\textbf{B}_{3,\xi}^{\prime}$ is $n\times (2n)$.

In (\ref{augmentedexample}), the conditional distribution of $\by, \bm{\beta},$ and $\bm{\eta}$ given $\textbf{q}_{\beta} = \bm{0}_{2n}$, $\textbf{q}_{\eta} = \bm{0}_{2n}$, and $\textbf{q}_{\xi} = \bm{0}_{2n}$ is proportional to the likelihood in (\ref{noaugmentedexample}). That is, in (\ref{augmentedexample}) replace $\textbf{q}_{\beta}$ with $\bm{0}_{2n}$, $\textbf{q}_{\eta}$ with $\bm{0}_{2n}$, and $\textbf{q}_{\xi}$ with $\bm{0}_{2n}$ to obtain a likelihood {proportional to} (\ref{noaugmentedexample}). Thus, a strategy to sample from a likelihood \textit{proportional to}  (\ref{noaugmentedexample}), would be to sample from $f(\bm{\beta},\bm{\eta},\bm{\xi}\vert \by, \textbf{q}_{\beta} = \bm{0}_{2n}, \textbf{q}_{\eta} = \bm{0}_{2n},\textbf{q}_{\xi} = \bm{0}_{2n})$. This can be easily implemented using the following collapsed Gibbs sampler \citep{liu1994collapsed} as outlined in Pseudo-Code (\ref{euclid2}).
\begin{algorithm}[H]\caption{Pseudo-Code 2: Collapsed Gibbs sampler for the model in (\ref{augmentedexample})}\label{euclid2}
	\begin{algorithmic}[1]
	\item Set $b = 1$ and initialize $\bm{\beta}^{[0]}$, $\bm{\eta}^{[0]}$, and $\bm{\xi}^{[0]}$.
	\item Sample $\bm{\beta}^{[b]}$ from $f(\bm{\beta}\vert \by, \bm{\eta}^{[b-1]},\bm{\xi}^{[b-1]}, \textbf{q}_{\eta} = \bm{0}_{2n},\textbf{q}_{\xi} =\bm{0}_{2n})$.
\item Sample $\bm{\eta}^{[b]}$ from $f(\bm{\eta}\vert \by, \bm{\beta}^{[b]},\bm{\xi}^{[b-1]}, \textbf{q}_{\beta} = \bm{0}_{2n},\textbf{q}_{\xi} = \bm{0}_{2n})$.
\item Sample $\bm{\xi}^{[b]}$ from $f(\bm{\xi}\vert \by, \bm{\beta}^{[b]},\bm{\eta}^{[b]}, \textbf{q}_{\beta} = \bm{0}_{2n},\textbf{q}_{\eta} = \bm{0}_{2n})$.
	\item Repeat Steps 2, 3, and 4 until $b = B$ for a prespecified value of $B$,
\end{algorithmic}
\end{algorithm}

\begin{figure}[t!]
		\begin{center}
			\begin{tabular}{c}
				\includegraphics[width=15.5cm,height=7cm]{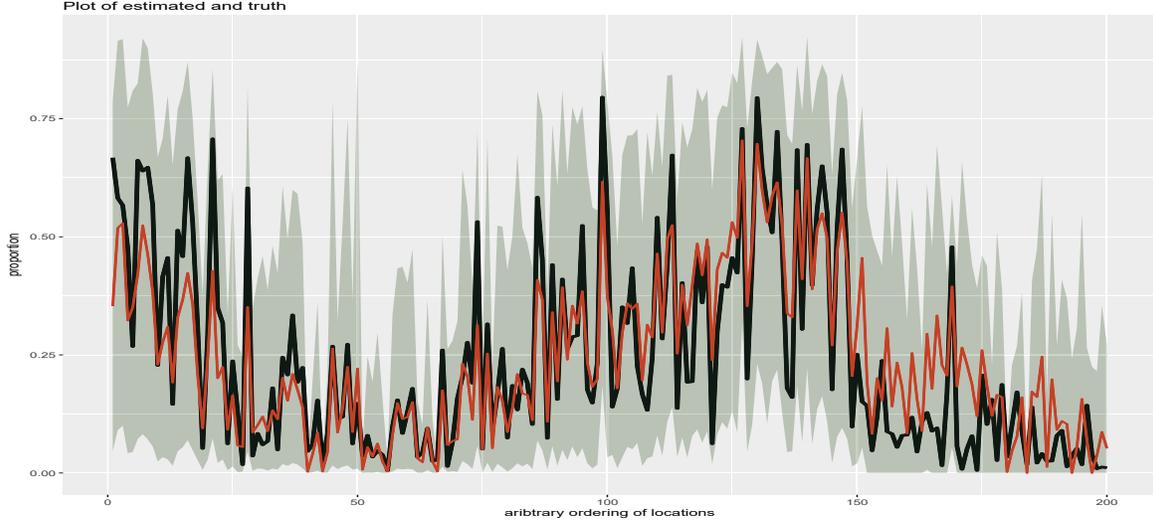}
			\end{tabular}
			\caption{\baselineskip=10pt{A small illustration of the true values of $\{\pi_{kit}\}$ and the corresponding predicted values from $n = 200$ simulated data values. The black line represents the true proportions, red line represents the estimated proportions, and the dark-green area signifies point-wise 95$\%$ credible intervals. The details behind the simulation model are provided in Appendix B. R code is used to implement Pseudo-Code 2 with the additional assumptions that $\alpha \sim Gamma(1,1)$, $\alpha_{\eta} \sim Gamma(1,1)$, $\alpha_{\xi} \sim Gamma(1,1)$, $\kappa\vert \alpha \sim Gamma(1,1)I(\kappa>\alpha)$, $\kappa_{\eta}\vert \alpha_{\eta} \sim Gamma(1,1)I(\kappa_{\eta}>\alpha_{\eta})$, and $\kappa_{\xi}\vert \alpha_{\xi} \sim Gamma(1,1)I(\kappa_{\xi}>\alpha_{\xi})$, where $I(\cdot)$ is an indicator function.}}\label{fig1}
		\end{center}
	\end{figure}

In Pseudo-Code 2, we have collapsed across the event $\{\textbf{q}_{\beta} = \bm{0}_{2n}\}$ in Step 2, $\{\textbf{q}_{\eta} = \bm{0}_{2n}\}$ in Step 3, and $\{\textbf{q}_{\xi} = \bm{0}_{2n}\}$ in Step 4. To do this, we first find $f(\bm{\beta},\textbf{q}_{\beta}\vert \by, \bm{\eta}, \bm{\xi},\textbf{q}_{\eta} = \bm{0}_{2n},\textbf{q}_{\xi} = \bm{0}_{2n})$ and then marginalize across $\textbf{q}_{\beta}$. That is,
\begin{align*}
\nonumber
&f(\bm{\beta},\textbf{q}_{\beta}\vert \by, \bm{\eta},\textbf{q}_{\eta} = \bm{0}_{2n},\textbf{q}_{\xi} = \bm{0}_{2n})\\
\nonumber
 & \propto \mathrm{exp}\left[ \textbf{y}^{\prime}\textbf{X}\bm{\beta} + (\textbf{y}+\bm{\epsilon})^{\prime}\textbf{B}_{1,\beta}\textbf{q}_{\beta} + \textbf{y}^{\prime}\textbf{B}_{2,\beta}\textbf{q}_{\beta} - \textbf{n}^{\prime}\mathrm{log}(\bm{1}_{n} + \mathrm{exp}(\textbf{X}\bm{\beta} + \textbf{B}_{1,\beta}\textbf{q}_{\beta}+ \bm{\Phi}\bm{\eta} + \bm{\xi})\right]\\
 &\mathrm{exp}\left[\bm{\epsilon}^{\prime}\textbf{X}\bm{\beta} + \bm{\epsilon}^{\prime}\textbf{B}_{2,\beta}\textbf{q}_{\beta} - \bm{\delta}^{\prime}\mathrm{log}(\bm{1}_{n} + \mathrm{exp}(\sigma\textbf{X}\bm{\beta} + \textbf{B}_{2,\beta}\textbf{q}_{\beta}))\right]\\
  &\mathrm{exp}\left[ \alpha\bm{1}_{p}^{\prime}\bm{\beta} + \alpha\bm{1}_{p}^{\prime}\textbf{B}_{3,\beta}\textbf{q}_{\beta}- \kappa\bm{1}_{p}^{\prime}\mathrm{log}(\bm{1}_{p} + \mathrm{exp}(\bm{\beta} + \textbf{B}_{3,\beta}\textbf{q}_{\beta}))\right]\\
&{\propto \mathrm{MLB}\left\lbrace \bm{\beta},\textbf{q}_{\beta}\hspace{5pt}\Bigg\vert \hspace{5pt}
\bm{\mu}^{*} = \textbf{V}^{-1}\left(\begin{array}{c}-\bm{\Phi}\bm{\eta} - \bm{\xi}\\
\bm{0}_{n}\\\bm{0}_{p}\end{array}\right),\textbf{V} = (\textbf{H}_{\beta}^{*},\textbf{B}_{\beta}),\bm{\alpha}_{\beta}^{*} ,\bm{\kappa}_{\beta}^{*}\right\rbrace}.
\end{align*}
\noindent
It follows from Proposition 1 that $(\textbf{H}_{\beta}^{*\prime}\textbf{H}_{\beta}^{*})^{-1}\textbf{H}_{\beta}^{*\prime}\textbf{w}$ is a sample from $f(\bm{\beta}\vert \by, \bm{\eta},\bm{\xi},\textbf{q}_{\eta} = \bm{0}_{2n},\textbf{q}_{\xi} = \bm{0}_{2n})$, where $\textbf{w}$ is distributed as $\mathrm{MLB}(\textbf{c}_{\beta},\textbf{I}_{2n+p}, \bm{\alpha}_{\beta}^{*},\bm{\kappa}_{\beta}^{*})$. Similar algebra shows that $(\textbf{H}_{\eta}^{*\prime}\textbf{H}_{\eta}^{*})^{-1}\textbf{H}_{\eta}^{*\prime}\textbf{w}$ is a sample from $f(\bm{\eta}\vert \by, \bm{\beta},\bm{\xi},\textbf{q}_{\beta} = \bm{0}_{2n},\textbf{q}_{\xi} = \bm{0}_{2n})$, where $\textbf{w}$ is distributed as $\mathrm{MLB}(\textbf{c}_{\eta},\textbf{I}_{2n+r}, \bm{\alpha}_{\eta}^{*},\bm{\kappa}_{\eta}^{*})$. Likewise, $(\textbf{H}_{\xi}^{*\prime}\textbf{H}_{\xi}^{*})^{-1}\textbf{H}_{\xi}^{*\prime}\textbf{w}$ is a sample from $f(\bm{\xi}\vert \by, \bm{\beta},\bm{\eta},\textbf{q}_{\beta} = \bm{0}_{2n},\textbf{q}_{\eta} = \bm{0}_{2n})$, where $\textbf{w}$ is distributed as $\mathrm{MLB}(\textbf{c}_{\xi},\textbf{I}_{3n}, \bm{\alpha}_{\xi}^{*},\bm{\kappa}_{\xi}^{*})$. Figure (\ref{fig1}) shows an example of $\{\pi_{kit}\}$ and the corresponding predicted values from data simulated with this choice of  $\{\pi_{kit}\}$ (details behind this small simulation are given in Appendix B). The estimated values appear to be trending the true values quite well in this small illustration.

Let $\bm{\epsilon} = \bm{\delta} = \bm{0}_{n}$. Then, it is possible to have a shape parameter equal to zero, since it is possible to observe a zero count. Similarly, if $\bm{\epsilon} = \bm{\delta} = \bm{0}_{n}$ and $\rho = 1$, it is possible to have the two shape parameters equal to each other, because it is possible to observe a $y_{kit} = m_{it}$. Thus, the careful attention required to introduce strictly positive $\bm{\epsilon}$, $\bm{\delta}$, and $\rho \in (0,1)$ allows us to use Proposition 1 to directly sample $\bm{\beta}$ in a collapsed Gibbs sampler. 

This example model illustrates the need for $\bm{\epsilon}$, $\bm{\delta}$, $\rho$, and Proposition 1. However, a consequence of this is that there is a considerable amount of bookkeeping needed. Despite the amount of attention required to implement such a collapsed Gibbs sampler, this approach allows one to directly sample from the full-conditional distribution in straightforward manner; and hence, avoids tuning proposal densities in a Metropolis-Hastings algorithm. We demonstrates the computational gains of this type of approach in Section 4.1.


\section{Spatio-Temporal Logistic Regression with Latent Multivariate Logit-Beta Random Effects} The incorporation of low-dimensional MLB random is a significant contribution to the CTM literature. However, modern datasets often exhibit more features than what the illustrative model in Section 2.3 can capture. In particular, many high-dimensional datasets exhibit dependent variation over time (i.e., dynamics). As such, additional model details are required to match this feature often found in modern datasets, and hence, in Section 3 we offer a dynamic modeling version of the illustrative model presented in Section 2.3. We refer to this dynamic model as the MN-STM. A complete summary of the MN-STM, and additional discussion on basic properties of the MN-STM are presented in Appendix A of the Supplementary Material. The full-conditional distributions associated with the MN-STM are derived in Appendix B of the Supplementary Material.

\subsection{The Process Model} Similar to Section 2.3, we make the following assumption on $\{{\nu}_{kit}\}$:
\begin{equation}\label{me}
	v_{kit} = \textbf{x}_{kit}^{\prime}\bm{\beta}_{t} + \bm{\phi}_{kit}^{\prime}\bm{\eta}_{t} + {\xi}_{kit}; \hspace{5pt} k = 1,\ldots, K-1, i = 1,\ldots, N, t = 1,\ldots, T,
\end{equation}
\noindent
where the $p_{t}$-dimensional vector of covariates $\textbf{x}_{kit}$ are assumed known, and $\bm{\beta}_{t} \in \mathbb{R}^{p_{t}}$ is an associated unknown $p_{t}$-dimensional vector. The $r$-dimensional real-valued vector $\bm{\phi}_{kit}$ is known, and $\bm{\eta}_{t}$ is an $r$-dimensional random vector with elements that are possibly correlated. The elements of $\{\xi_{kit}\}$ are assumed to be independent and identically distributed. 

The prior for the $p = \sum_{t=1}^{T}p_{t}$ dimensional vector $\bm{\beta} = (\bm{\beta}_{1}^{\prime},\ldots, \bm{\beta}_{T}^{\prime})^{\prime}$ is assumed to be $f_{MLB}$ with location-parameter-zero; has unknown shape parameters, $\bm{\alpha}_{\beta} = ((\epsilon/\sigma)\bm{1}_{n}^{\prime},\alpha_{\beta}\bm{1}_{p}^{\prime})^{\prime}$, $\epsilon>0$, $\alpha_{\beta}>\epsilon$,  $\bm{\kappa}_{\beta} = (\bm{\delta}^{\prime},\kappa_{\beta}\bm{1}_{p}^{\prime})^{\prime}$, and the elements of the $n$-dimensional vector $\bm{\delta}$ are positive, where $\kappa_{\beta}>\alpha_{\beta}$; and precision parameter $\textbf{H}_{\beta} = \left(\sigma\textbf{X}^{\prime}, \textbf{I}_{p}^{\prime}\right)^{\prime}$, where the $(K-1)N_{t}\times p$ matrix $\textbf{X}_{t} = \left(\textbf{x}_{kit}^{\prime}: k = 1,\ldots, K-1, A_{i}\right.$ $\left.\in D_{t}\right)$, the $n\times p$ block diagonal matrix $\textbf{X} = \mathrm{diag}\left(\textbf{X}_{1},\ldots, \textbf{X}_{T}\right)$, and $\textbf{I}_{p}$ is a $p\times p$ identity matrix. The $r$-dimensional random vector $\bm{\eta}_{t}$ is assumed to be $f_{MLB}$ with location-parameter $(\bm{\eta}_{t-1}^{\prime}\textbf{M}_{t}\bm{\Phi}_{t}^{\prime},\bm{\eta}_{t-1}^{\prime}\textbf{M}_{t}^{\prime}\textbf{V}_{t}^{\prime})^{\prime}$ and unknown shape parameters $\bm{\alpha}_{t} = ((\epsilon/\sigma)\bm{1}_{N_{t}}^{\prime},\alpha_{t}\bm{1}_{r}^{\prime})^{\prime}$, $\alpha_{t}>\epsilon$, and $\bm{\kappa}_{t} =(\bm{\delta}_{t}^{\prime},\kappa_{t}\bm{1}_{r}^{\prime})^{\prime}$, where $\kappa_{t}>\alpha_{t}$, the $(K-1)N_{t}$-dimensional vectors $\bm{\delta}_{t}$ stack to produce $\bm{\delta} = (\bm{\delta}_{1}^{\prime},\ldots, \bm{\delta}_{T}^{\prime})^{\prime}$, the precision parameter $\textbf{H}_{t} = \left(\bm{\Phi}_{t}^{\prime}, \textbf{V}_{t}^{\prime}\right)^{\prime}$, and $\textbf{V}_{t} \in \mathbb{R}^{r}\times \mathbb{R}^{r}$ is unknown. For a given $t$, define the $N_{t}$-dimensional vector $\bm{\xi}_{t} = (\xi_{kit}: k = 1,\ldots, K-1, A_{i}\in D_{t})$ follow a $f_{MLB}$ with zero location parameter, precision parameter $\textbf{H}_{\xi,t} = (\textbf{I}_{N_{t}}^{\prime},\textbf{I}_{N_{t}}^{\prime})^{\prime}$, and constant (across $k$ and $i$) shape parameters $\alpha_{\xi,t}>0$ and $\kappa_{\xi,t}>\alpha_{\xi,t}$.

The specification of $\{\bm{\phi}_{kit}\}$ is extremely important. If $\{\bm{\phi}_{kit}\}$ is chosen poorly, then the fixed effects $\bm{\beta}_{t}$ and the random effects $\bm{\eta}_{t}$ may be confounded \citep[e.g., see][and the references therein]{reich2}. Consider the extreme case where $\textbf{x}_{kit} = \bm{\phi}_{kit}$. Under this specification there is no way to disentangle the fixed effect $\bm{\beta}_{t}$ and the random effect $\bm{\eta}_{t}$ when estimating the expected values of $\{\nu_{kit}\}$. One solution to this problem is to specify $\bm{\phi}_{kit}$ so that it belongs to the orthogonal column space of $\{\textbf{x}_{kit}\}$ \citep[see,][]{griffith2000, griffith2002, griffith2004, griffith2007}. An approach used for areal data is to consider a spatially weighted version of the orthogonal complement. This is referred to as the so-called ``Moran's I'' basis functions \citep[see,][]{hughes,aaronp,burden2015sar,bradleyMSTM}. This spatially weighted orthogonal complement is the classical Moran's I operator \citep[see,][]{MoranI, hughes}.  Specifically, let the $(K-1)N\times p$ matrix $\textbf{X}_{t}^{\mathrm{P}} = \left(\textbf{x}_{kit}^{\prime}: k = 1,\ldots, K-1, i = 1,\ldots, N\right)$, where ``P'' represents ``prediction locations'' and delineates between $\textbf{X}_{t}$ which only stacks the covariates over observed regions. The Moran's I operator is written as,
\begin{equation}\label{mi}
	\textbf{MI}(\textbf{X}_{t}^{\mathrm{P}},\textbf{A}) \equiv\left(\textbf{I}_{(K-1)N}- \textbf{X}_{t}^{\mathrm{P}}\left(\textbf{X}_{t}^{\mathrm{P}\prime}\textbf{X}_{t}^{\mathrm{P}}\right)^{-1}\textbf{X}_{t}^{\mathrm{P}\prime}\right)\textbf{A}\left(\textbf{I}_{(K-1)N}- \textbf{X}_{t}^{\mathrm{P}}\left(\textbf{X}_{t}^{\mathrm{P}\prime}\textbf{X}_{t}^{\mathrm{P}}\right)^{-1}\textbf{X}_{t}^{\mathrm{P}\prime}\right); \hspace{5pt} t = 1,\ldots, T,
\end{equation}
where $\textbf{I}_{(K-1)N}$ is an $(K-1)N\times (K-1)N$ identity matrix, and $\textbf{A}$ is the $(K-1)N\times (K-1)N$ adjacency matrix associated with the edges formed by $\{A_{i}: i = 1,\ldots,N\}$. One choice for the adjacency matrix is to set the $(i,j)$-th element of the matrix $\textbf{A}$ equal to one if the areal unit $A_{i}$ is a neighbor of $A_{j}$ \citep{cressie,banerjee-etal-2004}. Let $\bm{\Psi}_{t}\bm{\Lambda}_{t}\bm{\Psi}_{t}^{\prime}$ be the spectral decomposition of the $(K-1)N\times (K-1)N$ matrix $\textbf{MI}(\textbf{X}_{t}^{\mathrm{P}},\textbf{A}_{t})$. Then set $\bm{\Phi}_{t}^{\mathrm{P}} = \left(\bm{\phi}_{kit}^{\prime}: k = 1,\ldots,K-1, i = 1,\ldots, N\right)$ equal to the first $r$ columns of $\bm{\Psi}_{t}$. 


For a given $i$ and $t$ the model in Equation (\ref{me}) is a type of spatial mixed effects model \citep[e.g., see][among others]{johan}. The random vector $\bm{\eta}_{t}$ is the ``shared random effect'' discussed in the Introduction, meaning the expressions of $\nu_{kit}$ and $\nu_{mjt}$ using Equation (\ref{me}) both contain the \textit{same} random vector $\bm{\eta}_{t}$ for $k\ne m$ and $i \ne j$. This shared random effect induces cross correlations among different categories (i.e., $k$ and $m$) and different spatial regions (i.e., $i$ and $j$), since
\begin{equation*}
	\mathrm{cov}(\nu_{kit},\nu_{mjt}\vert \bm{\beta},\bm{\theta}) = \bm{\phi}_{kit}^{\prime}\mathrm{cov}(\bm{\eta}_{t}\vert \bm{\beta},\bm{\theta})\bm{\phi}_{mjt},
\end{equation*}
\noindent
which is not equal to zero. As discussed in the Introduction, there are several examples of this technique in the spatial, multivariate spatial, spatio-temporal, and multivariate spatio-temporal literature.

\subsection{Spatio-Temporal Dynamics} Both \citet{blei2006_2} and \citet{nipps} consider the use of a VAR(1) model to incorporate dynamics within multinomial data. In this article, we consider incorporating a specific type of VAR(1) model used in \citet{bradleyMSTM} and \citet{bradleyPMSTM}. 

The VAR(1) assumption for the $r$-dimensional non-spatially referenced random vector $\bm{\eta}_{t}$ is given by \citep[e.g., see][Chap. 7]{cressie-wikle-book},
\begin{align}\label{autoregressive}
	\bm{\eta}_{t} &= \textbf{M}_{t}\bm{\eta}_{t-1}+ \bu_{t};\hspace{5pt}t = 2,3,\ldots,T,
\end{align}
\noindent
where $\textbf{M}_{t}$ is a $r \times r$ \textit{known} propagator matrix (see discussion below), and $\bu_{t}$ is an $r$-dimensional random vector. 
We use the so-called Moran's I propagator matrix introduced in \citet{bradleyMSTM}. The derivation of this matrix starts by substituting the VAR(1) representation into the mixed effects model in (\ref{me}), at all prediction locations, to obtain
\begin{equation}\label{dynamics}
	\bm{\nu}_{t}^{\mathrm{P}} = \textbf{X}_{t}^{\mathrm{P}}\bm{\beta} + \bm{\Phi}_{t}^{\mathrm{P}}\textbf{M}_{t}\bm{\eta}_{t-1} + \bm{\Phi}_{t}^{\mathrm{P}}\textbf{u}_{t} + \bm{\xi}_{t}^{\mathrm{P}}.
\end{equation}
\noindent
Then, straightforward algebra leads to,
\begin{equation*}
	\bm{\Phi}_{t}^{\prime}(\bm{\nu}_{t}^{\mathrm{P}}-\bm{\xi}_{t}^{\mathrm{P}}) = \textbf{B}_{t}\bm{\zeta}_{t} + \textbf{M}_{t}\bm{\eta}_{t-1};\hspace{5pt} t = 2,\ldots, T,
\end{equation*}
where the $r\times(p+r)$ matrix $\textbf{B}_{t}\equiv (\bm{\Phi}_{t}^{\mathrm{P}\prime}\textbf{X}_{t}^{\mathrm{P}}, \textbf{I})$ and the $(p+r)$-dimensional random vector $\bm{\zeta}_{t} \equiv (\bm{\beta}^{\prime},\bu_{t}^{\prime})^{\prime}$. Similar to the derivation of $\bm{\Phi}_{t}$, we let $\textbf{M}_{t}$ equal the $r$ eigenvectors of $\textbf{MI}(\textbf{B}_{t},\textbf{U}_{t})$, where in general $\textbf{U}_{t}$ is a real-valued $r\times r$ ``weight'' matrix. In Section~5, we set $\textbf{U}_{t}\equiv \textbf{I}_{r}$.

When introducing a VAR(1) model for forecasting, it is especially important to verify that the Wold representation of the VAR(1) exists \citep[e.g., see][for a standard reference]{andersonts}. This implies that the VAR(1) model is stable. The word ``stable'' means that the variance of the random effects $\bm{\eta}_{t}$ does not drift to infinity as $t$ increases. Our primary use of the VAR(1) model is to incorporate time-dependence to aid in predicting over the range times that were observed (i.e., smoothing), and not forecasting. However, it may be of interest to use our VAR(1) model for forecasting in future research. Thus, we provide a result that gives the necessary conditions for the MI propagator to be used to define a stable process \citep[e.g., see][pg. 688]{newintrotimeseries}.\\

\noindent
\textit{Proposition 2: Consider the VAR(1) model in (\ref{autoregressive}). Denote the first $r$ eigenvectors of $\textbf{MI}(\textbf{B}_{t},\textbf{U}_{t})$ with $\bm{\Psi}_{t}$ and let $\textbf{U}_{t}$ be a generic $r \times r$ real-valued matrix. For $t = 1,\ldots, t^{*}$, let $\textbf{M}_{t}=\bm{\Psi}_{t}$. Also, for $t = t^{*}+1,\ldots,$ let $\textbf{M}_{t} = \rho \bm{\Psi}_{t}$ for some $\rho \in (0,1)$. Then, the Wold representation of the VAR(1) model in (\ref{autoregressive}) converges in $L_{2}$ to a stable sequence of $r$-dimensional random vectors.}\\

\noindent
\textit{Proof:} See Appendix A.\\

\noindent
In Section 5, we set $t^{*} = T$ because we are only interested in smoothing. However, if forecasting is of interest, one might place a prior distribution on $\rho$ and set $t^{*} = 1$.

\subsection{Marginal Covariance of the Random Effects} For computational reasons, we do not include all random effects (i.e., $r\ll n$) and confounded random effects in (\ref{me}). Thus, we choose $\{\textbf{V}_{t}\}$ so that the spatially dependent term in (\ref{me}), namely $\bm{\Phi}_{t}^{\mathrm{P}}\bm{\eta}_{t}$, has a precision (i.e., inverse of the covariance matrix) similar to the precision from an intrinsic conditional autoregressive (ICAR) model, which is given by $\frac{1}{\sigma_{\eta}^{2}}(\textbf{I} - \textbf{A})$ for $\sigma_{\eta}^{2}>0$. It is well-known that ICAR models induces covariances that are functions of neighborhoods and are not functions of distances between locations; thus, the ICAR model enforces nonstationary spatial dependence \citep{besag-74}. 

We let $\textbf{V}_{t} = \textbf{V}_{t}^{*}(\textbf{A})$, where
\begin{equation}\label{Kstar}
\textbf{V}_{t}^{*}(\textbf{A})=\underset{\textbf{V}_{t}}{\mathrm{arg\hspace{5pt}min}}\left\lbrace \Big\Vert\frac{1}{\sigma_{\eta}^{2}}(\textbf{I} - \textbf{A}) - \mathrm{cov}(\bm{\Phi}_{t}^{\mathrm{P}}\bm{\eta}_{t}\vert \bm{\eta}_{t-1},\textbf{V}_{t},\alpha_{t}, \kappa_{t})^{-}\Big\Vert_{\mathrm{F}}^{2}\right\rbrace;\hspace{5pt} t = 1,\ldots,T.
\end{equation}
\noindent
For a square real-valued matrix $\textbf{C}$ let $\textbf{C}^{-}$ be the generalized inverse of $\textbf{C}$ and let the Frobenius norm $||\textbf{C}||_{\mathrm{F}}^{2} = \mathrm{trace}\left(\textbf{C}^{\prime}\textbf{C}\right)$. In (\ref{Kstar}), we minimize the Frobenius norm across the space of all real-valued matrices. The general strategy in (\ref{Kstar}) is that same as in \citet{bradleyMSTM} and \citet{bradleyPMSTM}; however, the term $\mathrm{cov}(\bm{\Phi}^{\mathrm{P}}\bm{\eta}_{t}\vert \bm{\eta}_{t-1}, \textbf{V}_{t},\alpha_{t}, \kappa_{t})$ is different from the associated covariance terms in \citet{bradleyMSTM} and \citet{bradleyPMSTM} because we use the MLB distribution. \\

\noindent
\textit{Proposition 3: Let the model in (\ref{me}) hold, and let $\textbf{P}$ be a generic $(K-1)N\times (K-1)N$ positive semi-definite matrix. Then the values of $\sigma_{\eta}^{2}$ and $\textbf{V}_{t}$ that minimize,
	\begin{equation}\label{KstarP}
	\underset{\textbf{V}_{t}}{\mathrm{arg\hspace{5pt}min}}\left\lbrace  \Big\Vert\frac{1}{\sigma_{\eta}^{2}}\textbf{P} - \mathrm{cov}(\bm{\Phi}_{t}^{\mathrm{P}}\bm{\eta}_{t}\vert \bm{\eta}_{t-1}, \textbf{V}_{t},\alpha_{t}, \kappa_{t})^{-}\Big\Vert_{\mathrm{F}}^{2}\right\rbrace;\hspace{5pt} t = 1,\ldots,T.
	\end{equation}
	are given by:
	\vspace{-40pt}
	\begin{align*}
	\sigma_{\eta}^{2} &= g(\alpha_{t}) + g(\kappa_{t}-\alpha_{t}),\\
	\textbf{V}_{t}&= \bm{\Lambda}_{\Sigma}^{1/2}\bm{\Psi}_{\Sigma},
	\end{align*}
	where $\bm{\Sigma}^{*}\equiv \mathcal{A}^{+}\left\lbrace \bm{\Phi}^{\mathrm{P}\prime}\textbf{P}\bm{\Phi}^{\mathrm{P}}-\left(\bm{\Phi}^{\prime}\bm{\Phi}\right)\right\rbrace$, $\mathcal{A}(\textbf{C})$ is the best positive approximate \citet{Higham} of the matrix $\textbf{C}$, $\bm{\Psi}_{\Sigma}\bm{\Lambda}_{\Sigma}\bm{\Psi}_{\Sigma}^{\prime}$ is the spectral decomposition of $\bm{\Sigma}^{*}$, and $g(\cdot)$ is the trigamma function \citep{varlogit}.\\
}

\noindent
\textit{Proof:} See Appendix A.\\

\noindent
We set $\textbf{P}$ in (\ref{KstarP}) to $\textbf{I} - \textbf{A}$. One could also choose to specify the joint precision of $\bm{\eta}\equiv \left(\bm{\eta}_{1}^{\prime},\ldots, \bm{\eta}_{T}^{\prime}\right)^{\prime}$ to be close to a spatio-temporal autoregressive model. However, the act of stacking the basis functions $\bm{\Phi}_{t}$ over $t$ quickly leads to memory issues. Consequently, we specify $\textbf{V}_{t}$ marginally over each $t = 1,\ldots, T$ using Proposition 3. Additionally, Proposition 3 shows that the shape parameters define the variances of the random effects $\bm{\eta}_{t}$, and consequently, we apply gamma priors to every shape parameter. Thus we include shape parameter priors for $\bm{\beta}$ and $\xi_{t}$ (see Appendix A of the Supplementary Material for more details).


\subsection{A Latent Gaussian Process Model Representation of the MN-STM} The latent Gaussian process model is a well established method used to analyze spatial and spatio-temporal data \citep[e.g., see][among others]{gs}, and hence, one might be hesitant to adopt a new distributional framework. Motivated by this, we show that the latent MLB model can be written as a latent Gaussian process model. Our strategy is to augment the MLB likelihood using P\'{o}ly-gamma random variables similar to \citet{polson}. The main difference between our result and \citet{polson} is that \citet{polson} augments the binomial pmf, while we augment the MLB distribution.\\

\noindent
\textit{{Proposition 4:} Let $\bm{\zeta}=\left(\alpha_{1}-\kappa_{1}/2,\ldots, \alpha_{M}-\kappa_{M}/2\right)$, $\bm{\gamma} = \left(\textbf{H}^{\prime}\bm{\Omega}\textbf{H}\right)^{-1}\textbf{H}^{\prime}\left(\textbf{H}\bm{\mu}_{1}-\bm{\zeta}\right)$, and $\bm{\Omega} = \mathrm{diag}\left(\omega_{1},\ldots, \omega_{M}\right)$, where $\omega_{i}$ is a P\'{o}ly-gamma random variable with density $p(\omega\vert b)$. The definition of $p(\omega\vert b)$ is given in Appendix A. Denote the joint pdf of the $M$ independent P\'{o}ly-gamma random variables $\omega_{1},\ldots, \omega_{M}$ with $p(\bm{\omega}\vert \bm{\kappa}) = \prod_{i = 1}^{M}p(\omega_{i}\vert \kappa_{i})$. Then,
	\begin{align*}
	& f_{MLB}(\textbf{q}_{1}\vert \bm{\mu}_{1}, \textbf{V} = (\textbf{H},\textbf{B}), \bm{\alpha}, \bm{\kappa}) \\
	&=\int g(\bm{\Omega},\bm{\mu}_{1}, \textbf{H}, \bm{\alpha}, \bm{\kappa})\frac{\mathrm{det}(\textbf{H}^{\prime}\bm{\Omega}\textbf{H})^{1/2}}{(2\pi)^{M/2}} \mathrm{exp}\left[ - (\textbf{q}_{1}-\bm{\gamma})^{\prime}\textbf{H}^{\prime}\bm{\Omega}\textbf{H}(\textbf{q}_{1}-\bm{\gamma})/2\right] p(\bm{\omega}\vert \bm{\kappa})d\bm{\omega},
	\end{align*}
	where $g$ is a strictly postive real-valued function of $\bm{\Omega},\bm{\mu}_{1}, \textbf{H}, \bm{\alpha},$ and $\bm{\kappa}$.\\
}

\noindent
\textit{Proof:} See Appendix A.\\

\noindent
Proposition 4 suggests that the MLB distribution can be seen as an infinite mixture of normal densities (i.e., using the Reimann sum expression of the integral). This implies that the MN-STM can be augmented by random variables so that it can be re-expressed as a latent Gaussian process model (see Appendix C  of the Supplementary Material for this expression). 
	\begin{figure}[!b]
		\begin{center}
			\begin{tabular}{lll}
				\includegraphics[width=5.5cm,height=5cm]{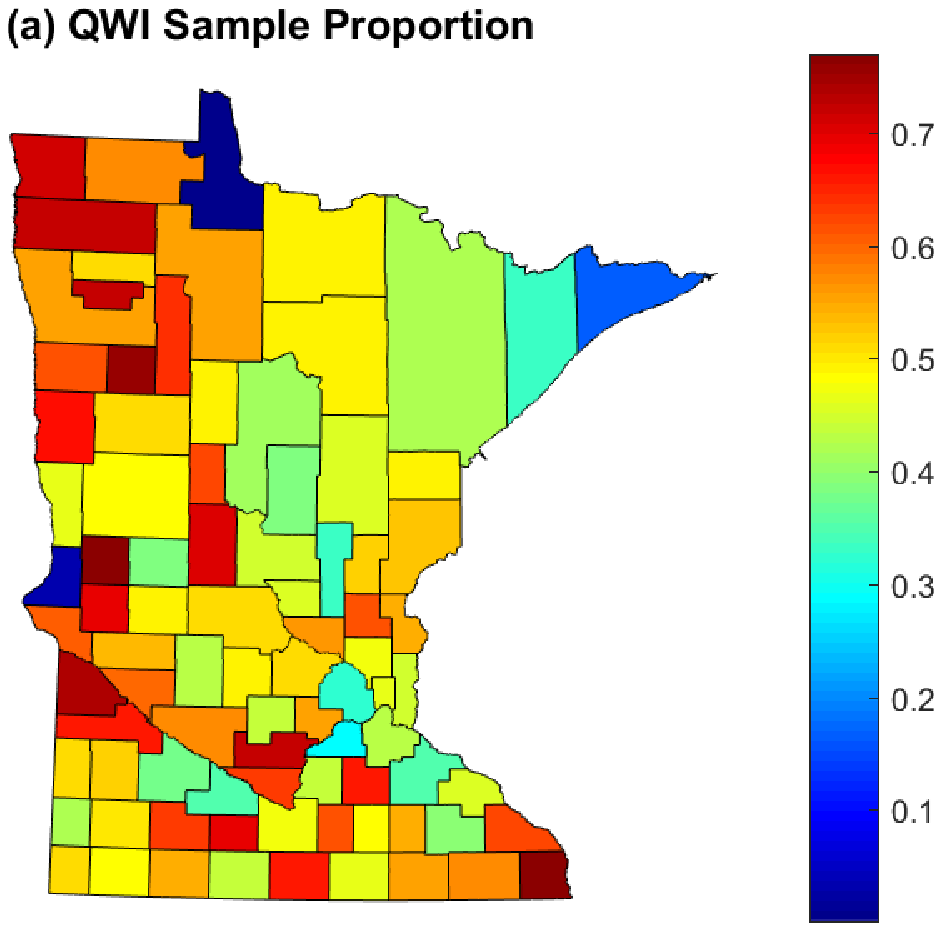}&
				\includegraphics[width=5.5cm,height=5cm]{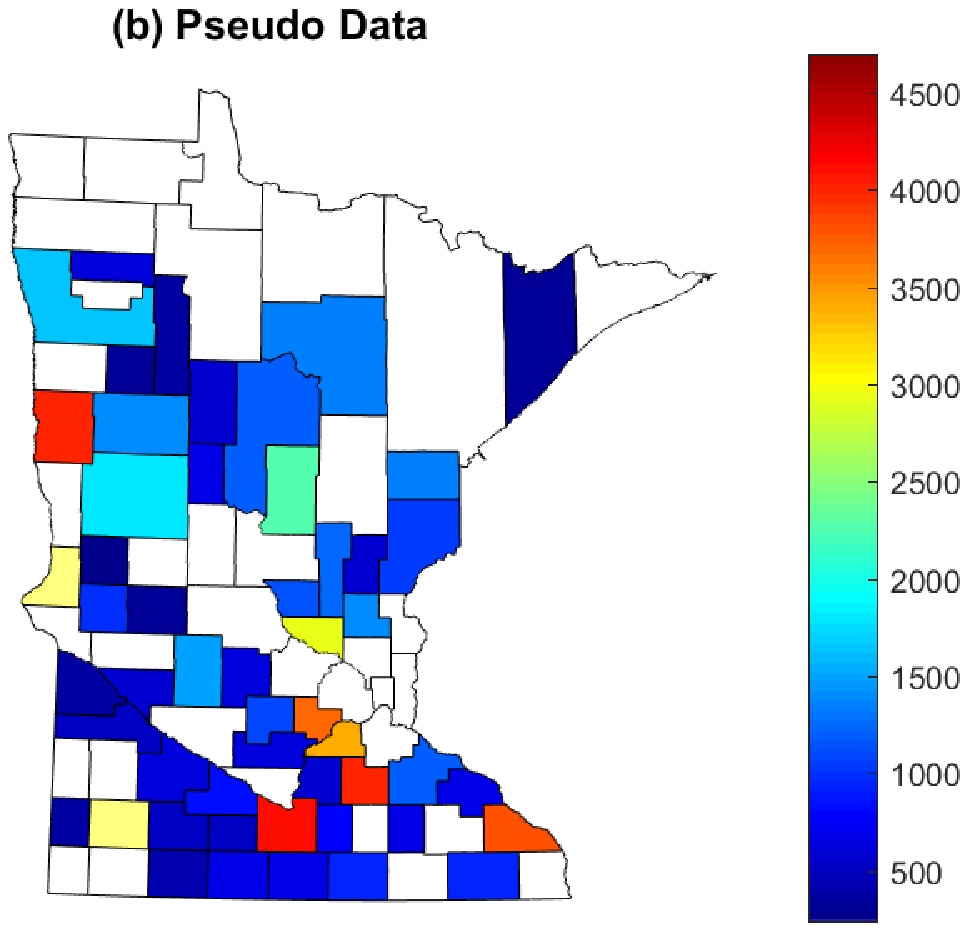}& \includegraphics[width=5.5cm,height=5cm]{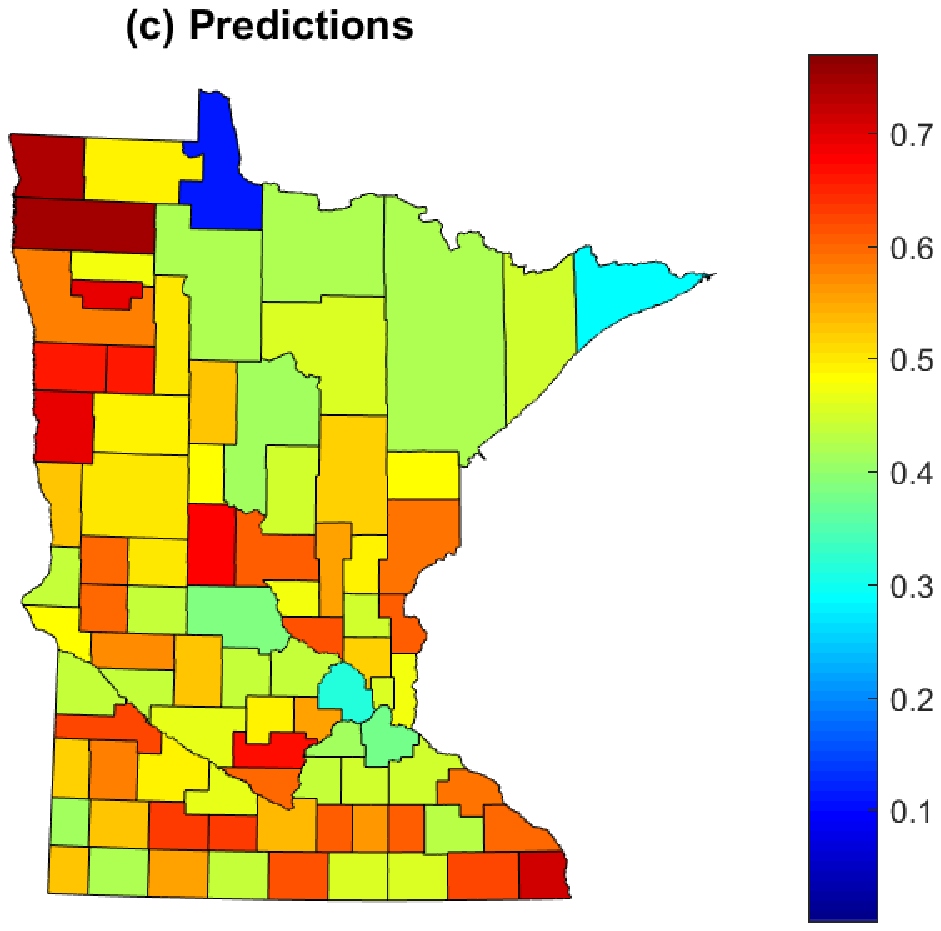}
			\end{tabular}
			\caption{\baselineskip=10pt{In (a), we plot the LEHD estimated proportion of individuals employed in the beginning of the 4-th quarter of 2013 within the information industry in Minnesota, and the predicted values. In (b), we plot the corresponding pseudo-data, in (c) we plot the corresponding predicted values. The colorbars in Panels (a) and (c) are the same, but are different from Panel (b).}}\label{fig2}
		\end{center}
	\end{figure}
	\section{Empirical Results} 
		\subsection{A Simulation Study}Both \citet{bradleyMSTM} and \citet{bradleyPMSTM} conduct an \textit{empirical simulation analysis} to understand the out-sample performance of their model if they were able to obtain independent replicates of QWIs. Specifically, the unknown parameters in the data model are replaced by public-use QWIs. Simulated values are then  generated from this empirical data model, and the simulated data are used for prediction. The main motivation of this approach is that it yields simulated values that are similar to what might be seen in practice. We adopt this strategy to build our simulation model.
		
		Let $y_{kit}$ be the beginning of quarter employment QWI, at quarter $t$, for industry $k$, and Minnesota county $A_{i}$. We simulate data, denoted by $R_{kit}$, from a multinomial distribution with probability $\frac{y_{kit}+1}{m_{it}+3}$ for $k = 1,2,3,$ $t = 76,\ldots,96,$ and $A_{i} \in D_{\mathrm{MN},t}$. Let $D_{\mathrm{MN},t}$ consist of counties in Minnesota (MN) that have available QWIs. Here, $k = 1$ denotes the information industry, $k = 2$ represents the professional, scientific and technical services industry, and $k = 3$ represents the finance industry. We add 1 in the numerator and 3 in the denominator of the probability of success so that the proportions are greater than zero and sum to one.  We randomly select 65$\%$ of the areal units in $D_{\mathrm{MN},t}$ to be ``observed.'' The covariates $\textbf{x}_{kit} = (1,I(k = 1),\ldots,I(k = 3), I(t = 1),\ldots,I(t = 1,\ldots,95))^{\prime}$, where $I(\cdot)$ is the indicator function. These covariates were the same used in the simulation study in \citet{bradleyPMSTM}. We set $r$ equal to approximately the top $10\%$ of the available basis functions. In this case, we set $r = 36$.

\begin{figure}[t!]
		\begin{center}
			\begin{tabular}{c}
				\includegraphics[width=8.5cm,height=5cm]{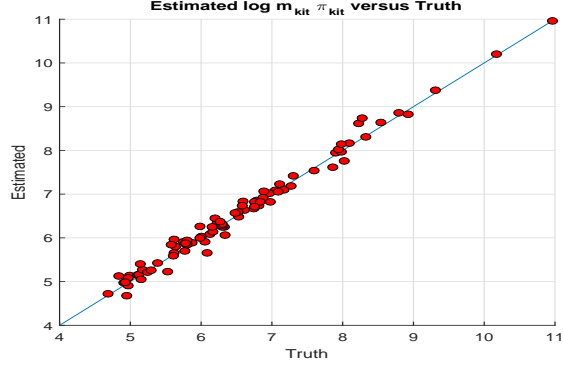}
			\end{tabular}
			\caption{\baselineskip=10pt{The log of the estimated $m_{kit}\pi_{kit}$ versus the true log $m_{kit}\pi_{kit}$, for $t = 10$ and $k = 2$. Plots of the remaining time points display similar patterns.}}\label{fig3}
		\end{center}
	\end{figure}

In {Figure (\ref{fig2})}, we present sample proportions using the QWIs, the simulated data, and the predictor $\sum_{b}\pi_{kit}^{[b]}/B$ (see Section 3.5). The predictions are fairly accurate (i.e., the predictions are close to the sample proportions). The high quality of these predictions are further supported by Figure (\ref{fig3}), where we display a scatterplot of the true proportions versus the estimated proportions. 

\begin{figure}[t!]
		\begin{center}
			\begin{tabular}{c}
				\includegraphics[width=8.5cm,height=5cm]{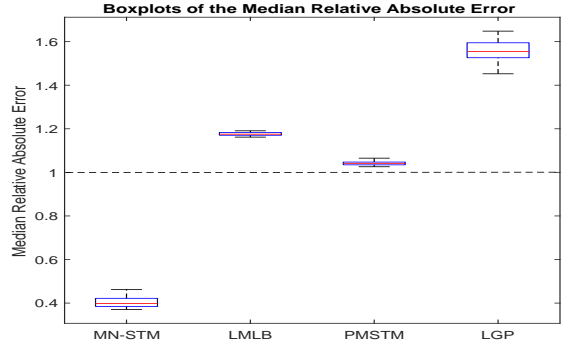}
			\end{tabular}
			\caption{\baselineskip=10pt{The median relative absolute error in (\ref{MRAE}) over 50 independent replicates of $\{R_{kit}\}$ for four methods of estimation, including the MN-STM, the model in Section 2.3 labled as the latent MLB (LMLB) model, the PMSTM, and a latent Gaussian process model (LGP). Values close to one show that the estimated value produces estimates that are comparable to the variability of the data at observed locations, and values less (greater) than one show an increased (decreased) performance relative to the variability of the data at observed locations. The dashed line indicates the a median relative absolute error of one.}}\label{fig4}
		\end{center}
	\end{figure}

We repeated this example over 50 independent replications of the set $\{R_{kit}\}$, and compare to the performance of the estimates using the MN-STM, PMSTM, and the latent Gaussian process (LGP) model considered in \citet{bradleyPMSTM}. In Figure (\ref{fig4}) we plot boxplots of the median relative absolute error,
\begin{equation}\label{MRAE}
\underset{kit}{\mathrm{median}}\left\lbrace\frac{\mathrm{abs}\left(m_{kit}\hat{\pi}_{kit} - m_{kit}{\pi}_{kit} \right)}{ m_{kit}{\pi}_{kit}(1- {\pi}_{kit})}\right\rbrace,
\end{equation}
\noindent
where ``median'' represents the median function over $k$, $i$, and $t$ (including $k$, $i$, and $t$ for unobserved $R_{kit}$), ``abs'' denotes the absolute value function, and $\hat{\pi}_{kit}$ represents a generic estimate of ${\pi}_{kit}$. Values close to one show that the estimated value produces estimates that are comparable to the variability of the data at observed locations, and values less (greater) than one show an increased (decreased) performance relative to the variability of the data at observed locations. In Figure (\ref{fig4}), we see that both the latent MLB (LMLB) model introduced in Section 2.3, and the P-MSTM performs reasonably well in terms of median relative absolute error with values close 1, while the LGP performs considerably worse than the P-MSTM. The MN-STM clearly outperforms the three competitors with median absolute errors close to 0.4. Both the PMSTM and the MN-STM appear to outperform the MLB, which does not explicitly model the dynamics through a VAR(1) model.

\begin{figure}[t!]
		\begin{center}
			\begin{tabular}{cc}
				\includegraphics[width=7.5cm,height=5cm]{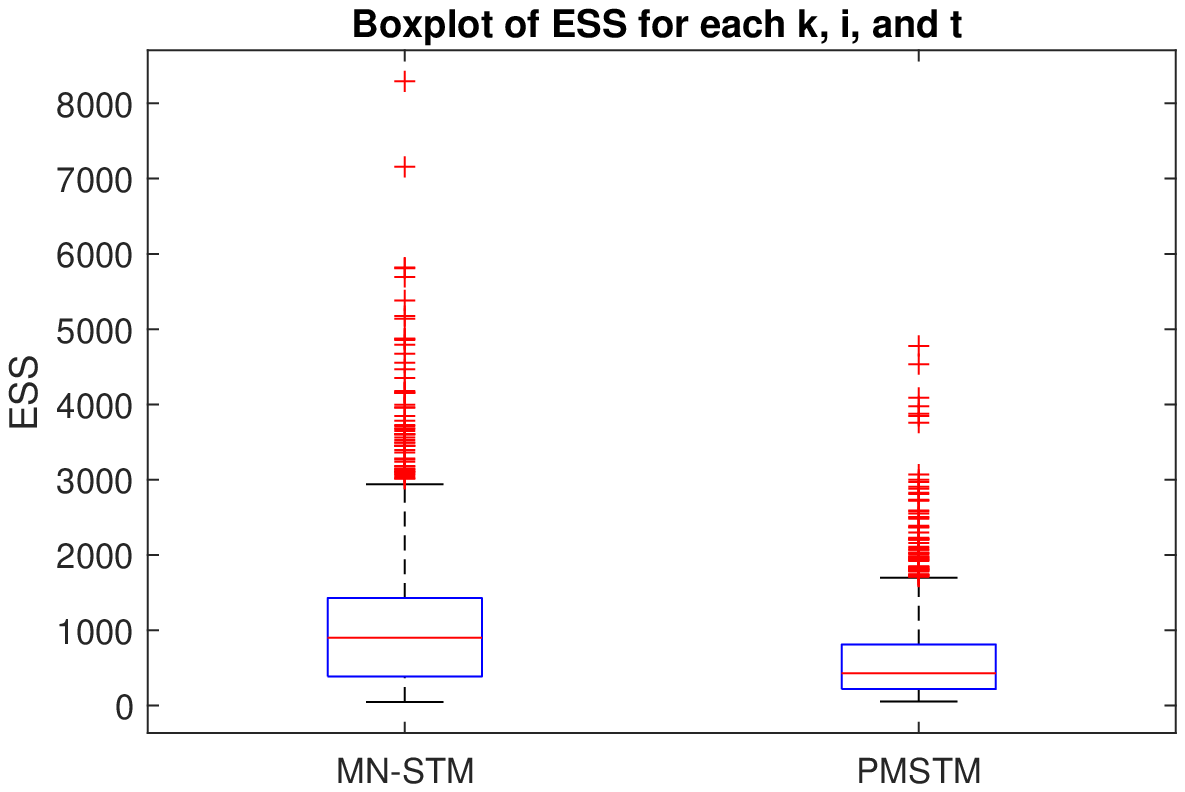}&\includegraphics[width=7.5cm,height=5cm]{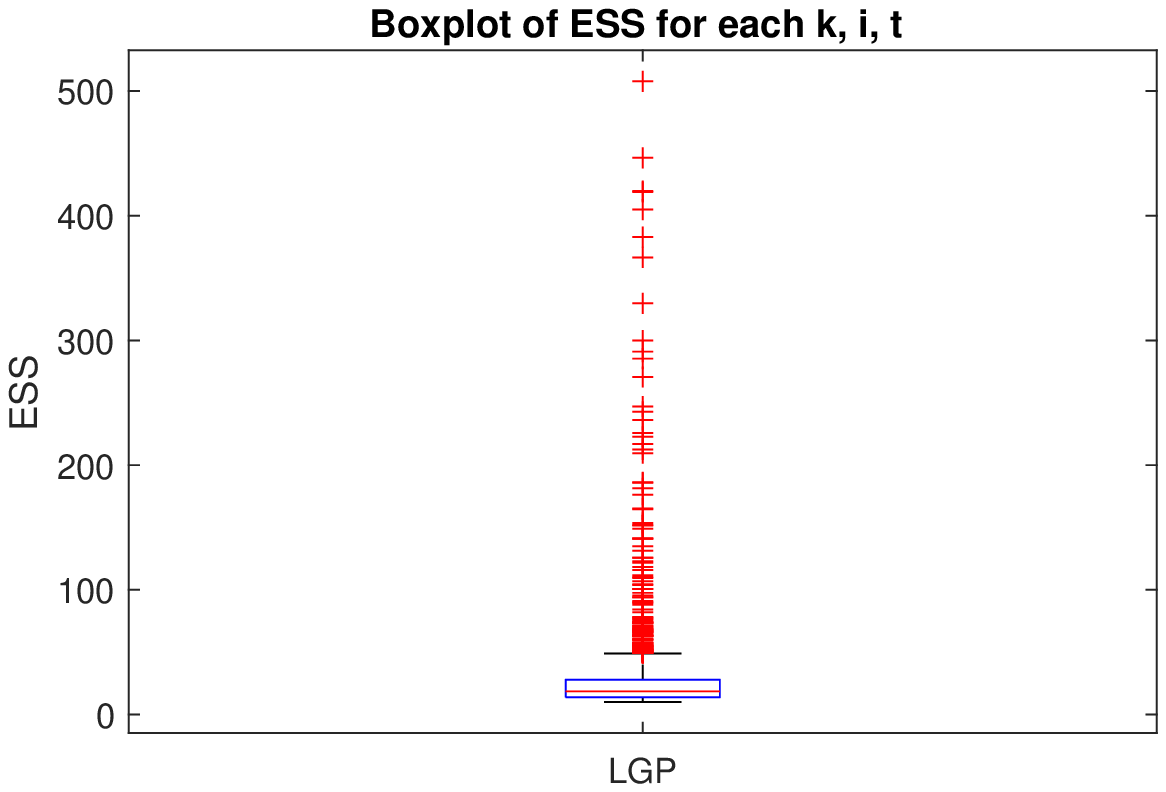}
			\end{tabular}
			\caption{\baselineskip=10pt{Boxplot of ESS ( over $k$, $i$, and $t$), for a single run of the MCMC (post burnin, $B = 1,000$) and for a single generation of $\{R_{kit}\}$. The left panel presents the ESS for MN-STM and PMSTM, and the right panel presents the ESS associated with a LGP. The axes between the two panels have a different scale.}}\label{fig5}
		\end{center}
	\end{figure}

The computational performance of the MN-STM is also important to investigate. We use the effective sample size (ESS) to assess the performance of the MCMC chain, which is computed as the length of the MCMC ($B$) times the ratio of the within chain variance and the between chain variance. Small (large) values of ESS suggest that the MCMC chain has positive (negative) correlations between values in the chain. This has allowed many to use ESS as a measure of the efficiency of the MCMC, since ESS close to or larger than the length of the MCMC implies an efficient MCMC (standard references include \citet{Kass}, \citet{Liu}, \citet{Robert}, and \citet{Gong} for component-wise ESS, and \citet{Vats} for a multivariate ESS). In Figure (\ref{fig5}), we show a boxplot of ESS ( over $k$, $i$, and $t$), for a single run of the MCMC (post burnin, $B = 1,000$) and for a single generation of $\{R_{kit}\}$. The component-wise ESS of MN-STM and P-MSTM are comparable with the MN-STM outperforming the P-MSTM (the median ESS values are 899 and 430, respectively), and indicates computationally efficient Markov chains. The ESS of the LGP model is strikingly smaller than 1,000 suggesting that the Metropolis-within-Gibbs implementation of this LGP is extremely inefficient (the median ESS is approximately 19).

	\subsection{Application: Predicting the Probability of Employment in a NAICS Sector}
	
	\begin{figure}[t!] 
		\begin{tabular}{ccc}\includegraphics[width=5.5cm,height=5cm]{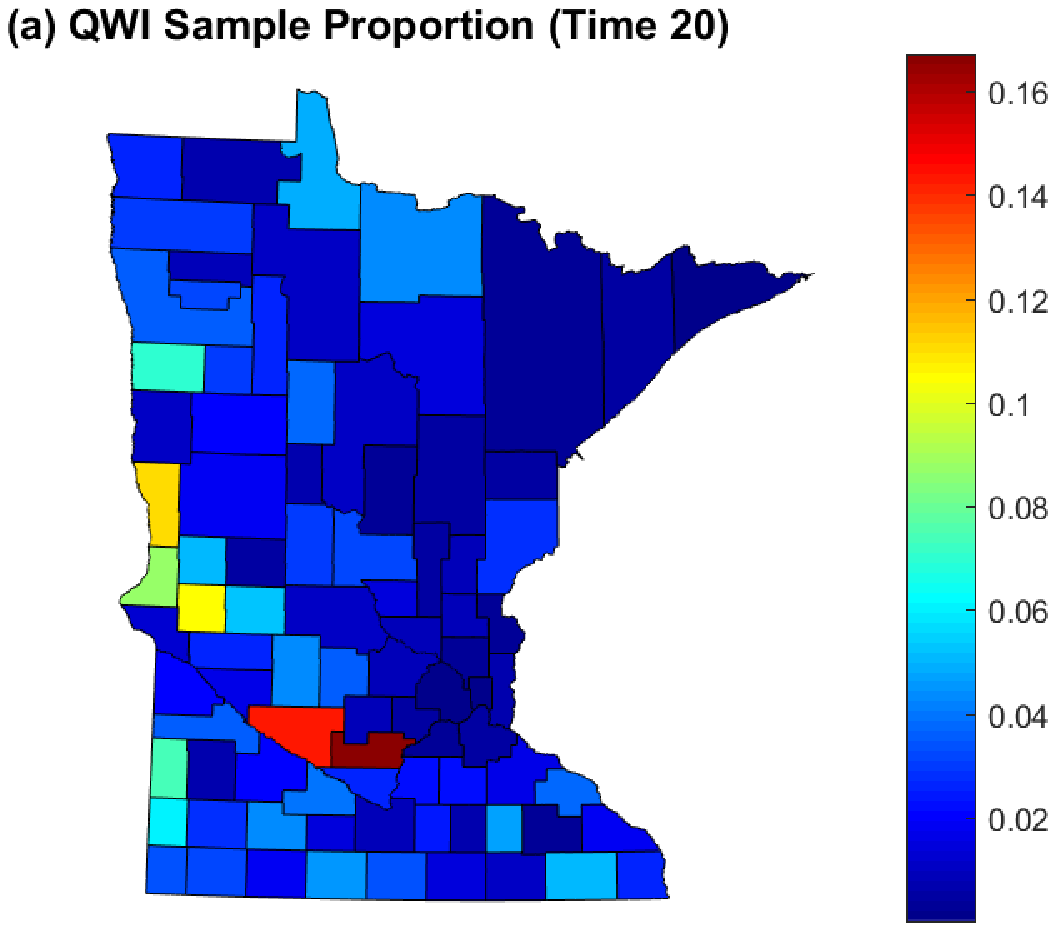}&
			\includegraphics[width=5.5cm,height=5cm]{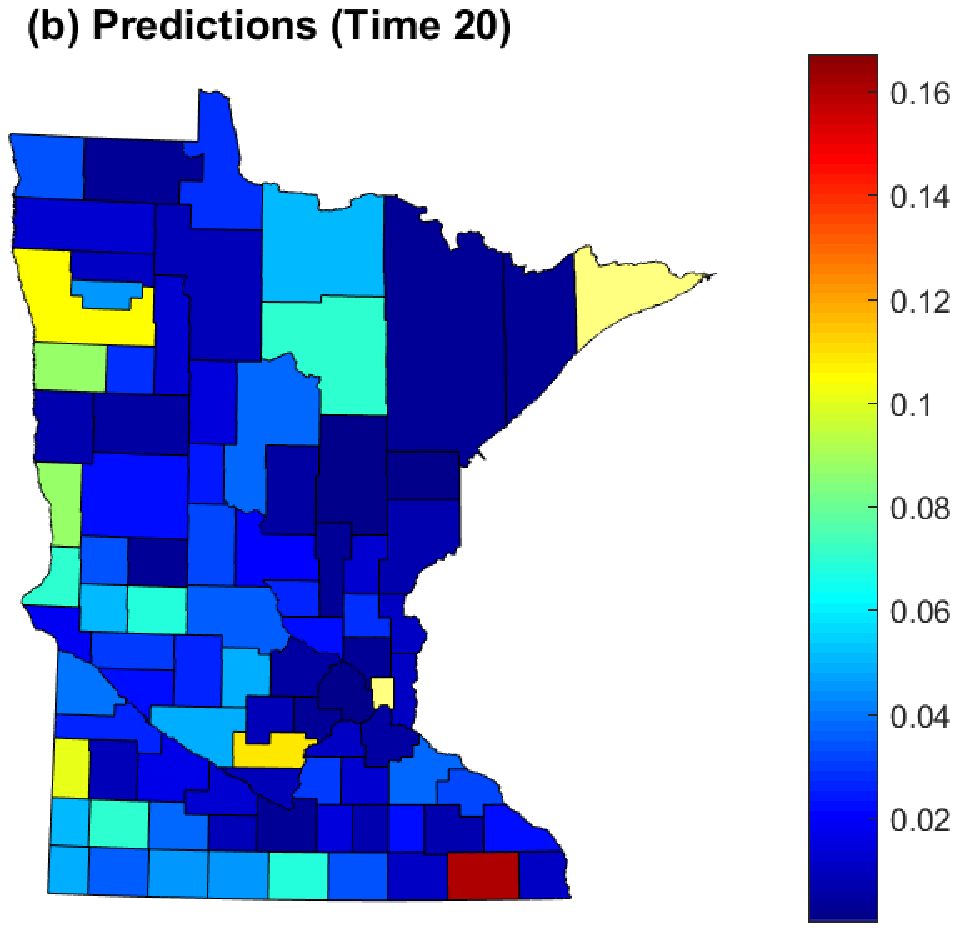}&
			\includegraphics[width=5.5cm,height=5cm]{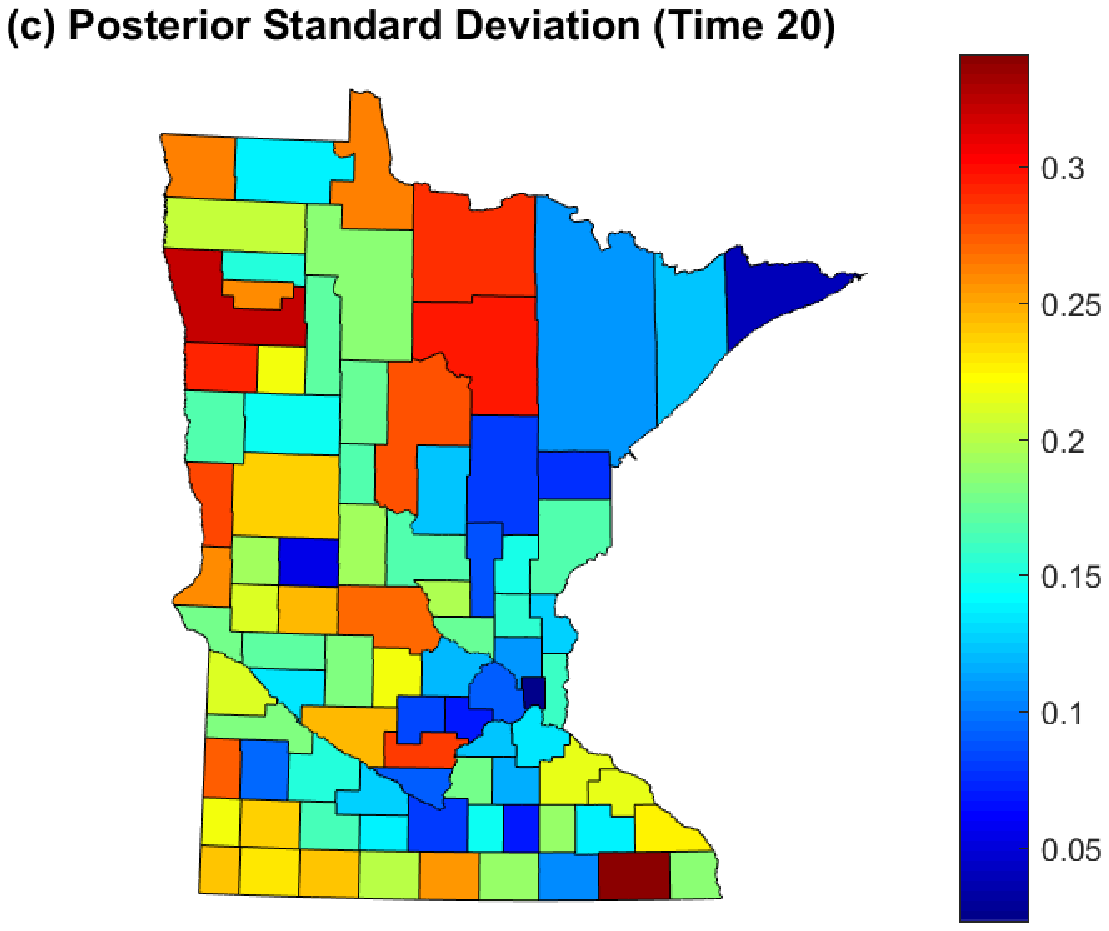}\\
			\includegraphics[width=5.5cm,height=5cm]{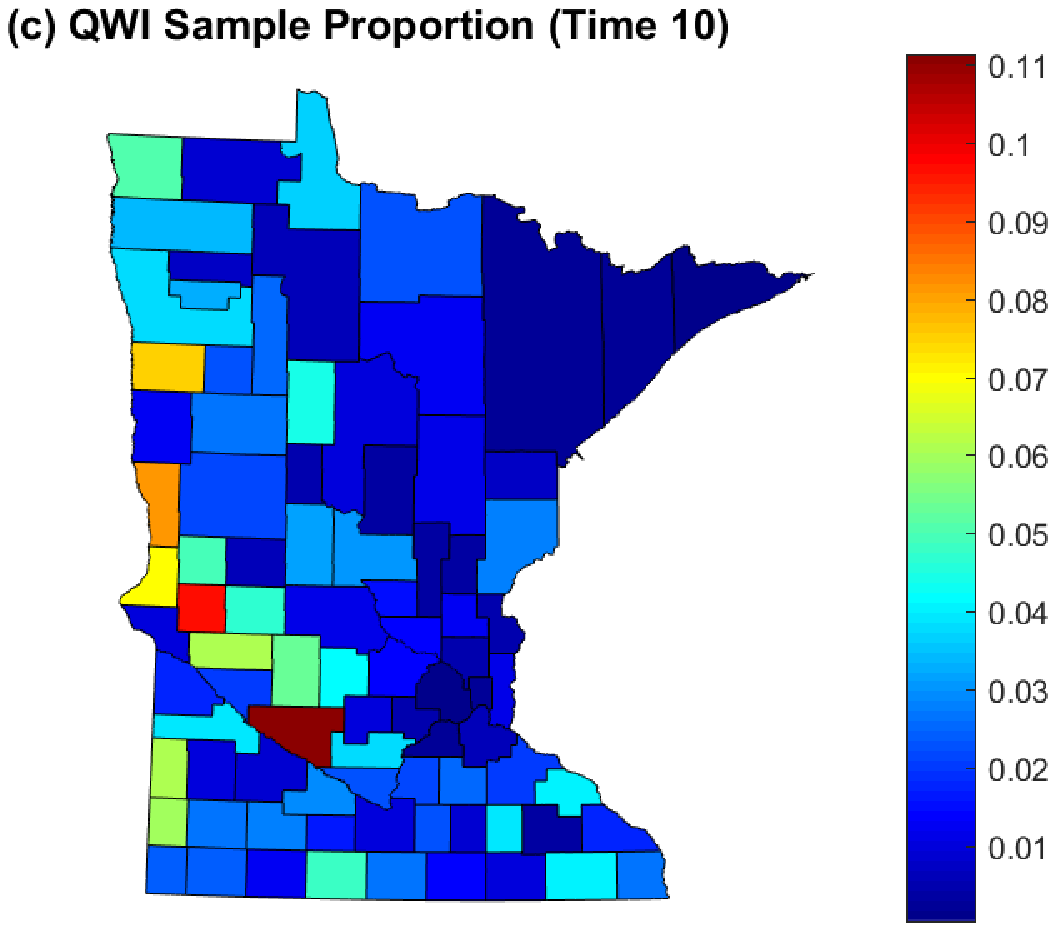}&
			\includegraphics[width=5.5cm,height=5cm]{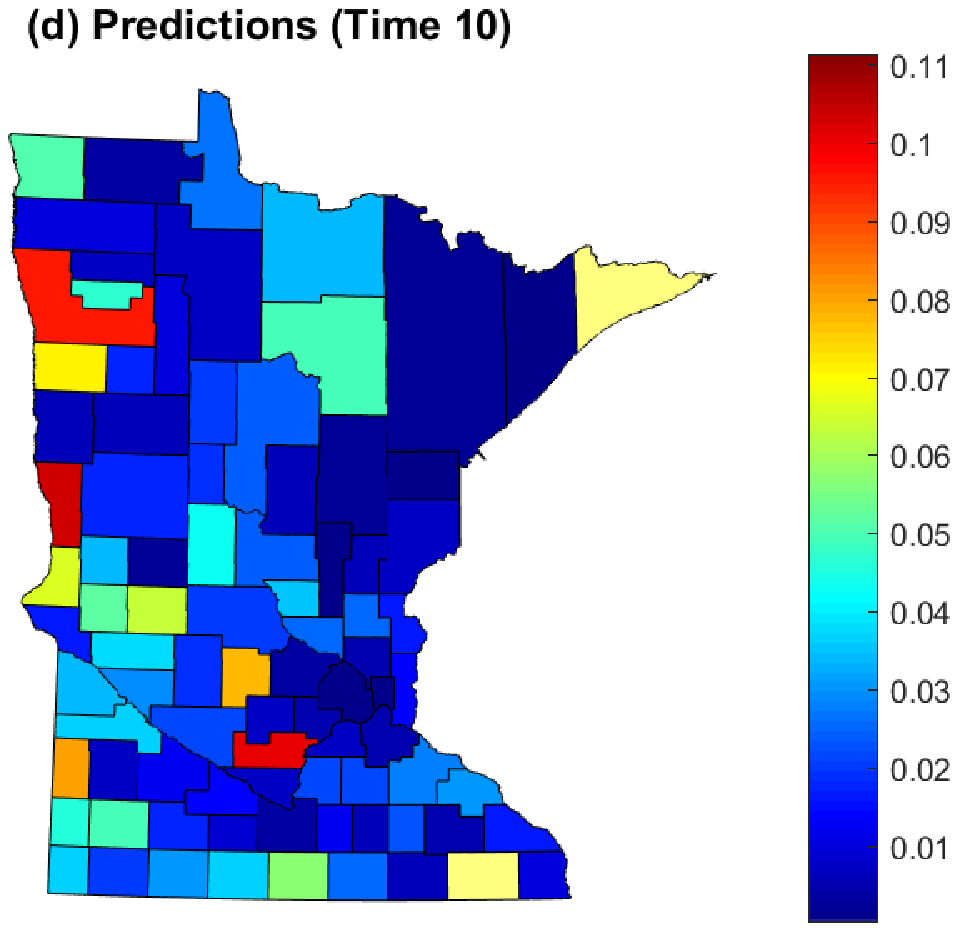}&
			\includegraphics[width=5.5cm,height=5cm]{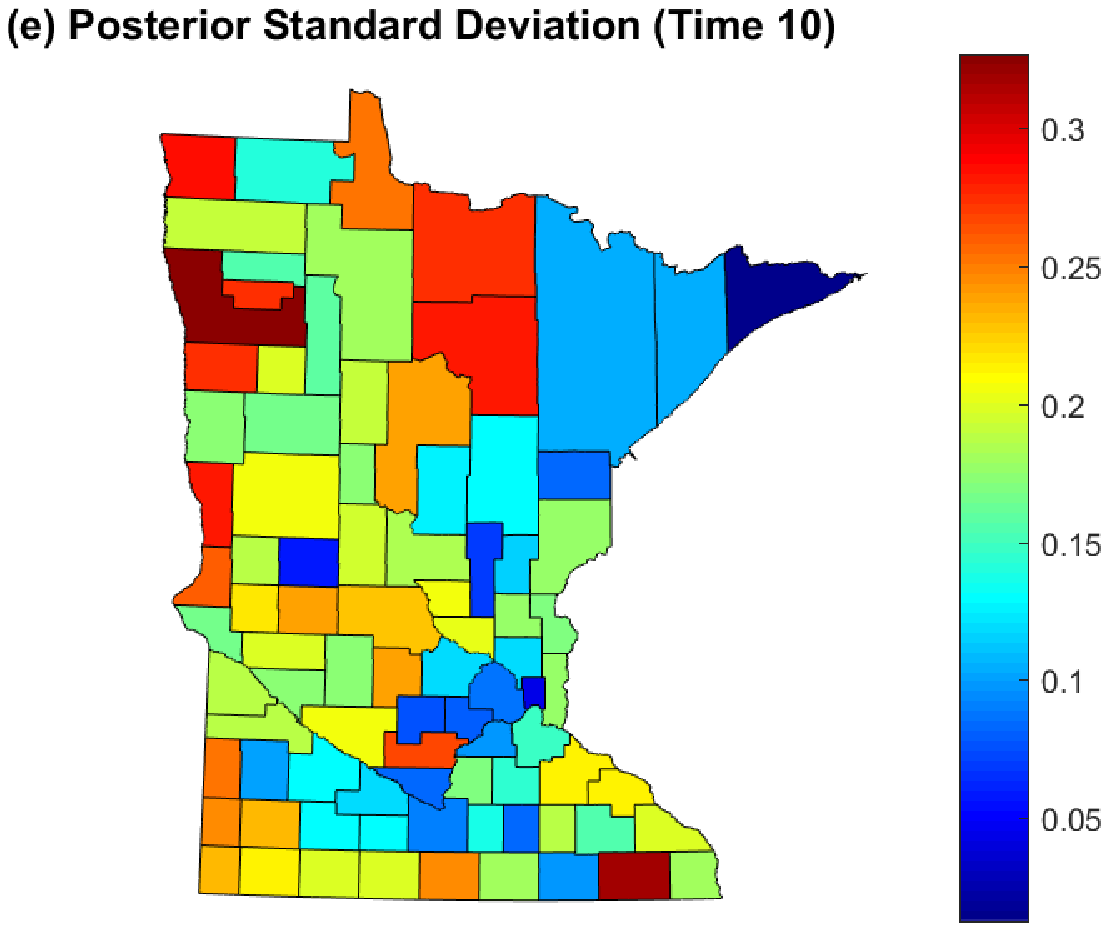}
		\end{tabular}
		\caption{\baselineskip=10pt{(a,d), Map of the LEHD estimated proportion number of individuals employed in the beginning of the 4-th quarter of 2013, within the education industry, and in the beginning of the 2-nd quarter of 2010, within the education industry. In (b,e) and (c,f), we present the predictions and standard deviations, respectively. The colorbars for Panels (a) and (b) are the same, and differ from the colorbar of Panel (c). Likewise, the colorbars for Panels (c) and (d) are the same, and differ from the colorbar of Panel (e).}}\label{fig6}
	\end{figure}
We now compute predictions of the probability of employment in a NAICS Sector. The main purpose of this demonstration is to show that it is possible to jointly analyze such a high-dimensional dataset. We observe data over all $K = 20$ NAICS sectors, $i = 1,\ldots, N=3,145$ U.S. counties, and $T = 96$ quarters, using the high-dimensional QWI dataset of size 2, 247, 586 (see {Figure (\ref{fig2})}). We use the same covariates from  \citet{bradleyPMSTM} and fit the Gibbs sampler based on the full-conditionals given in Appendix B of the Supplementary Material. Specifically, we let $\textbf{x}_{kit} = (1,I(k = 1),\ldots,I(\ell = 19), |A|, I(t = 1),\ldots,I(t = 1,\ldots,95),\mathrm{population}(A))^{\prime}$, where $\mathrm{population}(A)$ is the 2010 decennial Census value of the population of county $A$ and $I(\cdot)$ is the indicator function. We consider $r = 100$, where the first 100 eigenvalues of the Moran's I operator makes up 99.5$\%$ of the total variability of the Moran's I operator. The Gibbs sampler ran for 10,000 iterations, and had a burn-in of 1,000. Trace plots of the sample chains were used to check for convergence of the Gibbs sampler, and no lack of convergence was detected. The latent Gaussian process model did not mix well enough to provide any meaningful comparisons, and the model in Section 2.3 lead to storage errors.

	In {Figure (\ref{fig6})}, we (partially) plot the sample proportion of individuals employed at the beginning of the quarter, and the corresponding predictions and posterior standard deviation. In {Figure (\ref{fig6})}, we see that the predictions roughly track the patterns of the LEHD (QWI) proportions. We are also able to produce measures of uncertainty (i.e., posterior standard deviations), which are also presented in Figure (\ref{fig6}). These results are consistent across different subsets of the predictions. 
	
\section{Discussion} 

We have introduced methodology for jointly modeling dependent multinomial spatio-temporal data within the Bayesian framework. We refer to our model as the multinomial spatio-temporal mixed effects model (or MN-STM), which provides an avenue to model complex dependencies including nonstationarity, non-separability, and asymmetries. The proposed model is extremely parsimonious as we incorporate the Moran's I propagator and introduce a marginal multivariate-spatial precision chosen to be close to an intrinsically conditional autoregressive model. We show that this VAR(1) model is stable, which is an important contribution to the Moran's I propagator literature. Furthermore, the MN-STM is a computationally efficient, since the random effects are projected onto a reduced dimensional space. 

A critical contribution of the MN-STM is the use of the MLB distribution \citep{BradleyLCM} offers an exciting model to consider in the correlated topics model literature \citep{blei2006,blei2006_2,nipps}. The MLB distribution is conjugate, which allows one to avoid expensive data augmentation schemes. In particular, the introduction of a P\'{o}ly-gamma random variable is not necessary to obtain conjugacy. We show that the MLB can be expressed as an infinite mixture of multivariate normal distributions, and hence can be expressed as a more conventional latent Gaussian process model.

Simulation studies showed that the MN-STM based on latent MLB random vectors performs extremely well. Specifically, the simulated examples all indicated very small out-of-sample error of the MN-STM model, and outperforms the P-MSTM and an latent Gaussian process model. Additionally, the computational performance of the MN-STM was shown to be efficient in terms of effective sample size. Furthermore, the MN-STM performed quite well even though the simulated data were not generated from the MN-STM, which suggests that our model is robust to departures from model assumptions. We illustrated that the MN-STM is scalable to the high-dimensional settings by analyzing a big count-valued dataset (of 2, 247, 586 observations) consisting of public-use QWIs made available by the U.S. Census Bureau's LEHD program.

\section*{Acknowledgments} This research was partially supported by the U.S. National Science Foundation (NSF) and the U.S. Census Bureau under NSF grant SES-1132031, funded through the NSF-Census Research Network (NCRN) program. This article is released to inform interested parties of ongoing research and to encourage discussion. The views expressed on statistical issues are those of the authors and not necessarily those of the NSF or U.S. Census Bureau.

\section*{Appendix A: Proofs} 
\renewcommand{\theequation}{A.\arabic{equation}}
\setcounter{equation}{0}
%
%
%

\noindent
\textit{Proof of Proposition 1:} \normalsize
The distribution of $\textbf{q}^{*}$ is equal to $f_{MLB}(\textbf{q}_{1}^{*}\vert \textbf{c}^{*} = -\textbf{B}^{*}\textbf{q}_{2}^{*}+\bm{\mu}^{*}, \textbf{V}^{*} = (\textbf{H}^{*},\textbf{B}^{*}), \bm{\alpha}^{*}, \bm{\kappa}^{*})$ $g(\textbf{q}_{2}^{*}\vert \bm{\mu}^{*},\textbf{V}^{*} = (\textbf{H}^{*},\textbf{B}^{*}), \bm{\alpha}^{*}, \bm{\kappa}^{*})$, where recall we have reparameterized $\textbf{c}^{*} = -\textbf{B}^{*}\textbf{q}_{2}^{*}+\bm{\mu}^{*}$ (see the discussion below Equation (\ref{cmlbhelp})). Thus,
\begin{align*}
f(\textbf{q}_{1}^{*},\textbf{q}_{2}^{*}\vert\bm{\mu}^{*}, \textbf{V}^{*},\bm{\alpha}^{*},\bm{\kappa}^{*})&\propto \mathrm{exp}\left[ \bm{\alpha}^{*\prime}\textbf{H}^{*}\textbf{q}_{1}^{*} + \bm{\alpha}^{*\prime}\textbf{B}^{*}\textbf{q}_{2}^{*} - \bm{\alpha}^{*\prime}\bm{\mu}^{*} -\bm{\kappa}^{*\prime}\mathrm{log}\left\lbrace\bm{1}_{M}+\mathrm{exp}\left(\textbf{H}^{*}\textbf{q}_{1}^{*} + \textbf{B}\textbf{q}_{2}^{*} -\bm{\mu}^{*})\right)\right\rbrace\right] \\
&=\mathrm{exp}\left[ \bm{\alpha}^{*\prime}\textbf{V}^{*}(\textbf{q}^{*} - \textbf{V}^{*-1}\bm{\mu}^{*}) -\bm{\kappa}^{*\prime}\mathrm{log}\left\lbrace\bm{1}_{M}+\mathrm{exp}\left(\textbf{V}^{*}(\textbf{q}^{*} - \textbf{V}^{*-1}\bm{\mu}^{*})\right)\right\rbrace\right].
\end{align*}
Integrating out $\textbf{q}_{2}$ we obtain,
\begin{align}\label{int}
f(\textbf{q}_{1}^{*}\vert\bm{\mu}^{*}, \textbf{V}^{*},\bm{\alpha}^{*},\bm{\kappa}^{*})\propto
&\int \mathrm{exp}\left[ \bm{\alpha}^{*\prime}\textbf{V}^{*}(\textbf{q}^{*} - \textbf{V}^{*-1}\bm{\mu}^{*}) -\bm{\kappa}^{*\prime}\mathrm{log}\left\lbrace\bm{1}_{M}+\mathrm{exp}\left(\textbf{V}^{*}(\textbf{q}^{*} - \textbf{V}^{*-1}\bm{\mu}^{*})\right)\right\rbrace\right] d\textbf{q}_{2}^{*}.
\end{align}
Thus, $\textbf{q}_{1}^{*}$ is the marginal random vector associated with $f(\textbf{q}^{*}\vert \textbf{V}^{*-1}\bm{\mu}^{*}, \textbf{V}^{*} = (\textbf{H}^{*},\textbf{B}^{*}), \bm{\alpha}^{*}, \bm{\kappa}^{*})$ for $f$ defined in (\ref{mlb}). Thus, we are left to show that $\textbf{q}_{1}^{*}= (\textbf{H}^{*\prime}\textbf{H}^{*})^{-1}\textbf{H}^{*\prime}\textbf{w}^{*}$ is a sample from this marginal distribution.

Denote the QR decomposition of $\textbf{H}^{*} = \textbf{Q}\textbf{R}$,
where the $M\times r$ matrix $\textbf{Q}$ satisfies $\textbf{Q}^{\prime}\textbf{Q} = \textbf{I}_{r}$ and $\textbf{R}$ is a $r\times r$ upper triangular matrix. Now recall the definition of the $M\times (M-r)$ matrix $\textbf{B}^{*}$, which satisfies $\textbf{B}^{*\prime}\textbf{B}^{*} = \textbf{I}_{M-r}$ and $\textbf{B}^{*\prime}\textbf{Q} = \bm{0}_{M-r,r}$. Then $\textbf{V}^{*}$ can be written as
\begin{align}\label{special_precision}
& \textbf{V}^{*}=\left[
\begin{array}{cc}
\textbf{Q} & \textbf{B}^{*}
\end{array}\right]
\left[
\begin{array}{cc}
\textbf{R} & \bm{0}_{r,M-r} \\ 
\bm{0}_{M-r,r} & \textbf{I}_{M-r},
\end{array}\right].\\
\nonumber
\end{align}
\noindent
It follows that
\begin{equation*}
\textbf{V}^{*-1}=
\left[
\begin{array}{cc}
\textbf{R}^{-1} & \bm{0}_{r,M-r} \\ 
\bm{0}_{M-r,r} & \textbf{I}_{M-r},
\end{array}\right]\left[
\begin{array}{c}
\textbf{Q}^{\prime} \\ 
\textbf{B}^{*\prime},
\end{array}\right]=
\left[\begin{array}{c}
(\textbf{H}^{*\prime}\textbf{H}^{*})^{-1}\textbf{H}^{*\prime} \\ 
\textbf{B}^{*\prime}
\end{array}\right],
\end{equation*}
\noindent
where the last equality in the above can be verified by substituting $\textbf{H}^{*} = \textbf{Q}\textbf{R}$ into $(\textbf{H}^{*\prime}\textbf{H}^{*})^{-1}\textbf{H}^{*\prime}$. From (\ref{chov}), $\textbf{q}^{*}$ distributed according to $f(\textbf{q}^{*}\vert \textbf{V}^{*-1}\bm{\mu}^{*}, \textbf{V}^{*} = (\textbf{H}^{*},\textbf{B}^{*}), \bm{\alpha}^{*}, \bm{\kappa}^{*})$ can be written as 
\begin{equation}\label{margstep1}
\left[
\begin{array}{c}
\textbf{q}_{1}^{*} \\ 
\textbf{q}_{2}^{*}
\end{array}\right] =   \left[
\begin{array}{c}
(\textbf{H}^{*\prime}\textbf{H}^{*})^{-1}\textbf{H}^{*\prime}\bm{\mu}^{*} \\ 
\textbf{B}^{*\prime}\bm{\mu}^{*}
\end{array}\right] +  \left[
\begin{array}{c}
(\textbf{H}^{*\prime}\textbf{H}^{*})^{-1}\textbf{H}^{*\prime}\textbf{w} \\ 
\textbf{B}^{*\prime}\textbf{w}
\end{array}\right],
\end{equation}
\noindent
where the $n$-dimensional random vector $\textbf{w}$ is distributed according to $f(\textbf{w}\vert \bm{0}_{M}, \textbf{V}^{*} =\textbf{I}_{M}, \bm{\alpha}^{*}, \bm{\kappa}^{*})$ where $f$ is defined in  (\ref{mlb}). Multiplying both sides of (\ref{margstep1}) by $[\textbf{I}_{r},\bm{0}_{r,M-r}]$ we have
\begin{equation}\label{marg}
\textbf{q}_{1}^{*} =(\textbf{H}^{*\prime}\textbf{H}^{*})^{-1}\textbf{H}^{*\prime}\bm{\mu}^{*}+ (\textbf{H}^{*\prime}\textbf{H}^{*})^{-1}\textbf{H}^{*\prime}\textbf{w} = (\textbf{H}^{*\prime}\textbf{H}^{*})^{-1}\textbf{H}^{*\prime}\textbf{w}^{*},
\end{equation}
\noindent
and hence the distribution associated with $(\textbf{H}^{*\prime}\textbf{H}^{*})^{-1}\textbf{H}^{*\prime}\textbf{w}^{*}$ is the marginal distribution associated with\\ $f(\textbf{q}^{*}\vert \textbf{V}^{*-1}\bm{\mu}^{*}, \textbf{V}^{*} = (\textbf{H}^{*},\textbf{B}^{*}), \bm{\alpha}^{*}, \bm{\kappa}^{*})$ as desired.\\

\noindent
\textit{Proof of Proposition 2:}  Without loss of generality let $t^{*} = 1$. Then iterating the VAR(1) expression gives:
\begin{equation*}
\bm{\eta}_{T} = \sum_{t = 0}^{T-1}\textbf{B}_{j}\textbf{u}_{T-j},
\end{equation*}
where
\begin{align*}
&\textbf{B}_{0} = \textbf{I}_{r},\\
&\textbf{B}_{j} = \prod_{k = T - j + 1}^{T}\textbf{M}_{k} = \rho^{(j - 1)}\prod_{k = T - j + 1}^{T}\bm{\Psi}_{k}.
\end{align*}
\noindent
For the Wold representation to hold, we have from \citet[][pg., 29 $\--$ 31]{Fuller1976} that we need to show two items:
\begin{align}\label{abssum}
&\sum_{k = 1}^{\infty}\mathrm{trace}\left(\textbf{B}_{k}^{\prime}\textbf{B}_{k}\right) < \infty\\
\label{finitevar}
&E(\textbf{u}_{t}^{\prime}\textbf{u}_{t}\vert \textbf{H}_{t},\alpha_{t},\kappa_{t}) < \infty;\hspace{8pt} t = 1,2,\ldots \hspace{2pt}.
\end{align}
To show (\ref{abssum}), we have
\begin{align*}
&\sum_{k = 1}^{T}\mathrm{trace}\left(\textbf{B}_{j}^{\prime}\textbf{B}_{j}\right)=r + \sum_{k = 2}^{T}\rho^{2(k - 1)}\mathrm{trace}\left\lbrace \left(\prod_{t = T - k + 1}^{T}\bm{\Psi}_{t}\right)^{\prime}\left(\prod_{t = T - k + 1}^{T}\bm{\Psi}_{t}\right) \right\rbrace\\
&=r + r\sum_{k = 2}^{T}\rho^{2(k - 1)},
\end{align*}
\noindent
where $\mathrm{trace}\left\lbrace \left(\prod_{t = T - k + 1}^{T}\bm{\Psi}_{t}\right)^{\prime}\left(\prod_{t = T - k + 1}^{T}\bm{\Psi}_{t}\right)\right\rbrace = r$ for every $k$, since  $\bm{\Psi}_{t}$ is orthonormal for every $t$. Now, upon taking the limit at $T$ goes to infinity,
\begin{align*}
&\sum_{k = 1}^{\infty}\mathrm{trace}\left(\textbf{B}_{j}^{\prime}\textbf{B}_{j}\right)=r-\frac{r}{\rho^{2}}  - r + r\rho^{-2}\sum_{k = 0}^{\infty}\rho^{2k}\\
&=r-\frac{r}{\rho^{2}}  -r + \frac{r}{(1-\rho^{2})\rho^{2}} <\infty.
\end{align*}
\noindent
To show (\ref{finitevar}), first recall that the mean of a logit-beta random variable is $g(\alpha) - g(\kappa-\alpha)$, where $g$ is the digamma function. Also, the variance is given by $g_{1}(\alpha) - g_{1}(\kappa-\alpha)$, where $g_{1}$ is the trigamma function \citep{varlogit}. From (\ref{trans}),
\begin{align*}
&E(\textbf{u}_{t}^{\prime}\textbf{u}_{t}\vert \textbf{H}_{t},\alpha_{t},\kappa_{t})=E\left[\mathrm{trace}\left\lbrace(\textbf{H}_{t}^{\prime}\textbf{H}_{t})^{-1}\textbf{w}_{t}\textbf{w}_{t}^{\prime}\right\rbrace\vert \textbf{H}_{t},\alpha_{t},\kappa_{t}\right],
\end{align*}
where the $\textbf{w}_{t}$ consists i.i.d. logit-beta random variables with shape $\alpha_{t}$ and $\kappa_{t}$. Thus, $E(\textbf{w}_{t}\textbf{w}_{t}^{\prime}\vert \textbf{H}_{t},\alpha_{t},\kappa_{t}) = \textbf{D}$, where $\textbf{D} = \mathrm{diag}(E(w_{i})^{2} + var(w_{i}: i = 1,\ldots, N_{t}+r) = \textbf{D} = \mathrm{diag}(g(\alpha_{t}) + g_{1}(\alpha_{t}) -g(\kappa_{t}-\alpha_{t}) - g(\kappa_{t}-\alpha_{t}): i = 1,\ldots, N_{t}+r)$. Thus,
\begin{align*}
&E(\textbf{u}_{t}^{\prime}\textbf{u}_{t}\vert \textbf{H}_{t},\alpha_{t},\kappa_{t})=\mathrm{trace}\left\lbrace \textbf{D}^{1/2}(\textbf{H}_{t}^{\prime}\textbf{H}_{t})^{-1}\textbf{D}^{1/2}\right\rbrace <\infty,
\end{align*}
\noindent
provided that $\textbf{H}_{t}^{\prime}\textbf{H}_{t}$ is invertible.\\

{
\noindent
\textit{Proof of Proposition 3:} Augment $\bm{\eta}_{t}$ with an $(K-1)N_{t}$-dimensional random vector $\textbf{q}_{t}$ as done in Proposition 1, where the $((K-1)N_{t}+r)\times (K-1)N_{r}$ matrix $\textbf{B}_{t}$ be the orthogonal complement of $\textbf{H}_{t}$. Then, from (\ref{margstep1}),
\begin{align*}
\left[
\begin{array}{c}
\bm{\eta}_{t} \\ 
\textbf{q}_{t}
\end{array}\right] &=   \left[
\begin{array}{c}
(\textbf{H}_{t}^{\prime}\textbf{H}_{t})^{-1}\textbf{H}_{t}^{\prime} \\ 
\textbf{B}_{t}^{\prime}
\end{array}\right]\textbf{H}_{t}\textbf{M}_{t}\bm{\eta}_{t-1} +  \left[
\begin{array}{c}
(\textbf{H}_{t}^{\prime}\textbf{H}_{t})^{-1}\textbf{H}_{t}^{\prime}\textbf{w}_{t}\\ 
\textbf{B}_{t}^{\prime}\textbf{w}_{t}\end{array}\right]\\
 &=\left[
 \begin{array}{c}
\textbf{M}_{t}\bm{\eta}_{t-1} \\ 
 \bm{0}_{(K-1)N_{t}}
 \end{array}\right] +  \left[
 \begin{array}{c}
 (\textbf{H}_{t}^{\prime}\textbf{H}_{t})^{-1}\textbf{H}_{t}^{\prime}\textbf{w}_{t}\\ 
 \textbf{B}_{t}^{\prime}\textbf{w}_{t},
\end{array}\right],
\end{align*}
where $\textbf{w}_{t}$ consists of i.i.d. logit-beta random variables with the $i$-th element of $\textbf{w}_{t}$ is denoted with $\mathrm{logit}(\gamma_{i})$. It follows that $\mathrm{cov}(\bm{\eta}_{t}) = \left(\textbf{H}_{t}^{\prime}\textbf{H}_{t}\right)^{-1}\textbf{H}_{t}^{\prime}\mathrm{cov}(\textbf{w}_{t}\vert \alpha_{t}, \kappa_{t})\textbf{H}_{t}\left(\textbf{H}_{t}^{\prime}\textbf{H}_{t}\right)^{-1}$. This gives us,
\begin{equation}\label{cov2}
\mathrm{cov}(\bm{\Phi}^{\mathrm{P}}\bm{\eta}_{t}\vert \bm{\eta}_{t-1}, \textbf{V}_{t},\alpha_{t}, \kappa_{t})=\bm{\Phi}^{\mathrm{P}}\left(\textbf{H}_{t}^{\prime}\textbf{H}_{t}\right)^{-1}\textbf{H}_{t}^{\prime}\mathrm{cov}(\textbf{w}_{t}\vert \alpha_{t}, \kappa_{t})\textbf{H}_{t}\left(\textbf{H}_{t}^{\prime}\textbf{H}_{t}\right)^{-1}\bm{\Phi}^{\mathrm{P}\prime}.
\end{equation}
\noindent
Hence,
\begin{equation}\label{unit}
\mathrm{cov}(\textbf{w}_{t}\vert \alpha_{t}, \kappa_{t}) = \mathrm{var}\left\lbrace\mathrm{logit}(\gamma_{1})\right\rbrace\textbf{I}.
\end{equation}
Substituting (\ref{unit}) into (\ref{cov2}), and a few lines of algebra, leads to 
\begin{align} 
\nonumber
&\mathrm{cov}(\bm{\Phi}^{\mathrm{P}}\bm{\eta}_{t}\vert \textbf{V}_{t},\alpha_{t}, \kappa_{t})^{-}=\mathrm{var}\left\lbrace\mathrm{logit}(\gamma_{1})\right\rbrace\bm{\Phi}^{\mathrm{P}}\left(\textbf{H}_{t}^{\prime}\textbf{H}_{t}\right)\bm{\Phi}^{\mathrm{P}\prime}\\
\nonumber
&=\mathrm{var}\left\lbrace\mathrm{logit}(\gamma_{1})\right\rbrace\bm{\Phi}^{\mathrm{P}}\left(\bm{\Phi}^{\prime}\bm{\Phi} + \textbf{V}_{t}^{\prime}\textbf{V}_{t}\right)\bm{\Phi}^{\mathrm{P}\prime}\\
\label{finalform}
&=\frac{1}{\mathrm{var}\left\lbrace\mathrm{logit}(\gamma_{1})\right\rbrace}\bm{\Phi}^{\mathrm{P}}\left(\bm{\Phi}^{\prime}\bm{\Phi}\right)\bm{\Phi}^{\mathrm{P}\prime} + \frac{1}{\mathrm{var}\left\lbrace\mathrm{logit}(\gamma_{1})\right\rbrace}\bm{\Phi}^{\mathrm{P}}\left( \textbf{V}_{t}^{\prime}\textbf{V}_{t}\right)\bm{\Phi}^{\mathrm{P}\prime}
\end{align}
\noindent
Substituting (\ref{finalform}) into (\ref{KstarP}), we obtain
	\begin{equation}\label{KstarP2}
	\underset{\textbf{V}_{t}}{\mathrm{arg\hspace{5pt}min}}\left\lbrace \Big\Vert\frac{1}{\sigma_{\eta}^{2}}\textbf{P} -\frac{1}{\mathrm{var}\left\lbrace\mathrm{logit}(\gamma_{1})\right\rbrace}\bm{\Phi}^{\mathrm{P}}\left(\bm{\Phi}^{\prime}\bm{\Phi}\right)\bm{\Phi}^{\mathrm{P}\prime} - \frac{1}{\mathrm{var}\left\lbrace\mathrm{logit}(\gamma_{1})\right\rbrace}\bm{\Phi}^{\mathrm{P}}\left( \textbf{V}_{t}^{\prime}\textbf{V}_{t}\right)\bm{\Phi}^{\mathrm{P}\prime}\Big\Vert_{\mathrm{F}}^{2}\right\rbrace;\hspace{5pt} t = 1,\ldots,T.
	\end{equation}
	\noindent
	Let $\sigma_{\eta}^{2} = \mathrm{var}\left\lbrace\mathrm{logit}(\gamma_{1})\right\rbrace$ so that minimizing (\ref{KstarP2}) is that same as minimizing,
		\begin{equation}\label{KstarP3}
		\underset{\textbf{V}_{t}}{\mathrm{arg\hspace{5pt}min}}\left\lbrace \Big\Vert\textbf{P} -\bm{\Phi}^{\mathrm{P}}\left(\bm{\Phi}^{\prime}\bm{\Phi}\right)\bm{\Phi}^{\mathrm{P}\prime} - \bm{\Phi}^{\mathrm{P}}\left( \textbf{V}_{t}^{\prime}\textbf{V}_{t}\right)\bm{\Phi}^{\mathrm{P}\prime}\Big\Vert_{\mathrm{F}}^{2}\right\rbrace;\hspace{5pt} t = 1,\ldots,T.
		\end{equation}
	It follows from Proposition 1 in \citet{bradleyMSTM} that the value of $\textbf{V}^{\prime}\textbf{V}_{t}$ that minimizes (\ref{KstarP3}) is 
	\begin{equation*}
	\bm{\Sigma}^{*}\equiv \mathcal{A}^{+}\left\lbrace \bm{\Phi}^{\mathrm{P}\prime}\textbf{P}\bm{\Phi}^{\mathrm{P}}-\left(\bm{\Phi}^{\prime}\bm{\Phi}\right)\right\rbrace.
	\end{equation*}
\noindent
Then, it follows that $\textbf{V}_{t} = \bm{\Lambda}_{\Sigma}^{1/2}\bm{\Psi}_{\Sigma}$ minimizes the Frobenius norm in (\ref{KstarP}). Finally, it is well known that \citep{varlogit}
\begin{equation*}
\mathrm{var}\left\lbrace\mathrm{logit}(\gamma_{1})\right\rbrace = g(\alpha_{t}) + g(\kappa_{t}-\alpha_{t}),
\end{equation*}
\noindent
which completes the proof.\\

\noindent
\textit{Proof of Proposition 4:} Let $g_{k}\sim Gamma(b,1)$. Then the P\'olya-Gamma random variable is defined to be,
\begin{align*}
\frac{1}{2\pi}\sum_{k = 1}^{\infty}\frac{g_{k}}{(k - 1/2)^{2}},
\end{align*}
has a density $p(\omega\vert b)$, which does not have a closed-form expression. We make use of the follow integral identity \citep[e.g., see][]{polson}:
\begin{equation*}
\frac{\mathrm{exp}(ah)}{(1+\mathrm{exp}(h))^{b}} = 2^{-b}\mathrm{exp}\left\lbrace \left(a - b/2\right)h\right\rbrace\int_{0}^{\infty}\mathrm{exp}\left( -\omega h^{2}/2\right) p(\omega\vert b)d\omega.
\end{equation*}
\noindent
 Let $\textbf{w} = \textbf{H}(\textbf{q}_{1} - \bm{\mu}_{1})$. Denote the normalizing constant of $f_{MLB}$ with $\mathcal{B}(\textbf{H},\bm{\mu}_{1}, \bm{\alpha}, \bm{\kappa})$, which is known to be finite \citep{BradleyLCM}. Then, 
\begin{align*}
& f_{MLB}(\textbf{q}_{1}\vert \bm{\mu}, \textbf{V}, \bm{\alpha}, \bm{\kappa}) = \frac{1}{\mathcal{B}(\textbf{H},\bm{\mu}_{1}, \bm{\alpha}, \bm{\kappa})}\mathrm{exp}\left[\bm{\alpha}^{\prime} \textbf{H}(\textbf{q}_{1} - \bm{\mu}_{1}) - \bm{\kappa}^{\prime}\mathrm{log}\left\lbrace \bm{1}_{M} + \mathrm{exp}(\textbf{H}(\textbf{q}_{1} - \bm{\mu}_{1}))\right\rbrace\right]\\
&= \frac{1}{\mathcal{B}(\textbf{H},\bm{\mu}_{1}, \bm{\alpha}, \bm{\kappa})}\mathrm{exp}\left[\bm{\alpha}^{\prime}\textbf{w} - \bm{\kappa}^{\prime}\mathrm{log}\left\lbrace\bm{1} +\mathrm{exp}\left(\textbf{h}\right)\right\rbrace\right] = \frac{1}{\mathcal{B}(\textbf{H},\bm{\mu}_{1}, \bm{\alpha}, \bm{\kappa})}\prod_{i = 1}^{M}\frac{\mathrm{exp}(\alpha_{i}w_{i})}{(1+\mathrm{exp}(w_{i}))^{\kappa_{i}}}\\
&=\frac{2^{-\sum_{i = 1}^{M}\kappa_{i}}}{\mathcal{B}(\textbf{H},\bm{\mu}_{1}, \bm{\alpha}, \bm{\kappa})} \int \mathrm{exp}\left\lbrace \sum_{i = 1}^{M}\left(\alpha_{i}-\kappa_{i}/2\right)w_{i} \right\rbrace \mathrm{exp}\left(-\sum_{i = 1}^{m}\omega_{i}w_{i}^{2}/2\right)p(\bm{\omega}\vert \bm{\kappa})d\bm{\omega}\\
&=\frac{2^{-\sum_{i = 1}^{M}\kappa_{i}}}{\mathcal{B}(\textbf{H},\bm{\mu}_{1}, \bm{\alpha}, \bm{\kappa})} \int \mathrm{exp}\left\lbrace \sum_{i = 1}^{M}\left(\alpha_{i}/\omega_{i}-\kappa_{i}/2\omega_{i}\right)^{2} \right\rbrace \mathrm{exp}\left[-\sum_{i = 1}^{m}\omega_{i}\left\lbrace w_{i} - \left(\alpha_{i}/\omega_{i}-\kappa_{i}/2\omega_{i}\right)\right\rbrace^{2}/2\right]p(\bm{\omega}\vert \bm{\kappa})d\bm{\omega}\\
&=\frac{2^{-\sum_{i = 1}^{M}\kappa_{i}}}{\mathcal{B}(\textbf{H},\bm{\mu}_{1}, \bm{\alpha}, \bm{\kappa})} \int  \mathrm{exp}\left\lbrace \sum_{i = 1}^{M}\left(\alpha_{i}/\omega_{i}-\kappa_{i}/2\omega_{i}\right)^{2} \right\rbrace\mathrm{exp}\left[-(\textbf{w}-\bm{\zeta})^{\prime}\bm{\Omega}(\textbf{w}-\bm{\zeta})/2\right]p(\bm{\omega}\vert \bm{\kappa})d\bm{\omega},
\end{align*}
which after a few lines of algebra becomes,
\begin{align*}
& =\int g(\bm{\Omega},\bm{\mu}, \textbf{H}, \bm{\alpha}, \bm{\kappa})\frac{\mathrm{det}(\textbf{H}^{\prime}\bm{\Omega}\textbf{H})^{1/2}}{(2\pi)^{M/2}} \mathrm{exp}\left[ - (\textbf{q}_{1}-\bm{\gamma})^{\prime}\textbf{H}^{\prime}\bm{\Omega}\textbf{H}(\textbf{q}_{1}-\bm{\gamma})/2\right] p(\bm{\omega}\vert \bm{\kappa})d\bm{\omega}.,
\end{align*}
where, 
\begin{equation*}
g(\bm{\Omega},\bm{\mu}, \textbf{H}, \bm{\alpha}, \bm{\kappa}) = \frac{2^{-\sum_{i = 1}^{M}\kappa_{i}}}{\mathcal{B}(\textbf{H},\bm{\mu}_{1}, \bm{\alpha}, \bm{\kappa})}\mathrm{exp}\left\lbrace \sum_{i = 1}^{M}\left(\alpha_{i}/\omega_{i}-\kappa_{i}/2\omega_{i}\right)^{2} + \bm{\gamma}^{\prime}\bm{\gamma}\right\rbrace \frac{(2\pi)^{M/2}}{\mathrm{det}(\textbf{H}^{\prime}\bm{\Omega}\textbf{H})^{1/2}}.
\end{equation*}
This completes the result. As a small side-note, notice that $g(\bm{\Omega},\bm{\mu}, \textbf{H}, \bm{\alpha}, \bm{\kappa})$ implies that
\begin{equation*}
\mathcal{B}(\textbf{H},\bm{\mu}_{1}, \bm{\alpha}, \bm{\kappa}) = 2^{-\sum_{i = 1}^{M}\kappa_{i}} E\left[\mathrm{exp}\left\lbrace \sum_{i = 1}^{M}\left(\alpha_{i}/\omega_{i}-\kappa_{i}/2\omega_{i}\right)^{2} +\bm{\gamma}^{\prime}\bm{\gamma}\right\rbrace \frac{(2\pi)^{M/2}}{\mathrm{det}(\textbf{H}^{\prime}\bm{\Omega}\textbf{H})^{1/2}}\right],
\end{equation*}
\noindent
where the expectation is taken with respect to $p(\bm{\omega}\vert \bm{\kappa})$.\\

\section*{Appendix B: Specifications of the Small Simulated Example in Section 2.3}
\renewcommand{\theequation}{B.\arabic{equation}}
\setcounter{equation}{0}

To produce Figure (\ref{fig1}), we set $T$ = 1, $N = 50$, $K = 5$, and $\text{g}_{kit} = (1,sin\left[\pi \left\lbrace ki + (k-1)N\right\rbrace\right])^{\prime}$. We define $\textbf{X} = \{\text{x}_{kit}\}$ to be the orthogonalization of $\{\text{g}_{kit}\}$ and set $\bm{\Phi}$ equal to the first $r = 125$ columns of the orthogonal complement of $\textbf{X}$. Set $\bm{\beta} = (0.01,-2)^{\prime}$, the elements of $\bm{\eta}$ to be values generated from a normal distribution with mean zero and variance 1, and generate the elements of $\bm{\xi}$ to be from a normal distribution with mean zero and variance 0.33. 

We run the MCMC for $2000$ iterations, and trace plots do not indicate a lack of convergence. The value of $\bm{\delta} = \textbf{n}$, $\delta = 0.0001$, $\bm{\epsilon}_{1} = 0.05\bm{1}_{250}$, $\rho = 0.9$, $\bm{\epsilon}_{2} = \sigma\rho \by - (1-\rho)\by/\sigma + 0.05\sigma\bm{1}_{250}$, and $\bm{\epsilon} = \sigma\rho \by - (1-\rho)\by/\sigma+ (0.05+\sigma0.05)\bm{1}_{250}$. In our experience, the performance of Pseudo-Code 2 improves as the elements of $(1-\rho)\by/\sigma + \bm{\epsilon}_{2}$ get close to $n_{kit}p_{kit}$. Thus, our specification of $\bm{\epsilon}_{2}$ might be interpreted as an empirical Bayes specification of Pseudo-Code 2, where we substitute an estimate of $\bm{\epsilon}_{2}$.

}

\newpage
\chapter{}

\thispagestyle{empty} \baselineskip=28pt

\thispagestyle{empty} \baselineskip=28pt

\begin{center}
	{\LARGE{\bf Supplementary Appendix:  {Spatio-Temporal Models for Big Multinomial Data using the Conditional Multivariate Logit-Beta Distribution}}}
\end{center}

\baselineskip=12pt

%
%
%
\vskip 2mm
\begin{center}
	Jonathan R. Bradley\footnote{(\baselineskip=10pt to whom correspondence should be addressed) Department of Statistics, Florida State University, 117 N. Woodward Ave., Tallahassee, FL 32306-4330, jrbradley@fsu.edu},
	Christopher K. Wikle\footnote{\baselineskip=10pt  Department of Statistics, University of Missouri, 146 Middlebush Hall, Columbia, MO 65211-6100}, and
	Scott H. Holan$^{2,}$ \footnote{\baselineskip=10pt U.S. Census Bureau, 4600 Solver Hill Road, Washington D. C. 20233-9100}\\
\end{center}

\pagenumbering{arabic}

\baselineskip=24pt

\section*{Introduction} In this Supplementary Appendix, a summary and further description of the MN-STM (Supplementary Appendix A), the derivation of the Gibbs sampler for the MN-STM (Supplementary Appendix B), and an alternative latent Gaussian process expression of the MN-STM (Supplementary Appendix C).

\section*{Appendix A: Summary of the MN-STM}
\renewcommand{\theequation}{A.\arabic{equation}}
\setcounter{equation}{0}
We choose to write the model for $\textbf{y}_{it}$ using the ``data model,'' ``process model,'' and ``parameter model'' notation that is commonly used in the spatio-temporal statistics literature \citep[e.g., see][for an early reference]{berlinhier}. This terminology is useful because it allows one to discuss highly parameterized joint distributions in terms of more manageable ``pieces'' (i.e., conditional and marginal distributions). The ``data model'' refers to the conditional distribution $f(y_{kit}\vert \nu_{it})$, where $\{\nu_{kit}\}$ is an unobserved random variable assumed to be correlated across the indexes $k$, $i$, and $t$, and $f$ will be used to denote a probability density function/probability mass function (pdf/pmf). Similarly, $f({\nu}_{kit}\vert \bm{\theta})$ is referred to as the process model, where $\bm{\theta}$ is a generic real-valued parameter vector. A parameter model, $f(\bm{\theta})$, is assumed for $\bm{\theta}$. Together, the  ``data model,'' ``process model,'' and ``parameter model'' define the joint distribution that is used for statistical inference. That is,
\begin{equation*}
\underset{kit}{\prod}f(y_{kit},\nu_{kit},\bm{\theta}) = \underset{kit}{\prod} f(y_{kit}\vert \nu_{kit})f({\nu}_{kit}\vert \bm{\theta})f(\bm{\theta}).
\end{equation*}
\noindent
Each component of this hierarchical model has been discussed in the main text. Specifically, Section 2.1 describes the multinomial ``data model,'' the ``process model'' is defined in Sections 3.1 and 3.2, and the ``parameter models are defined in Section 3.3. For a general review of the hierarchical modeling strategy see \citet{cressie-wikle-book}.

\subsection*{Appendix A.i: Model Summary: The MN-STM} We now organize all the levels of the hierarchical model discussed in Sections 2.1$\--$2.3 and 3.1 $\--$ 3.3. The MN-STM is \textit{proportional} to the product of the following conditional and marginal distributions:
\begin{align}\label{fullsummary}
\nonumber
&\mathrm{Data\hspace{5pt}Model:}\hspace{5pt} \textbf{y}_{t}\vert \bm{\beta},\bm{\eta}_{t}, \bm{\xi}_{t},\left\lbrace m_{it}\right\rbrace \ind f_{MN}(\textbf{y}_{t}\vert \bm{\nu}_{t} = \textbf{X}_{t}\bm{\beta} + \bm{\Phi}\bm{\eta}_{t} + \bm{\xi}_{t}, \left\lbrace m_{it}\right\rbrace);\hspace{15pt} t = 1,\ldots,T;\\
\nonumber
&\mathrm{Process\hspace{5pt}Model\hspace{5pt}1:}\hspace{5pt} \bm{\eta}_{t}\vert \bm{\eta}_{t-1},\textbf{H}_{t}, \alpha_{t},\kappa_{t}\sim f_{MLB}\left(\bm{\eta}_{t}\vert (\bm{\eta}_{t-1}^{\prime}\textbf{M}_{t}^{\prime}\bm{\Psi}_{t}^{\prime},\bm{\eta}_{t-1}^{\prime}\textbf{M}_{t}^{\prime}\textbf{V}_{t}^{\prime})^{\prime},\textbf{H}_{t}, \bm{\alpha}_{t}, \bm{\kappa}_{t}\right); \hspace{15pt} t = 2,\ldots, T;\\
\nonumber
&\mathrm{Process\hspace{5pt}Model\hspace{5pt}2:}\hspace{5pt} \bm{\eta}_{1}\vert \textbf{H}_{1}, \alpha_{1},\kappa_{1}\sim f_{MLB}(\bm{\eta}_{1}\vert \bm{0}_{(K-1)N_{t}+r},\textbf{H}_{1},\bm{\alpha}_{1}, \bm{\kappa}_{1});\\
\nonumber
&\mathrm{Process\hspace{5pt}Model\hspace{5pt}3:}\hspace{5pt} \bm{\xi}_{t}\vert \alpha_{\xi,t},\kappa_{\xi,t}\sim f_{MLB}(\xi_{t}\vert \bm{0}_{2(K-1)N_{t}},\textbf{H}_{\xi}, \bm{\alpha}_{\xi,t}, \bm{\kappa}_{\xi,t}); \hspace{15pt} t = 1,\ldots, T;\\
\nonumber
&\mathrm{Parameter\hspace{5pt}Model\hspace{5pt}1:}\hspace{5pt} \bm{\beta}\vert \alpha_{\beta,t},\kappa_{\beta,t}\sim f_{MLB}(\beta\vert \bm{0}_{n+p},\textbf{H}_{\beta}, \bm{\alpha}_{\beta}, \bm{\kappa}_{\beta});
\\
\nonumber
&\mathrm{Parameter\hspace{5pt}Model\hspace{5pt}2:}\hspace{5pt}\kappa_{\beta}\vert \alpha_{\beta}\sim f_{G}(a_{\beta},\tau_{\beta})I(\kappa_{\beta}>\alpha_{\beta});\\
\nonumber
&\mathrm{Parameter\hspace{5pt}Model\hspace{5pt}3:}\hspace{5pt}\kappa_{t}\vert \alpha_{t}\sim f_{G}(a_{\eta},\tau_{\eta})I(\kappa_{t}>\alpha_{t});\hspace{15pt} t = 1,\ldots, T;\\
\nonumber
&\mathrm{Parameter\hspace{5pt}Model\hspace{5pt}4:}\hspace{5pt}\kappa_{\xi,t}\vert \alpha_{\xi,t}\sim f_{G}(a_{\xi},\tau_{\xi})I(\kappa_{\xi,t}>\alpha_{\xi,t});\hspace{15pt} t = 1,\ldots, T,\\
\nonumber
&\mathrm{Parameter\hspace{5pt}Model\hspace{5pt}5:}\hspace{5pt}\alpha_{\beta}\sim f_{G}(a_{\beta,1},\tau_{\beta,1});\\
\nonumber
&\mathrm{Parameter\hspace{5pt}Model\hspace{5pt}6:}\hspace{5pt}\alpha_{t}\sim f_{G}(a_{\eta,1},\tau_{\eta,1});\hspace{15pt} t = 1,\ldots, T;\\
\nonumber
&\mathrm{Parameter\hspace{5pt}Model\hspace{5pt}7:}\hspace{5pt}\alpha_{\xi,t}\sim f_{G}(a_{\xi,1},\tau_{\xi,1});\hspace{15pt} t = 1,\ldots, T,\\
\end{align}
\noindent
where ``$f_{G}(\alpha,\tau)$ denotes a gamma density with shape $\alpha>0$ and rate $\tau>0$. To use (\ref{fullsummary}), we need to specify {$a_{\beta}>0$, $\tau_{\beta}>0$, $a_{\eta}>0$, $\tau_{\eta}>0$, $a_{\xi}>0$, $\tau_{\xi}$, $a_{\beta,1}>0$, $\tau_{\beta,1}>0$, $a_{\eta,1}>0$, $\tau_{\eta,1}>0$, $a_{\xi,1}>0$, and $\tau_{\xi,1}$.} We choose these hyperparameters so that the prior is ``flat.'' Specifically, set $a_{\beta}=\tau_{\beta}=a_{\eta}=\tau_{\eta}=a_{\xi}=\tau_{\xi}=a_{\beta,1}=\tau_{\beta,1}=a_{\eta,1}=\tau_{\eta,1}=a_{\xi,1}=\tau_{\xi,1}=1$ \citep{gelmanprior}. However, in general, these hyperparameters do not have to be the same, and one should consider alternate specifications. The full-conditional distributions for the shape parameters are computationally feasible to sample from using the adaptive rejection algorithm \citep{adrej}, which is possible due to Proposition 4. Although there are many levels in this statistical model, (\ref{fullsummary}) is fairly parsimonious compared to the (latent Gaussian process model) MSTM. Both the MSTM and MN-STM includes $p+rT+n$ parameters for regression parameters and random effects. The MN-STM only requires an additional $4T+2$ shape parameters to estimate.

Using the collapsed Gibbs sampler we obtain $B$ replicates from the posterior distribution of $\bm{\beta}$, $\bm{\eta}_{t}$, and $\bm{\xi}_{t}$, which we denote with $\bm{\beta}^{[b]}$, $\bm{\eta}_{t}^{[b]}$, and $\bm{\xi}_{t}^{[b]}$, respectively; $b = 1,\ldots, B$ and $t = 1,\ldots, T$. Then, the posterior replicate of $p_{kit}$ is given by,
\begin{equation*}
p_{kit}^{[b]} = \frac{\mathrm{exp}\left(\textbf{x}_{kit}^{\prime}\bm{\beta}^{[b]} + \bm{\phi}_{kit}^{\prime}\bm{\eta}_{t}^{[b]} + \xi_{kit}^{[b]}\right)}{1 + \mathrm{exp}\left(\textbf{x}_{kit}^{\prime}\bm{\beta}^{[b]} + \bm{\phi}_{kit}^{\prime}\bm{\eta}_{t}^{[b]} + \xi_{kit}^{[b]}\right)}; \hspace{15pt}k = 1,\ldots, K-1, i = 1,\ldots, N, t = 1,\ldots, T.
\end{equation*}
\noindent
Finally, to obtain posterior replicates of $\pi_{kit}$, first note that $\pi_{1it} = p_{1it}$, so that $\pi_{1it}^{[b]} = p_{1it}^{[b]}$. Then, we have
\begin{equation*}
p_{2it} = \frac{\pi_{2it}}{1 - \pi_{1it}}\hspace{15pt}i = 1,\ldots, N, t = 1,\ldots, T,
\end{equation*}
\noindent
which implies $\pi_{2it} = (1 - \pi_{1it})p_{2it}$, and a posterior replicate can be found by computing $\pi_{2it}^{[b]} = (1 - \pi_{1it}^{[b]})p_{2it}^{[b]}$. Continuing in this manner we obtain $\pi_{kit}^{[b]} = (1 - \underset{j<k}{\sum}\pi_{jit}^{[b]})p_{kit}^{[b]}$ for $k = 2,\ldots, K-1$, and $\pi_{Kit}^{[b]} = 1 - \underset{j<K}{\sum}\pi_{jit}^{[b]}$. Finally, we compute averages and variances of $\pi_{kit}^{[b]}$ (across $b$) to perform inference on $\pi_{kit}$.

One could also consider an empirical Bayesian version of (\ref{fullsummary}) by substituting estimates of parameters and removing levels of the hierarchical model. Besides possibly simplifying computations, this approach may help address the issue of confounding without the need for the Moran's I basis functions. In particular, empirical Bayesian hierarchical models treat $\bm{\beta}_{t}$ as fixed, which may help with the issue of confounding.
%

%
%

\subsection*{Appendix A.ii: Basic Properties of the MN-STM Covariances} The parsimonious model presented in (\ref{fullsummary}) allows for spatio-temporal dynamics, nonstationarity in space (see first paragraph of Section 3.3), nonstationarity in time, and asymmetric covariances. In this section, we clarify what is meant by spatio-temporal dynamics, non-stationary in time, and symmetric/asymmetric covariances.

We use the term ``spatio-temporal dynamics'' to refer to changes over time of the $(K-1)N$-dimensional vector,
\begin{equation*}
\bm{\nu}_{t}^{\mathrm{P}} = \textbf{X}_{t}^{\mathrm{P}}\bm{\beta} + \bm{\Phi}_{t}^{\mathrm{P}}\bm{\eta}_{t} + \bm{\xi}_{t}^{\mathrm{P}},
\end{equation*}
where $\bm{\nu}_{t}^{\mathrm{P}} = \left(v_{kit}: k = 1,\ldots, K-1, i = 1,\ldots, N\right)^{\prime}$ and $\bm{\xi}_{t}^{\mathrm{P}} = \left(\xi_{kit}: k = 1,\ldots, K-1, i = 1,\ldots, N\right)^{\prime}$. Consider the ``random walk'' model as an example of a model \textit{without} spatio-temporal dynamics. A random walk assumes $\bm{\nu}_{t}^{\mathrm{P}} = \bm{\nu}_{t-1}^{\mathrm{P}}+\bm{\epsilon}_{t}$ with $\bm{\epsilon}_{t}$ mutually independent of $\bm{\nu}_{t}^{\mathrm{P}}$ and $\bm{\nu}_{t-1}^{\mathrm{P}}$. The conditional expected value $E[\bm{\nu}_{t}^{\mathrm{P}} - \bm{\nu}_{t-1}^{\mathrm{P}}\vert \bm{\nu}_{t-1}^{\mathrm{P}}]$ is equal to an $(K-1)N$-dimensional vector of zeros.

The MN-STM incorporates spatio-temporal dynamics in the term $\bm{\Phi}_{t}^{\mathrm{P}}\bm{\eta}_{t}$, since for example,
\begin{equation}
E[\bm{\Phi}_{t}^{\mathrm{P}}\bm{\eta}_{t} - \bm{\Phi}_{t-1}^{\mathrm{P}}\bm{\eta}_{t-1}\vert \bm{\eta}_{t-1}] = (\bm{\Phi}_{t}^{\mathrm{P}}\textbf{M}_{t} - \bm{\Phi}_{t-1}^{\mathrm{P}})\bm{\eta}_{t-1} + \bm{\Phi}_{t}^{\mathrm{P}} E[\textbf{u}_{t}].
\end{equation}
\noindent
This expression is not identically equal to zero, and hence, we say that the MN-STM expresses spatio-temporal dynamics. Here the dynamics are determined by the propagator matrix $\textbf{M}_{t}$.

Stationarity in time refers to the case where the covariance between multivariate spatial fields, $\bm{\nu}_{t}^{\mathrm{P}}$ and $\bm{\nu}_{d}^{\mathrm{P}}$ depend only on their temporal lag. We have that,
\begin{equation}
\mathrm{cov}\left(\bm{\nu}_{t}^{\mathrm{P}},\bm{\nu}_{d}^{\mathrm{P}}\vert \{\textbf{H}_{t}\},\bm{\beta}\right) =\bm{\Phi}_{t}^{\mathrm{P}} \left(\prod_{j = d+1}^{t}\textbf{M}_{j}\right)\mathrm{cov}(\bm{\eta}_{d}\vert \{\textbf{H}_{t}\})\bm{\Phi}_{d}^{\mathrm{P}\prime},
\end{equation}
\noindent
which is not a function of $t - d$, where we have assumed that $t>d$. Stationarity is a special case, and occurs when $\textbf{M}_{1} = \cdots =  \textbf{M}_{T}$ and $\mathrm{cov}(\bm{\eta}_{1}\vert \{\textbf{H}_{t}\}) =\cdots = \mathrm{cov}(\bm{\eta}_{T}\vert \{\textbf{H}_{t}\})$, so that
\begin{equation*}
\mathrm{cov}\left(\bm{\nu}_{t}^{\mathrm{P}},\bm{\nu}_{d}^{\mathrm{P}}\vert \{\textbf{H}_{t}\},\bm{\beta}\right) = \bm{\Phi}_{t}^{\mathrm{P}}\textbf{M}_{1}^{t-d}\mathrm{cov}(\bm{\eta}_{1}\vert \{\textbf{H}_{t}\},\bm{\beta})\bm{\Phi}_{d}^{\mathrm{P}\prime},
\end{equation*}
for every $t$ and $d$ such that $t>d$. However, our specifications in Section 2.3 imply that $\textbf{M}_{j} \ne \textbf{M}_{k}$ and $\mathrm{cov}(\bm{\eta}_{i}\vert \{\textbf{H}_{t}\}) \ne \mathrm{cov}(\bm{\eta}_{j}\vert \{\textbf{H}_{t}\})$ for $j \ne k$ and $i \ne w$, and hence, nonstationarity holds in time. 

Symmetric covariances imply that the covariance between two spatial regions stay the same regardless of which variables/time points that are specified. In practice, the assumption of symmetry is unrealistic \citep[see the discussion in][pg. 234]{cressie-wikle-book}. A popular class of symmetric covariances are known as ``separable covariances'' \citep[e.g., see][and the references therein]{steinSep}. Also, the covariances implied by a linear models for coregionalization can be interpreted as a type of mixture of separable covariancs, and also implies symmetry \citep[e.g., see][among others]{jin,gmcar}. The MN-STM yields asymmetric covariances. Specifically, the expression of the covariance (given $\bm{\beta}$) between $\nu_{kit}$ and $\nu_{mjd}$ is,
\begin{equation*}
\mathrm{cov}\left( \nu_{kit}, \nu_{mjd}\vert \bm{\beta},\{\textbf{H}_{t}\}\right) = \bm{\phi}_{kit}^{\prime}\mathrm{cov}\left(\bm{\eta}_{t},\bm{\eta}_{d}\vert \{\textbf{H}_{t}\}\right)\bm{\phi}_{mjd} \ne \bm{\phi}_{kjd}^{\prime}\mathrm{cov}\left(\bm{\eta}_{t},\bm{\eta}_{d}\vert \{\textbf{H}_{t}\}\right)\bm{\phi}_{mit} = \mathrm{cov}\left( \nu_{kjd}, \nu_{mit}\vert \{\textbf{H}_{t}\}\right).
\end{equation*}
\noindent
The not equal to sign in the above holds because in our specification of the model in (\ref{fullsummary}), $\bm{\phi}_{kit}\ne \bm{\phi}_{kjd}$, $\bm{\phi}_{mjd}\ne \bm{\phi}_{mit}$, $\bm{\phi}_{kit}\ne \bm{\phi}_{mit}$, and $\bm{\phi}_{mjd}\ne \bm{\phi}_{kjd}$.

\subsection*{Appendix A.iii: The MN-STM and the PMSTM} One has a choice between using the MN-STM or the PMSTM from \citet{bradleyPMSTM} in practice. To see this, change the inverse logit function with the following:
\begin{equation*}
\pi_{kit} = \frac{\mathrm{exp}(\nu_{kit})}{\sum_{k}exp(\nu_{kit})}.
\end{equation*}
Substituting this expression into the multinomial pdf we have,
\begin{equation*}
f(\textbf{y}_{it}\vert \left\lbrace\pi_{kit}\right\rbrace, \left\lbrace m_{it}\right\rbrace) = \frac{m_{it}!}{Y_{1it}!\ldots  Y_{K,it}!}\mathrm{exp}(\nu_{1it})^{Y_{1it}}\ldots \mathrm{exp}(\nu_{Kit})^{Y_{Kit}}\left(\frac{1}{\sum_{k}exp(\nu_{kit})}\right)^{m_{it}};\hspace{5pt} t = 1,\ldots, T,\hspace{5pt} A_{i}\in D_{t}.
\end{equation*}
Then assuming $m_{kit}\vert \{\pi_{kit}\}\sim Poisson\left\lbrace \sum_{k}exp(\nu_{kit})\right\rbrace$ we have,
\begin{equation*}
f(\textbf{y}_{it}, m_{it}\vert \left\lbrace\pi_{kit}\right\rbrace) = \frac{1}{Y_{1it}!\ldots  Y_{K,it}!}\mathrm{exp}\left\lbrace\sum_{k}Y_{kit}\nu_{kit} - \sum_{k}exp(\nu_{kit})\right\rbrace;\hspace{5pt} t = 1,\ldots, T,\hspace{5pt} A_{i}\in D_{t},
\end{equation*}
which is proportional to a Poisson distribution as a function of $\textbf{y}_{it}$ and the multivariate log-gamma distribution as a function of $\{\nu_{kit}\}$. Thus, one could use the PMSTM from \citet{bradleyPMSTM} to analyze $\{y_{kit}\}$ in place of our MN-STM.

\section*{Appendix B: The Full-Conditional Distributions for the MN-STM}
\renewcommand{\theequation}{B.\arabic{equation}}
\setcounter{equation}{0}

Each full conditional distribution associated with MN-STM is listed below. 
\begin{enumerate}
	\item The full conditional distribution for $\bm{\beta}$ satisfies
	\begin{align}
	\nonumber
	f(\bm{\beta}\vert \cdot)&\propto \mathrm{exp}\left[\sum_{t = 1}^{T}\textbf{y}_{t}^{\prime}\textbf{X}_{t}\bm{\beta} - \sum_{t = 1}^{T}\textbf{n}_{t}^{\prime}\mathrm{log}\left\lbrace\bm{1} + \mathrm{exp}\left(\textbf{X}_{t}\bm{\beta} + \bm{\Phi}_{t}\bm{\eta}_{t} + \bm{\xi}_{t}\right)\right\rbrace\right]\\
	\nonumber
	&\hspace{5pt} \times \mathrm{exp}\left[\epsilon\bm{1}^{\prime}\textbf{X}\bm{\beta} +\alpha_{\beta}\bm{1}_{p}^{\prime}\bm{\beta}-\bm{\delta}^{\prime}\mathrm{log}\left\lbrace \bm{1} + \mathrm{exp}\left(\textbf{X}\bm{\beta}\right)\right\rbrace -\kappa_{\beta}\bm{1}_{p}^{\prime}\mathrm{log}\left\lbrace \bm{1} + \mathrm{exp}\left(\bm{\beta}\right)\right\rbrace\right].
	\end{align}
	Rearranging terms we have
	\begin{equation*}
	f(\bm{\beta}\vert \cdot) \propto \mathrm{exp}\left[\bm{\alpha}_{\beta}^{*\prime}\textbf{H}_{\beta}^{*}\bm{\beta}-\bm{\kappa}_{\beta}^{*\prime}\mathrm{log}\left\lbrace \bm{1} + \mathrm{exp}\left(\textbf{H}_{\beta}^{*}\bm{\beta} - \bm{\mu}_{\beta}^{*}\right)\right\rbrace\right],
	\end{equation*}
	\noindent
	which implies that $f(\bm{\beta}\vert \cdot)$ is MLB with mean $\bm{\mu}_{\beta}^{*}$, covariance $\textbf{H}_{\beta}^{*}$, shape $\bm{\alpha}_{\beta}^{*}$, and $\bm{\kappa}_{\beta}^{*}$, where 
	\begin{align*}
	\textbf{H}_{\beta}^{*} &= \left(\textbf{X}^{\prime},\sigma\textbf{X}^{\prime},\textbf{I}_{p}\right)^{\prime}\\
	\bm{\mu}_{\beta}^{*} &= \left(-\bm{\eta}_{1}^{\prime}\bm{\Phi}_{1}^{\prime}-\bm{\xi}_{1}^{\prime},\ldots, \bm{\eta}_{T}^{\prime}\bm{\Phi}_{T}^{\prime}-\bm{\xi}_{T}^{\prime},\bm{0}_{n+p}\right)^{\prime}\\
	\bm{\alpha}_{\beta}^{*} &= \left((\rho\textbf{y}^{\prime}+(\epsilon/2)\bm{1}_{n})^{\prime}, ((1-\rho)\textbf{y}^{\prime}/\sigma+(\epsilon/2\sigma)\bm{1}_{n}^{\prime})^{\prime}, \alpha_{\beta}\bm{1}_{p}\right)^{\prime},\\ \bm{\kappa}_{\beta}^{*}& = \left(\textbf{n}^{\prime}, \bm{\delta}^{\prime},\kappa_{\beta}\bm{1}_{p}^{\prime}\right)^{\prime}\\
	\textbf{y}& = \left(\textbf{y}_{1},\ldots, \textbf{y}_{T}\right)^{\prime}\\
	\textbf{n} &= \left(\textbf{n}_{1},\ldots, \textbf{n}_{T}\right)^{\prime},
	\end{align*}
	\noindent
	and where in our implementation we set $\rho = 0.9$, $\sigma = 1$, $\bm{\delta} = \textbf{n}$, and $\epsilon = 0.05$. Now, multiply the data model in (\ref{fullsummary}) by $\mathrm{exp}\left(\by^{\prime}\textbf{B}_{2,\beta}^{*}\textbf{q}_{\beta}^{*} + \epsilon\bm{1}_{n}^{\prime}\textbf{B}_{1,\beta}^{*}\textbf{q}_{\beta}^{*}\right)$, add $\textbf{B}_{1,\beta}^{*}\textbf{q}_{\beta}^{*}$ to $\bm{\nu}$, and set the location parameter of $\bm{\beta}$ in (\ref{fullsummary}) to be equal to $(-\textbf{q}_{\beta}^{*\prime}\textbf{B}_{2,\beta}^{*\prime},-\textbf{q}_{\beta}^{*\prime}\textbf{B}_{3,\beta}^{*\prime})^{\prime}$, where $f(\textbf{q}_{\beta}^{*}) = 1$ and $\textbf{B}_{\beta}^{*} = (\textbf{B}_{1,\beta}^{*\prime},\textbf{B}_{2,\beta}^{*\prime},\textbf{B}_{3,\beta}^{*\prime})^{\prime}$ is the $(2n+p)\times 2n$ orthonormal basis of the $(2n+p)\times p$ matrix $\textbf{H}_{\beta}^{*}$ immediately above. Let $\textbf{q}_{-\beta}$ be the stacked vectors of each augmented vector (with improper prior) defined at the end of each step in this Gibbs sampler  (not including $\textbf{q}_{\beta}$). Also let $\bm{\theta}$ contain each shape parameter in (\ref{fullsummary}). Then it follows from Proposition 2 that to simulate from $f(\bm{\beta}\vert \{\bm{\eta}_{t}\}, \{\bm{\xi}_{t}\},\bm{\theta},\by,\textbf{q}_{-\beta} = \bm{0})$, one can compute $(\textbf{H}_{\beta}^{*\prime}\textbf{H}_{\beta}^{*})^{-1}\textbf{H}_{\beta}^{*\prime}\textbf{w}$, where $\textbf{w}\sim \mathrm{MLB}(\bm{\mu}_{\beta}^{*},\textbf{I}_{2n+p},\bm{\alpha}_{\beta}^{*},\bm{\kappa}_{\beta}^{*})$. 
	\item For $T>1$ and $1 < t < T$, the full conditional distribution for $\bm{\eta}_{t}$ satisfies
	\begin{align}
	\nonumber
	f(\bm{\eta}_{t}\vert \cdot)&\propto \mathrm{exp}\left[\textbf{y}_{t}^{\prime}\bm{\Phi}_{t}\bm{\eta}_{t} - \textbf{n}_{t}^{\prime}\mathrm{log}\left\lbrace\bm{1} + \mathrm{exp}\left(\bm{\Phi}_{t}\bm{\eta}_{t}+\textbf{X}_{t}\bm{\beta}  + \bm{\xi}_{t}\right)\right\rbrace\right]\\
	\nonumber
	&\hspace{5pt} \times \mathrm{exp}\left[\epsilon\bm{1}_{(K-1)N_{t}}^{\prime}\bm{\Phi}_{t}\bm{\eta}_{t} +\alpha_{t}\bm{1}_{r}^{\prime}\bm{\eta}_{t}-\bm{\delta}_{t}^{\prime}\mathrm{log}\left\lbrace \bm{1} + \mathrm{exp}\left(\sigma\bm{\Phi}_{t}\bm{\eta}_{t} - \bm{\Phi}_{t}\textbf{M}_{t}\bm{\eta}_{t-1}\right)\right\rbrace\right.\\
	\nonumber
	&\hspace{50pt}\left.-\kappa_{t}\bm{1}_{r}^{\prime}\mathrm{log}\left\lbrace \bm{1} + \mathrm{exp}\left(\textbf{V}_{t}\bm{\eta}_{t} - \textbf{V}_{t}\textbf{M}_{t}\bm{\eta}_{t-1}\right)\right\rbrace\right]\\
	\nonumber
	&\hspace{5pt} \times\mathrm{exp}\left[-\epsilon\bm{1}_{(K-1)N_{t+1}}^{\prime}\bm{\Phi}_{t+1}\textbf{M}_{t+1}\bm{\eta}_{t} -\alpha_{t+1}\bm{1}_{r}^{\prime}\textbf{V}_{t+1}\textbf{M}_{t+1}\bm{\eta}_{t}\right.\\
	\nonumber
	&\hspace{50pt}\left.-\bm{\delta}_{t+1}^{\prime}\mathrm{log}\left\lbrace \bm{1} + \mathrm{exp}\left(\sigma\bm{\Phi}_{t+1}\bm{\eta}_{t+1} - \bm{\Phi}_{t+1}\textbf{M}_{t+1}\bm{\eta}_{t}\right)\right\rbrace \right.\\
	&\hspace{100pt}\left.-\kappa_{t+1}\bm{1}_{r}^{\prime}\mathrm{log}\left\lbrace \bm{1} + \mathrm{exp}\left(\textbf{V}_{t+1}\bm{\eta}_{t} - \textbf{V}_{t+1}\textbf{M}_{t+1}\bm{\eta}_{t}\right)\right\rbrace\right].
	\end{align}
	Rearranging terms we have
	\begin{equation*}
	f(\bm{\eta}_{t}\vert \cdot) \propto \mathrm{exp}\left[\bm{\alpha}_{t}^{*\prime}\textbf{H}_{t}^{*}\bm{\eta}_{t}-\bm{\kappa}_{t}^{*\prime}\mathrm{log}\left\lbrace \bm{1} + \mathrm{exp}\left(\textbf{H}_{t}^{*}\bm{\eta}_{t} - \bm{\mu}_{t}^{*}\right)\right\rbrace\right],
	\end{equation*}
	\noindent
	which implies that $f(\bm{\eta}_{t}\vert \cdot)$ is MLB with mean $\bm{\mu}_{t}^{*}$, covariance $\textbf{H}_{t}^{*}$, shape $\bm{\alpha}_{t}^{*}$, and $\bm{\kappa}_{t}^{*}$, where 
	\begin{align*}
	& \textbf{H}_{t}^{*} = \left(\bm{\Phi}_{t}^{\prime},\sigma\bm{\Phi}_{t}^{\prime},\textbf{V}_{t},-\textbf{M}_{t+1}^{\prime}\bm{\Phi}_{t+1}^{\prime},-\textbf{M}_{t+1}^{\prime}\textbf{V}_{t+1}^{\prime}\right)^{\prime}\\
	&\bm{\mu}_{t}^{*} = \left(-\bm{\beta}^{\prime}\textbf{X}_{t}^{\prime} - \bm{\xi}_{t}^{\prime},\bm{\Phi}\textbf{M}_{t}\bm{\eta}_{t-1},-\textbf{V}_{t}\textbf{M}_{t}\bm{\eta}_{t-1},-\sigma\bm{\Phi}_{t+1}\bm{\eta}_{t+1},-\textbf{V}_{t+1}\bm{\eta}_{t}\right)^{\prime}\\
	&\bm{\alpha}_{t}^{*} = \left((\rho\textbf{y}_{t}^{\prime} + (\epsilon/2) \bm{1}_{(K-1)N_{t}}^{\prime})^{\prime},  ((1-\rho)\textbf{y}_{t}^{\prime}/\sigma+(\epsilon/2\sigma)\bm{1}_{(K-1)N_{t}}^{\prime})^{\prime}, \alpha_{t}\bm{1}_{r}^{\prime},\epsilon\bm{1}_{(K-1)N_{t+1}}^{\prime},\alpha_{t+1}\bm{1}_{r}^{\prime}\right)^{\prime}
	\\ &\bm{\kappa}_{t}^{*} = \left(\textbf{n}_{t}, \bm{\delta}_{t}^{\prime},\kappa_{t}\bm{1}_{r}^{\prime},\bm{\delta}_{t+1}^{\prime},\kappa_{t+1}\bm{1}_{r}^{\prime}\right),
	\end{align*}
	and where in our implementation we set $\bm{\delta}_{t} = \textbf{n}_{t}$. Now, multiply the data model in (\ref{fullsummary}) by $\mathrm{exp}\left(\by_{t}^{\prime}\textbf{B}_{2,t}^{*}\textbf{q}_{t}^{*} + \epsilon\bm{1}_{(K-1)N_{t}}^{\prime}\textbf{B}_{1,t}^{*}\textbf{q}_{t}^{*}\right)$, add $\textbf{B}_{1,t}^{*}\textbf{q}_{t}^{*}$ to $\bm{\nu}_{t}$, add $(-\textbf{q}_{t}^{*\prime}\textbf{B}_{2,t}^{*\prime},-\textbf{q}_{t}^{*\prime}\textbf{B}_{3,t}^{*\prime})^{\prime}$ to the location parameter of $\bm{\eta}_{t}$ in (\ref{fullsummary}), and add $(-\textbf{q}_{t}^{*\prime}\textbf{B}_{4,t}^{*\prime},-\textbf{q}_{t}^{*\prime}\textbf{B}_{5,t}^{*\prime})^{\prime}$ to the location parameter of $\bm{\eta}_{t+1}$ in (\ref{fullsummary}), where $f(\textbf{q}_{t}^{*}) = 1$ and $\textbf{B}_{t}^{*} = (\textbf{B}_{1,t}^{*\prime},\textbf{B}_{2,t}^{*\prime},\textbf{B}_{3,t}^{*\prime},\textbf{B}_{4,t}^{*\prime},\textbf{B}_{5,t}^{*\prime})^{\prime}$ is the $(2(K-1)N_{t}+(K-1)N_{t+1}+2r)\times (2(K-1)N_{t}+(K-1)N_{t+1}+r)$ orthonormal basis of the $(2(K-1)N_{t}+(K-1)N_{t+1}+2r)\times r$ matrix $\textbf{H}_{t}^{*}$ immediately above. Here, $\textbf{B}_{1,t}^{*\prime}$ is $(K-1)N_{t}\times (2(K-1)N_{t}+(K-1)N_{t+1}+r)$, $\textbf{B}_{2,t}^{*\prime}$ is $(K-1)N_{t}\times (2(K-1)N_{t}+(K-1)N_{t+1}+r)$, $\textbf{B}_{3,t}^{*\prime}$ is $r\times (2(K-1)N_{t}+(K-1)N_{t+1}+r)$, $\textbf{B}_{4,t}^{*\prime}$ is $(K-1)N_{t+1}\times (2(K-1)N_{t}+(K-1)N_{t+1}+r)$, and $\textbf{B}_{5,t}^{*\prime}$ is $r\times (2(K-1)N_{t}+(K-1)N_{t+1}+r)$. Let $\textbf{q}_{-\eta,t}$ be the stacked vectors of each augmented vector (with improper prior) defined at the end of each step in this Gibbs sampler (not including $\textbf{q}_{t}$). Then it follows from Proposition 2 that to simulate from $f(\bm{\eta}_{t}\vert \bm{\beta},\{\bm{\eta}_{t}: j \ne t\}, \{\bm{\xi}_{t}\},\bm{\theta},\by,\textbf{q}_{-\eta,t} = \bm{0})$, one can compute $(\textbf{H}_{t}^{*\prime}\textbf{H}_{t}^{*})^{-1}\textbf{H}_{t}^{*\prime}\textbf{w}$, where $\textbf{w}\sim \mathrm{MLB}(\bm{\mu}_{t}^{*},\textbf{I}_{2(K-1)N_{t}+(K-1)N_{t+1}+2r},\bm{\alpha}_{t}^{*},\bm{\kappa}_{t}^{*})$. 
	
	\item If $T>1$, then the full conditional distribution for $\bm{\eta}_{1}$ is given by
	\begin{align}
	\nonumber
	f(\bm{\eta}_{1}\vert \cdot)&\propto \mathrm{exp}\left[\textbf{y}_{1}^{\prime}\bm{\Phi}_{1}\bm{\eta}_{1} - \textbf{n}_{1}^{\prime}\mathrm{log}\left\lbrace\bm{1} + \mathrm{exp}\left(\bm{\Phi}_{1}\bm{\eta}_{1}+\textbf{X}_{1}\bm{\beta}  + \bm{\xi}_{1}\right)\right\rbrace\right]\\
	\nonumber
	&\hspace{5pt} \times \mathrm{exp}\left[\epsilon\bm{1}_{(K-1)N_{1}}^{\prime}\bm{\Phi}_{1}\bm{\eta}_{1} +\alpha_{1}\bm{1}_{r}^{\prime}\bm{\eta}_{1}-\bm{\delta}_{1}^{\prime}\mathrm{log}\left\lbrace \bm{1} + \mathrm{exp}\left(\sigma\bm{\Phi}_{1}\bm{\eta}_{1}\right)\right\rbrace\right.\\
	\nonumber
	&\hspace{50pt}\left.-\kappa_{1}\bm{1}_{r}^{\prime}\mathrm{log}\left\lbrace \bm{1} + \mathrm{exp}\left(\textbf{V}_{1}\bm{\eta}_{1}\right)\right\rbrace\right]\\
	\nonumber
	&\hspace{5pt} \times\mathrm{exp}\left[-\epsilon\bm{1}_{(K-1)N_{2}}^{\prime}\bm{\Phi}_{2}\textbf{M}_{2}\bm{\eta}_{1} -\alpha_{2}\bm{1}_{r}^{\prime}\textbf{V}_{2}\textbf{M}_{2}\bm{\eta}_{1}\right.\\
	\nonumber
	&\hspace{50pt}\left.-\bm{\delta}_{2}^{\prime}\mathrm{log}\left\lbrace \bm{1} + \mathrm{exp}\left(\sigma\bm{\Phi}_{2}\bm{\eta}_{2} - \bm{\Phi}_{2}\textbf{M}_{2}\bm{\eta}_{1}\right)\right\rbrace \right.\\
	&\hspace{100pt}\left.-\kappa_{2}\bm{1}_{r}^{\prime}\mathrm{log}\left\lbrace \bm{1} + \mathrm{exp}\left(\textbf{V}_{2}\bm{\eta}_{1} - \textbf{V}_{2}\textbf{M}_{2}\bm{\eta}_{1}\right)\right\rbrace\right].
	\end{align}
	Rearranging terms we have
	\begin{equation*}
	f(\bm{\eta}_{1}\vert \cdot) \propto \mathrm{exp}\left[\bm{\alpha}_{1}^{*\prime}\textbf{H}_{1}^{*}\bm{\eta}_{1}-\bm{\kappa}_{1}^{*\prime}\mathrm{log}\left\lbrace \bm{1} + \mathrm{exp}\left(\textbf{H}_{1}^{*}\bm{\eta}_{t} - \bm{\mu}_{1}^{*}\right)\right\rbrace\right],
	\end{equation*}
	\noindent
	which implies that $f(\bm{\eta}_{1}\vert \cdot)$ is MLB with mean $\bm{\mu}_{1}^{*}$, covariance $\textbf{H}_{1}^{*}$, shape $\bm{\alpha}_{1}^{*}$, and $\bm{\kappa}_{1}^{*}$, where 
	\begin{align*}
	& \textbf{H}_{1}^{*} = \left(\bm{\Phi}_{1}^{\prime},\sigma\bm{\Phi}_{1}^{\prime},\textbf{V}_{1},-\textbf{M}_{2}^{\prime}\bm{\Phi}_{2}^{\prime},-\textbf{M}_{2}^{\prime}\textbf{V}_{2}^{\prime}\right)^{\prime}\\
	&\bm{\mu}_{1}^{*} = \left(-\bm{\beta}^{\prime}\textbf{X}_{1}^{\prime} - \bm{\xi}_{1}^{\prime},\bm{0}_{(K-1)N_{t}+r},-\sigma\bm{\Phi}_{2}\bm{\eta}_{2},-\textbf{V}_{2}\bm{\eta}_{1}\right)^{\prime}\\
	&\bm{\alpha}_{1}^{*} = \left((\rho\textbf{y}_{1}^{\prime} + (\epsilon/2) \bm{1}_{(K-1)N_{1}}^{\prime})^{\prime},  ((1-\rho)\textbf{y}_{1}^{\prime}/\sigma+(\epsilon/2\sigma)\bm{1}_{(K-1)N_{1}}^{\prime})^{\prime}, \alpha_{1}\bm{1}_{r}^{\prime},\epsilon\bm{1}_{(K-1)N_{2}}^{\prime},\alpha_{2}\bm{1}_{r}^{\prime}\right)^{\prime}
	\\ &\bm{\kappa}_{t}^{*} = \left(\textbf{n}_{1}, \bm{\delta}_{1}^{\prime},\kappa_{1}\bm{1}_{r}^{\prime},\bm{\delta}_{2}^{\prime},\kappa_{2}\bm{1}_{r}^{\prime}\right).
	\end{align*}
	Now, multiply the data model in (\ref{fullsummary}) by $\mathrm{exp}\left(\by_{1}^{\prime}\textbf{B}_{2,1}^{*}\textbf{q}_{t}^{*} + \epsilon\bm{1}_{(K-1)N_{1}}^{\prime}\textbf{B}_{1,1}^{*}\textbf{q}_{t}^{*}\right)$, add $\textbf{B}_{1,1}^{*}\textbf{q}_{1}^{*}$ to $\bm{\nu}_{1}$, add $(-\textbf{q}_{1}^{*\prime}\textbf{B}_{2,1}^{*\prime},-\textbf{q}_{1}^{*\prime}\textbf{B}_{3,1}^{*\prime})^{\prime}$ to the location parameter of $\bm{\eta}_{1}$ in (\ref{fullsummary}), and add $(-\textbf{q}_{1}^{*\prime}\textbf{B}_{4,1}^{*\prime},-\textbf{q}_{1}^{*\prime}\textbf{B}_{5,1}^{*\prime})^{\prime}$ to the location parameter of $\bm{\eta}_{2}$ in (\ref{fullsummary}), where $f(\textbf{q}_{1}^{*}) = 1$ and $\textbf{B}_{1}^{*} = (\textbf{B}_{1,1}^{*\prime},\textbf{B}_{2, 1}^{*\prime},\textbf{B}_{3,1}^{*\prime},\textbf{B}_{4,1}^{*\prime},\textbf{B}_{5,1}^{*\prime})^{\prime}$ is the $(2(K-1)N_{1}+(K-1)N_{2}+2r)\times (2(K-1)N_{1}+(K-1)N_{2}+r)$ orthonormal basis of the $(2(K-1)N_{1}+(K-1)N_{2}+2r)\times r$ matrix $\textbf{H}_{1}^{*}$ immediately above. Here, $\textbf{B}_{1,1}^{*\prime}$ is $(K-1)N_{1}\times (2(K-1)N_{1}+(K-1)N_{2}+r)$, $\textbf{B}_{2,1}^{*\prime}$ is $(K-1)N_{1}\times (2(K-1)N_{1}+(K-1)N_{2}+r)$, $\textbf{B}_{3,1}^{*\prime}$ is $r\times (2(K-1)N_{1}+(K-1)N_{2}+r)$, $\textbf{B}_{4,1}^{*\prime}$ is $(K-1)N_{2}\times (2(K-1)N_{1}+(K-1)N_{2}+r)$, and $\textbf{B}_{5,1}^{*\prime}$ is $r\times (2(K-1)N_{1}+(K-1)N_{2}+r)$. Let $\textbf{q}_{-\eta,1}$. Let $\textbf{q}_{-\eta,t}$ be the stacked vectors of each augmented vector (with improper prior) defined at the end of each step in this Gibbs sampler (not including $\textbf{q}_{1}$). Then it follows from Proposition 2 that to simulate from $f(\bm{\eta}_{1}\vert \bm{\beta},\{\bm{\eta}_{1}: j \ne 1\}, \{\bm{\xi}_{t}\},\bm{\theta},\by,\textbf{q}_{-\eta,1} = \bm{0})$, one can compute $(\textbf{H}_{1}^{*\prime}\textbf{H}_{1}^{*})^{-1}\textbf{H}_{1}^{*\prime}\textbf{w}$, where $\textbf{w}\sim \mathrm{MLB}(\bm{\mu}_{1}^{*},\textbf{I}_{2(K-1)N_{1}+(K-1)N_{2}+2r},\bm{\alpha}_{1}^{*},\bm{\kappa}_{1}^{*})$. 
	\item The full conditional distribution for $\bm{\eta}_{T}$ satisfies
	\begin{align}
	\nonumber
	f(\bm{\eta}_{T}\vert \cdot)&\propto \mathrm{exp}\left[\textbf{y}_{T}^{\prime}\bm{\Phi}_{T}\bm{\eta}_{T} - \textbf{n}_{T}^{\prime}\mathrm{log}\left\lbrace\bm{1} + \mathrm{exp}\left(\bm{\Phi}_{T}\bm{\eta}_{T}+\textbf{X}_{T}\bm{\beta}  + \bm{\xi}_{T}\right)\right\rbrace\right]\\
	\nonumber
	&\hspace{5pt} \times \mathrm{exp}\left[\epsilon\bm{1}_{(K-1)N_{T}}^{\prime}\bm{\Phi}_{T}\bm{\eta}_{T} +\alpha_{T}\bm{1}_{r}^{\prime}\bm{\eta}_{T}-\bm{\delta}_{T}^{\prime}\mathrm{log}\left\lbrace \bm{1} + \mathrm{exp}\left(\sigma\bm{\Phi}_{T}\bm{\eta}_{T} - \bm{\Phi}_{T}\textbf{M}_{T}\bm{\eta}_{T-1}\right)\right\rbrace\right. \\
	\nonumber
	&\hspace{50pt}\left.-\kappa_{T}\bm{1}_{r}^{\prime}\mathrm{log}\left\lbrace \bm{1} + \mathrm{exp}\left(\textbf{V}_{T}\bm{\eta}_{T} - \textbf{V}_{T}\textbf{M}_{T}\bm{\eta}_{T-1}\right)\right\rbrace\right].
	\end{align}
	Rearranging terms we have
	\begin{equation*}
	f(\bm{\eta}_{T}\vert \cdot) \propto \mathrm{exp}\left[\bm{\alpha}_{T}^{*\prime}\textbf{H}_{T}^{*}\bm{\eta}_{T}-\bm{\kappa}_{T}^{*\prime}\mathrm{log}\left\lbrace \bm{1} + \mathrm{exp}\left(\textbf{H}_{T}^{*}\bm{\eta}_{T} - \bm{\mu}_{T}^{*}\right)\right\rbrace\right],
	\end{equation*}
	\noindent
	which implies that $f(\bm{\eta}_{T}\vert \cdot)$ is MLB with mean $\bm{\mu}_{T}^{*}$, covariance $\textbf{H}_{T}^{*}$, shape $\bm{\alpha}_{T}^{*}$, and $\bm{\kappa}_{T}^{*}$, where 
	\begin{align}\label{etaTfull}
	& \textbf{H}_{T}^{*} = \left(\bm{\Phi}_{T}^{\prime},\sigma\bm{\Phi}_{T}^{\prime},\textbf{V}_{T},\right)^{\prime}\\
	&\bm{\mu}_{T}^{*} = \left(-\bm{\beta}^{\prime}\textbf{X}_{T}^{\prime} - \bm{\xi}_{T}^{\prime},\bm{\Phi}_{T}\textbf{M}_{T}\bm{\eta}_{T-1},-\textbf{V}_{T}\textbf{M}_{T}\bm{\eta}_{T-1}\right)^{\prime}\\
	&\bm{\alpha}_{T}^{*} = \left(\rho\textbf{y}_{T}^{\prime} + (\epsilon/2)\bm{1}_{(K-1)N_{t}}^{\prime}, (1-\rho)\textbf{y}_{T}^{\prime}/\sigma + (\epsilon/2\sigma)\bm{1}_{(K-1)N_{t}}^{\prime}, \alpha_{T}\bm{1}_{r}\right)^{\prime}
	\\ &\bm{\kappa}_{T}^{*} = \left(\textbf{n}_{T}, \bm{\delta}_{T}^{\prime},\kappa_{T}\bm{1}_{r}^{\prime}\right).
	\end{align}
	If $T=1$ then replace $\textbf{M}_{T}$ and $\bm{\eta}_{T-1}$ with $\bm{0}_{r,r}$ and $\bm{0}_{r}$ within the expressions in (\ref{etaTfull}).  Now, multiply the data model in (\ref{fullsummary}) by $\mathrm{exp}\left(\by_{T}^{\prime}\textbf{B}_{2,T}^{*}\textbf{q}_{T}^{*} + \epsilon\bm{1}_{(K-1)N_{T}}^{\prime}\textbf{B}_{1,T}^{*}\textbf{q}_{T}^{*}\right)$, add $\textbf{B}_{1,T}^{*}\textbf{q}_{T}^{*}$ to $\bm{\nu}_{T}$, and add $(-\textbf{q}_{T}^{*\prime}\textbf{B}_{2,T}^{*\prime},-\textbf{q}_{T}^{*\prime}\textbf{B}_{3,T}^{*\prime})^{\prime}$ to the location parameter of $\bm{\eta}_{T}$ in (\ref{fullsummary}), where $f(\textbf{q}_{T}^{*}) = 1$ and $\textbf{B}_{T}^{*} = (\textbf{B}_{1,T}^{*\prime},\textbf{B}_{2,T}^{*\prime},\textbf{B}_{3,T}^{*\prime})^{\prime}$ is the $(2(K-1)N_{t}+r)\times 2(K-1)N_{t}$ orthonormal basis of the $(2(K-1)N_{t}+r)\times r$ matrix $\textbf{H}_{T}^{*}$ immediately above. Let $\textbf{q}_{-\eta,T}$ be the stacked vectors of each augmented vector (with improper prior) defined at the end of each step in this Gibbs sampler  (not including $\textbf{q}_{T}$). Also let $\bm{\theta}$ contain each shape parameter in (\ref{fullsummary}). Then it follows from Proposition 2 that to simulate from $f(\bm{\eta}_{T}\vert \{\bm{\eta}_{t}: t<T\}, \{\bm{\xi}_{t}\},\bm{\theta},\by,\textbf{q}_{-\eta,T} = \bm{0})$, one can compute $(\textbf{H}_{T}^{*\prime}\textbf{H}_{T}^{*})^{-1}\textbf{H}_{T}^{*\prime}\textbf{w}$, where $\textbf{w}\sim \mathrm{MLB}(\bm{\mu}_{T}^{*},\textbf{I}_{2(K-1)N_{t}+r},\bm{\alpha}_{T}^{*},\bm{\kappa}_{T}^{*})$. 
	\item The full conditional distribution for $\bm{\xi}_{t}$ is given by
	\begin{align}
	\nonumber
	f(\bm{\xi}_{t}\vert \cdot)&\propto \hspace{5pt}\mathrm{exp}\left[\textbf{y}_{t}^{\prime}\bm{\xi}_{t}- \textbf{n}_{t}^{\prime}\mathrm{log}\left\lbrace\bm{1} + \mathrm{exp}\left(\bm{\xi}_{t}+\textbf{X}_{t}\bm{\beta}+\bm{\Phi}_{t}\bm{\eta}_{t}\right)\right\rbrace\right]\\
	\nonumber
	&\hspace{5pt} \times \mathrm{exp}\left[(\epsilon + \alpha_{\xi,t})\bm{1}_{(K-1)N_{t}}^{\prime}\bm{\xi}_{t}-(\bm{\delta}_{t}+\kappa_{\xi,t})\bm{1}_{(K-1)N_{t}}^{\prime}\mathrm{log}\left\lbrace \bm{1} + \mathrm{exp}\left(\bm{\xi_{t}}\right)\right\rbrace\right].
	\end{align}
	Rearranging terms we have
	\begin{equation*}
	f(\bm{\xi}_{t}\vert \cdot) \propto \mathrm{exp}\left[\bm{\alpha}_{\xi,t}^{*\prime}\textbf{H}_{\xi,t}^{*}\bm{\xi}_{t}-\bm{\kappa}_{\xi,t}^{*\prime}\mathrm{log}\left\lbrace \bm{1} + \mathrm{exp}\left(\textbf{H}_{\xi,t}^{*}\bm{\eta}_{T} - \bm{\mu}_{\xi,t}^{*}\right)\right\rbrace\right],
	\end{equation*}
	which implies that $f(\bm{\xi}_{t}\vert \cdot)$ is MLB with mean $\bm{\mu}_{\xi,t}^{*}$, covariance $\textbf{H}_{\xi,t}^{*}$, shape $\bm{\alpha}_{\xi,t}^{*}$, and $\bm{\kappa}_{\xi,t}^{*}$, where 
	\begin{align*}
	& \textbf{H}_{\xi,t}^{*} = \left(\textbf{I}_{N_{t}},\textbf{I}_{N_{t}},\textbf{I}_{N_{t}}\right)^{\prime}\\
	&\bm{\mu}_{\xi,t}^{*} = \left(-\bm{\beta}^{\prime}\textbf{X}_{t}^{\prime}  -\bm{\eta}_{t}^{\prime}\bm{\Phi}_{t}^{\prime},\bm{0}_{2(K-1)N_{t}}\right)^{\prime}\\
	&\bm{\alpha}_{\xi,t}^{*} = \left(\rho\textbf{y}_{t}^{\prime}+ (\epsilon/2)\bm{1}_{(K-1)N_{t}}^{\prime},(1-\rho)\textbf{y}_{t}^{\prime} + (\epsilon/2)\bm{1}_{(K-1)N_{t}}^{\prime},\alpha_{\xi,t}\bm{1}_{(K-1)N_{t}}^{\prime}\right)^{\prime}
	\\ &\bm{\kappa}_{\xi,t}^{*} = \left(\textbf{n}_{t}^{\prime},\bm{\delta}_{t}, \kappa_{\xi,t}\bm{1}_{(K-1)N_{t}}^{\prime}\right).
	\end{align*}
	Now, multiply the data model in (\ref{fullsummary}) by $\mathrm{exp}\left(\by_{t}^{\prime}\textbf{B}_{2,\xi,t}^{*}\textbf{q}_{t}^{*} + \epsilon\bm{1}_{(K-1)N_{T}}^{\prime}\textbf{B}_{1,\xi,t}^{*}\textbf{q}_{\xi,t}^{*}\right)$, add $\textbf{B}_{1,\xi,t}^{*}\textbf{q}_{\xi,t}^{*}$ to $\bm{\nu}_{t}$, and add $(-\textbf{q}_{\xi,t}^{*\prime}\textbf{B}_{2,\xi,t}^{*\prime},-\textbf{q}_{\xi,t}^{*\prime}\textbf{B}_{3,\xi,t}^{*\prime})^{\prime}$ to the location parameter of $\bm{\eta}_{\xi,t}$ in (\ref{fullsummary}), where $f(\textbf{q}_{\xi,t}^{*}) = 1$ and $\textbf{B}_{\xi,t}^{*} = (\textbf{B}_{1,\xi,t}^{*\prime},\textbf{B}_{2,\xi,t}^{*\prime},\textbf{B}_{3,\xi,t}^{*\prime})^{\prime}$ is the $(3(K-1)N_{t})\times 2(K-1)N_{t}$ orthonormal basis of the $(3(K-1)N_{t})\times N_{t}$ matrix $\textbf{H}_{\xi,t}^{*}$ immediately above. Let $\textbf{q}_{-\xi,t}$ be the stacked vectors of each augmented vector (with improper prior) defined at the end of each step in this Gibbs sampler  (not including $\textbf{q}_{\xi,t}$). Also let $\bm{\theta}$ contain each shape parameter in (\ref{fullsummary}). Then it follows from Proposition 2 that to simulate from $f(\bm{\xi}_{t}\vert \{\bm{\eta}_{t}\}, \{\bm{\xi}_{j}: j \ne t\},\bm{\theta},\by,\textbf{q}_{-\xi,t} = \bm{0})$, one can compute $(\textbf{H}_{\xi,t}^{*\prime}\textbf{H}_{\xi,t}^{*})^{-1}\textbf{H}_{\xi,t}^{*\prime}\textbf{w}$, where $\textbf{w}\sim \mathrm{MLB}(\bm{\mu}_{\xi,t}^{*},\textbf{I}_{3(K-1)N_{t}},\bm{\alpha}_{\xi,t}^{*},\bm{\kappa}_{\xi,t}^{*})$. 
	\item The full conditional distribution for $\alpha_{t}$ and $\kappa_{t}$, can be found as, 
	\begin{align}\label{alphaFC}
	\nonumber
	f(\alpha_{t}\vert \cdot)
	&\propto f(\bm{\eta}_{t},\textbf{q}_{t} = \bm{0}_{N_{t}}\vert \cdot)f(\alpha_{t})\\
	& \propto \left\lbrace \frac{\Gamma(\kappa_{t})}{\Gamma(\alpha_{t})\Gamma(\kappa_{t} - \alpha_{t})}\right\rbrace^{N_{t}+r} \mathrm{exp}\left(\alpha_{t}\bm{1}_{r}^{\prime}\textbf{V}_{t}\bm{\eta}_{1} - \alpha_{t}\bm{1}_{r}^{\prime}\textbf{M}_{t}\bm{\eta}_{t-1}\textbf{V}_{t}\bm{\eta}_{1}\right)f_{G}(\alpha_{t}),
	\end{align}
	\noindent
	where the $N_{t}$-dimensional vector $\textbf{q}_{t}$ is defined in Steps 2 - 4 above. Take the first and second derivative of the log of (\ref{alphaFC}) and obtain:
	\begin{equation}\label{derivs}
	\frac{d^{2}}{d^{2}\alpha_{t}}\mathrm{log}\left\lbrace f(\alpha_{t}\vert \cdot) \right\rbrace =\frac{d^{2}}{d^{2}\alpha_{t}}\left[ -r\hspace{2pt}\mathrm{log}\left\lbrace \frac{\Gamma(\kappa_{t})}{\Gamma(\alpha_{t})\Gamma(\kappa_{t} - \alpha_{t})}\right\rbrace\right] + \frac{d^{2}}{d^{2}\alpha_{t}}\mathrm{log}\left\lbrace f(\alpha_{t})\right\rbrace.
	\end{equation}
	\noindent
	It is well known that the log of the beta function is convex and the log of the gamma probability density function (with shape parameter greater than or equal to one) is concave \citep{Dragomir}. Thus, it follows from (\ref{derivs}) that the pdf for $\alpha_{t}$ is log-concave. The proofs for $\kappa_{t}$, $\alpha_{\beta}$, $\kappa_{\beta}$, $\alpha_{\xi,t}$, and $\kappa_{\xi,t}$ are similar.\\
	\item The full-conditional for $\alpha_{t}$ was derived in Step 5, and can be found in (\ref{alphaFC}). The derivation of the remaining shape parameters are straightforward and are as follows:
	\begin{align*}
	f(\alpha_{t}\vert \cdot) & \propto \left\lbrace \frac{1}{\Gamma(\alpha_{t})\Gamma(\kappa_{t} - \alpha_{t})}\right\rbrace^{r} \mathrm{exp}\left(\alpha_{t}\bm{1}^{\prime}\textbf{V}_{t}\bm{\eta}_{t} - \alpha_{t}\bm{1}_{r}^{\prime}\textbf{M}_{t}\bm{\eta}_{t-1}\right)f_{G}(\alpha_{t})\\
	f(\alpha_{1}\vert \cdot) & \propto \left\lbrace \frac{1}{\Gamma(\alpha_{1})\Gamma(\kappa_{1} - \alpha_{1})}\right\rbrace^{r} \mathrm{exp}\left(\alpha_{1}\bm{1}_{r}^{\prime}\textbf{V}_{1}\bm{\eta}_{1}\right)f_{G}(\alpha_{t})\\
	f(\alpha_{\beta}\vert \cdot) & \propto \left\lbrace \frac{1}{\Gamma(\alpha_{\beta})\Gamma(\kappa_{\beta} - \alpha_{\beta})}\right\rbrace^{p} \mathrm{exp}\left(\alpha_{\beta}\bm{1}_{p}^{\prime}\bm{\beta}\right)f_{G}(\alpha_{\beta})\\
	f(\alpha_{\xi,t}\vert \cdot) & \propto \left\lbrace \frac{1}{\Gamma(\alpha_{\xi,t}+\epsilon)\Gamma(\kappa_{\xi,t} - \alpha_{\xi,t})}\right\rbrace^{(K-1)N_{t}} \mathrm{exp}\left(\alpha_{\xi,t}\bm{1}_{(K-1)N_{t}}^{\prime}\bm{\xi}_{t}\right)f_{G}(\alpha_{\xi,t})\\
	f(\kappa_{t}\vert \cdot) & \propto \left\lbrace \frac{\Gamma(\kappa_{t})}{\Gamma(\kappa_{t} - \alpha_{t})}\right\rbrace^{r} \mathrm{exp}\left[-\kappa_{t}\bm{1}_{r}^{\prime}\mathrm{log}\left\lbrace \bm{1}_{r} + \mathrm{exp}\left(\textbf{V}_{t}\bm{\eta}_{t}-\textbf{M}_{t}\bm{\eta}_{t-1}\right)\right\rbrace\right]f(\kappa_{t})
	\end{align*}
	\begin{align*}
	f(\kappa_{1}\vert \cdot) & \propto \left\lbrace \frac{\Gamma(\kappa_{1})}{\Gamma(\kappa_{1} - \alpha_{1})}\right\rbrace^{r} \mathrm{exp}\left[-\kappa_{1}\bm{1}_{r}^{\prime}\mathrm{log}\left\lbrace \bm{1}_{r} + \mathrm{exp}\left(\textbf{V}_{1}\bm{\eta}_{1}\right)\right\rbrace\right]f(\kappa_{1})\\
	f(\kappa_{\beta}\vert \cdot) & \propto \left\lbrace \frac{\Gamma(\kappa_{\beta})}{\Gamma(\kappa_{\beta} - \alpha_{\beta})}\right\rbrace^{p} \mathrm{exp}\left[-\kappa_{\beta}\bm{1}_{p}^{\prime}\mathrm{log}\left\lbrace \bm{1}_{p} + \mathrm{exp}\left(\bm{\beta}\right)\right\rbrace\right]f(\kappa_{\beta})\\
	f(\kappa_{\xi,t}\vert \cdot) & \propto \left\lbrace \frac{\Gamma(\kappa_{\xi,t})}{\Gamma(\kappa_{\xi,t} - \alpha_{\xi,t})}\right\rbrace^{(K-1)N_{t}} \mathrm{exp}\left[-\kappa_{\xi,t}\bm{1}_{(K-1)N_{t}}^{\prime}\mathrm{log}\left\lbrace \bm{1}_{(K-1)N_{t}} + \mathrm{exp}\left(\bm{\xi}_{t}\right)\right\rbrace\right]f(\kappa_{t}).
	\end{align*}
	\noindent
	These full-conditional distributions are all computationally practical to simulate from using the adaptive rejection algorithm (see Step 5 for a proof of log-concavity).
\end{enumerate}

\section*{Appendix C: A Latent Gaussian Process Model Representation of the Latent MLB model}
\renewcommand{\theequation}{C.\arabic{equation}}
\setcounter{equation}{0}

Proposition 4 allows us to augment the model in (\ref{fullsummary}) as follows:
\begin{align}
\nonumber
&\mathrm{Data\hspace{5pt}Model:}\hspace{5pt} \textbf{y}_{t}\vert \bm{\beta},\bm{\eta}_{t}, \bm{\xi}_{t} \ind f_{MN}(\textbf{y}_{t}\vert \bm{\nu}_{t} = \textbf{X}_{t}\bm{\beta} + \bm{\Phi}\bm{\eta}_{t} + \bm{\xi}_{t});\hspace{15pt} t = 1,\ldots,T;\\
\nonumber
&\mathrm{Process\hspace{5pt}Model\hspace{5pt}1:}\hspace{5pt} \bm{\eta}_{t}\vert \bm{\eta}_{t-1},\textbf{H}_{t}, \alpha_{t},\kappa_{t}\sim f_{Gau}(\bm{\eta}_{t}\vert (\textbf{H}^{\prime}\bm{\Omega}\textbf{H})^{-1}(\textbf{H}\textbf{M}_{t}\bm{\eta}_{t-1} - \bm{\zeta}_{t}),\bm{\Omega}_{t}); \hspace{15pt} t = 2,\ldots, T;\\
\nonumber
&\mathrm{Process\hspace{5pt}Model\hspace{5pt}2:}\hspace{5pt} \bm{\eta}_{1}\vert \textbf{H}_{1}, \alpha_{1},\kappa_{1}\sim f_{Gau}(\bm{\eta}_{1}\vert -(\textbf{H}^{\prime}\bm{\Omega}\textbf{H})^{-1} \bm{\zeta}_{t},\bm{\Omega}_{1});\\
\nonumber
&\mathrm{Process\hspace{5pt}Model\hspace{5pt}3:}\hspace{5pt} \bm{\xi}_{t}\vert \alpha_{\xi,t},\kappa_{\xi,t}\sim f_{Gau}(\bm{\eta}_{t}\vert - \bm{\zeta}_{\xi,t},\bm{\Omega}_{\xi,t}); \hspace{15pt} t = 1,\ldots, T;\\
\nonumber
&\mathrm{Parameter\hspace{5pt}Model\hspace{5pt}1:}\hspace{5pt} \bm{\beta}\vert \alpha_{\beta,t},\kappa_{\beta,t}\sim f_{Gau}(\bm{\beta}\vert -(\textbf{H}_{\beta}^{\prime}\bm{\Omega}_{\beta}\textbf{H}_{\beta})^{-1}\bm{\zeta}_{\beta},\bm{\Omega}_{\beta});
\\
\nonumber
&\mathrm{Parameter\hspace{5pt}Model\hspace{5pt}2:}\hspace{5pt}\bm{\Omega}_{t}\vert \kappa_{t}\sim g(\bm{\Omega}_{t},\textbf{M}_{t}\bm{\eta}_{t-1}, \textbf{H}_{t}, {\alpha}_{t}\bm{1}_{N_{t}+r}, {\kappa}_{t}\bm{1}_{N_{t}+r})\frac{\mathrm{det}(\textbf{H}_{t}^{\prime}\bm{\Omega}_{t}\textbf{H}_{t})^{1/2}}{(2\pi)^{M/2}} p(\bm{\Omega}_{t}\vert {\kappa}_{t}\bm{1}_{N_{t}+r});\\
\nonumber
&\mathrm{Parameter\hspace{5pt}Model\hspace{5pt}3:}\hspace{5pt}\bm{\Omega}_{\xi,t}\vert \kappa_{\xi,t}\sim g(\bm{\Omega}_{\xi,t},\bm{0}_{N_{t}}, \textbf{H}_{\xi,t}, {\alpha}_{\xi,t}\bm{1}_{N_{t}}, {\kappa}_{\xi,t}\bm{1}_{N_{t}})\frac{\mathrm{det}(\textbf{H}_{\xi,t}^{\prime}\bm{\Omega}_{\xi,t}\textbf{H}_{\xi,t})^{1/2}}{(2\pi)^{M/2}} p(\bm{\Omega}_{\xi,t}\vert {\kappa}_{\xi,t}\bm{1}_{N_{t}});\\
\nonumber
&\mathrm{Parameter\hspace{5pt}Model\hspace{5pt}4:}\hspace{5pt}\bm{\Omega}_{\beta}\vert \kappa_{\beta}\sim g(\bm{\Omega}_{\beta},\bm{0}_{n+p}, \textbf{H}_{\beta}, {\alpha}_{\beta}\bm{1}_{n+p}, {\kappa}_{\beta}\bm{1}_{n+p})\frac{\mathrm{det}(\textbf{H}_{\beta}^{\prime}\bm{\Omega}_{\beta}\textbf{H}_{\beta})^{1/2}}{(2\pi)^{M/2}} p(\bm{\Omega}_{\beta}\vert {\kappa}_{\beta}\bm{1}_{n+p})\\
\nonumber
&\mathrm{Parameter\hspace{5pt}Model\hspace{5pt}5:}\hspace{5pt}\kappa_{\beta}\vert \alpha_{\beta}\sim f_{G}(a_{\beta,2},\tau_{\beta,2})I(\kappa_{\beta}>\alpha_{\beta});\\
\nonumber
&\mathrm{Parameter\hspace{5pt}Model\hspace{5pt}6:}\hspace{5pt}\kappa_{t}\vert \alpha_{t}\sim f_{G}(a_{\eta,2},\tau_{\eta,2})I(\kappa_{t}>\alpha_{t});\hspace{15pt} t = 1,\ldots, T;\\
\nonumber
&\mathrm{Parameter\hspace{5pt}Model\hspace{5pt}7:}\hspace{5pt}\kappa_{\xi,t}\vert \alpha_{\xi,t}\sim f_{G}(a_{\xi,2},\tau_{\xi,2})I(\kappa_{\xi,t}>\alpha_{\xi,t});\hspace{15pt} t = 1,\ldots, T,\\
\nonumber
&\mathrm{Parameter\hspace{5pt}Model\hspace{5pt}8:}\hspace{5pt}\alpha_{\beta}\sim f_{G}(a_{\beta,1},\tau_{\beta,1});\\
\nonumber
&\mathrm{Parameter\hspace{5pt}Model\hspace{5pt}9:}\hspace{5pt}\alpha_{t}\sim f_{G}(a_{\eta,1},\tau_{\eta,1});\hspace{15pt} t = 1,\ldots, T;\\
\label{gausummary}
&\mathrm{Parameter\hspace{5pt}Model\hspace{5pt}10:}\hspace{5pt}\alpha_{\xi,t}\sim f_{G}(a_{\xi,1},\tau_{\xi,1});\hspace{15pt} t = 1,\ldots, T,
\end{align}
\noindent
where $f_{Gau}(\cdot\vert \bm{\mu},\bm{\Sigma})$ is a Gaussian pdf with mean $\bm{\mu}$ and covariance matrix $\bm{\Sigma}$, $\bm{\zeta}_{t} =(\alpha_{t}-\kappa_{t}/2) \bm{1}_{r}$, $\bm{\zeta}_{\xi,t} =(\alpha_{\xi,t}-\kappa_{\xi,t}/2) \bm{1}_{N_{t}}$, and $\bm{\zeta}_{\beta} =(\alpha_{\beta}-\kappa_{\beta}/2) \bm{1}_{p}$. It follows from Proposition 3 that the latent MLB model in (\ref{fullsummary}) is proportional to the density in (\ref{gausummary}) after integrating out the $r\times r$ diagonal matrices $\bm{\Omega}_{t}$, the $N_{t}\times N_{t}$ diagonal matrices $\bm{\Omega}_{\xi,t}$, and the $p \times p$ diagonal matrix $\bm{\Omega}_{\beta}$.


\section*{Acknowledgments} This research was partially supported by the U.S. National Science Foundation (NSF) and the U.S. Census Bureau under NSF grant SES-1132031, funded through the NSF-Census Research Network (NCRN) program. This article is released to inform interested parties of ongoing research and to encourage discussion. The views expressed on statistical issues are those of the authors and not necessarily those of the NSF or U.S. Census Bureau.

\baselineskip=14pt \vskip 4mm\noindent

\singlespacing
\bibliographystyle{jasa}  
\bibliography{myref}

\end{document}